\documentclass[a4paper,11pt]{article}

\usepackage[latin1]{inputenc}
\usepackage{indentfirst}
\usepackage[T1]{fontenc}
\usepackage{amsmath}
\usepackage{amssymb}
\usepackage{amsfonts}
\usepackage{latexsym}
\usepackage[dvips]{graphics,graphicx,epsfig,color}
\makeatletter \@addtoreset{equation}{section}

\makeatother \makeatletter \@addtoreset{figure}{section}

\makeatother
\begin{document}
\thispagestyle{empty}
\title{\bf 2D ground motion at a soft viscoelastic layer/hard substratum
site in response to SH cylindrical seismic waves radiated by deep
and shallow line sources\\}
\author{
Jean-Philippe Groby\thanks{Laboratoire de M\'ecanique et
d'Acoustique, UPR 7051 du CNRS, 31 chemin Joseph Aiguier, 13402
Marseille cedex 20, France, ({\tt groby@lma.cnrs-mrs.fr})}
 \and Armand
Wirgin\thanks{LMA/CNRS, 31 chemin Joseph Aiguier, 13402 Marseille
cedex 20, France, ({\tt wirgin@lma.cnrs-mrs.fr})} }
\date{\today}
\maketitle
\newpage
\begin{abstract}
We show, essentially by theoretical means, that for a site with
the chosen simple geometry and mechanical properties (horizontal,
homogeneous, soft viscoelastic layer of infinite lateral extent
 overlying, and in welded contact with, a homogeneous, hard elastic substratum
 of half-infinite radial extent,
 shear-horizontal motion):
1) coupling to Love modes is all the weaker the farther the
seismic
 source (modeled as a line, assumed to lie in the substratum) is from
the lower boundary of the soft layer, 2) for a line source close
to the lower boundary of the soft layer, the ground response is
characterized by possible beating phenomena, and is of
significantly-longer duration than for excitation by cylindrical
waves radiated by deep sources. Numerical applications of the
theory show, for instance, that a line source, located 40m below
the lower boundary of a 60m thick soft layer in a hypothetical
Mexico City-like site, radiating a SH pulse of 4s duration,
produces substantial ground motion  during 200s, with marked
beating, at an epicentral distance of 3km. This response is in
some respects similar to that observed in real cities located at
soft-soil sites so that the model employed herein may help to
establish the causes and pinpoint the major contributing factors
of the devastating effects of earthquakes in such cities.
\end{abstract}
\newline
\newline
Abbreviated title: Seismic site response: a canonical problem
\newline
\newline
Corresponding author: Armand Wirgin, tel.: 33 4 91 16 40 50, fax:
33 4 91 22 08 75, e-mail: wirgin@lma.cnrs-mrs.fr
\newpage
\section{Introduction}\label{s1}
This investigation is relevant to several topics of broad interest
in seismic wave propagation:

(a) regional path effects in connection with seismic response in
urban environments \cite{cach03}, \cite{sior93}, \cite{shsi01},
\cite{shol02}, \cite{shol00}, \cite{ce04}, \cite{bako04},
\cite{pava01}, \cite{fasu94}, \cite{fapa94}

(b) effects of the underlying soil heterogeneities, lateral
variations of the underlying soil layer, and built environment on
seismic ground response at various (particularly urban) sites
\cite{chte98}, \cite{tswi03}, \cite{sedu00}, \cite{ce04},
\cite{sabo04}, \cite{robi96}, \cite{paro00a},
\cite{paro00b},\cite{bo03}, \cite{fasu94}, \cite{fapa94},
\cite{segu03}

(c) analysis of surface wave response on the ground to determine
the structure and composition of the crust \cite{chte98} and
underground fault zones \cite{igja02}, \cite{jaig02}

(d) analysis of surface wave response on the ground to identify
earthquake sources \cite{me77}, \cite{kagi81}.

Research on topic (a) was rekindled by efforts to explain some
puzzling features of the devastating Michoacan earthquake which
struck Mexico City in 1985. Other than the fact that the response
in downtown Mexico varied considerably in a spatial sense, was
quite intense and of very-long duration \cite{trbr75},
\cite{trwe76} (as much as $\sim$3 min) at certain locations, and
often took the form of a quasi-monochromatic signal with beating,
a remarkable feature of this earthquake was that such strong (in
the sense just mentioned) response could be caused by a seismic
disturbance so far (its epicenter was in the subduction zone off
the Pacific coast approximately 350 km) from the city
\cite{chba94}, \cite{chra95}, \cite{fasu94}, \cite{fapa94}. A part
of the cause of the large intensity and long-duration was
attributed in \cite{sior93} to multipathing between the source and
the site. This hypothesis was further explored in \cite{chba94},
\cite{chra95} while being associated with surface wave propagation
of the Rayleigh and Love types, presumably between the source and
the entry to the Mexico City basin, via the intervening crust.

In a rather complete (other than the neglect of attenuation) 3D
numerical study \cite{shol00}, the long duration and large
amplitude of response at various distances from subduction zone
earthquakes in Mexico were attributed to the entrapment of the
seismic disturbance in an acccretionary prism (wedge-shaped
heterogeneity) of the crust and its subsequent propagation to the
point of observation. The authors of this work later
\cite{shsi01}, \cite{shol02} stressed the role of higher-order
surface waves which propagate in the relatively-high Q layer of
the Trans Mexican Volcanic Belt (TMVB) underlying the soft clay
basin of Mexico City in addition to that of the accretionary prism
in producing large response (particularly with respect to
duration) in the city. More recently, an analysis \cite{cach03} of
seismograms recorded at various sites in central Mexico, for
earthquake sources located in the subduction zone off the Pacific
coast, have shown that the crustal structure (including that of
the TMVB, composed of low-velocity volcanic lava and tuff
overlying higher velocity limestone) between the source and
observation points "acts as a waveguide for surface waves coming
from distances greater than 200km", leading, by an unexplained
mechanism, to amplification and increase of duration of motion at
various sites, this being thought to account for at least part of
the anomalous response in Mexico City to remote seismic
disturbances. Numerical results obtained in earlier studies (e.g.,
\cite{fasu94}, \cite{fapa94}) with a rather complete, 2D hybrid
model of the propagation path between the source and the
Mexico-City basin, and of the action of the basin on the incident
wave, also stressed the important role of regional path effects on
anomalous response.

Anomalous response in other cities such as Beijing, Bucharest,
Rome, etc. has been studied in great detail, principally in
numerical manner, within the framework of the UNESCO-IGCP project
414 \cite{paro00b}, \cite{pava01}. The features of this response
were attributed to the specifics of the source parameters,
regional path effects, and the specifics of the soil distribution
and geometry in the urban basins (see next paragraph). These
findings have been substantiated in a more recent study
\cite{bo03}.

Topic (b) deals with a class of alternative or complementary
(so-called local) paradigms for explaining seismic motion in urban
sites built on soft soil. Even though the anomalous response in
1985 in Mexico City originated in a subduction zone source whose
epicentral distance was some 350 km from the city, it has been
common to seek explanations of this response (and others such as
in  Nice \cite{sedu00}, \cite{segu03} and Bucharest \cite{sabo04})
by employing models involving vertically-propagating or nearly
vertically-propagating plane waves. This requires that the focal
distance of the source to the surface be large and that the
epicentral distance from the source to the city be rather small.
Although both of these conditions are often not met in practice
(and, in particular, as concerns the 1985 Michoacan earthquake),
the vertically-propagating plane wave sollicitation usually
prevails in the theoretical/numerical studies \cite{sabo04},
\cite{bako04}, \cite{sepa04}, apparently because it simplifies the
analysis (another reason is that it facilitates comparison with
the so-called 1D model of normally-incident plane waves on a
vertically-layered half space). This has the effect of putting the
focus on what occurs in the structure vertically below the city,
namely, on the soft basin on which the earthquake-prone cities are
built. Thus, a considerable amount of studies (see \cite{chba94},
\cite{baee88}, \cite{sepa04} for comprehensive reviews) examine
the (local) effect of the soft basin on the incident wave, but at
present, it is thought that local effects account for only part of
the anomalous response \cite{chba94}, \cite{chra95},
\cite{paro00b}, \cite{pava01}, \cite{cach03}, \cite{bo03}. Another
idea that has been explored in the past few years is that the
buildings of the city, in interaction with the soft soil and with
each other, may also amplify and lengthen the duration of the
ground motion (see \cite{tswi03}, \cite{segu03} for reviews of
this subject). All of these studies (including or excluding the
buildings) point to the central role of surface waves, qualified
either as locally-launched surface (e.g., Love) waves (at the
basin edges or at heterogeneities of the soft soil) \cite{babo80},
\cite{baee88}, \cite{fasu94}, \cite{bo03} or as quasi-Love waves
(excited at the base of the buildings and re-amplified by
interaction with neighboring buildings \cite{tswi03},
\cite{grts04}, \cite{segu03}) as a possible causal agent of
anomalous response, but little \cite{wiko93}, if any, theoretical
evidence has been given to back up these assertions.

Topic (c) is classical in seismological geophysics \cite{ewja57}.
The seismic signals associated with various types of surface
(e.g., Love and Rayleigh) waves are oft-used tools for
reconstructing features of the earth's crust such as thickness,
composition (e.g., vertical layering characteristics
\cite{ewja57}, \cite{pa81}) and even lateral heterogeneities
\cite{wi88}, \cite{chte98}. More recently \cite{igja02},
\cite{jaig02}, it has been shown that seismic sources in the
neighborhood of fault zones (FZ, i.e., soft nearly-vertical layers
surrounded by relatively hard soil) excite surface waves
(qualified as "trapped") in the vicinity of the FZ which propagate
to the ground where they can be detected and used to furnish
information on the physical and geometrical characteristics of the
FZ. To treat these inverse problems in a fully unambiguous manner
requires a thorough understanding of the way the seismic source
interacts (notably how accurately one must know the position and
characteristics of the source) with the inhomogeneities.

Topic (d) is also a classical one in seismology, the main concern
being to localize and qualify (e.g., determine the moment tensor
of) earthquake sources \cite{me77}, \cite{kagi81}. As the seismic
wave, including its surface-wave components, travels laterally
(sometimes over long distances) in and along the crust before
reaching the measurement locations on the ground, the inverse
problem is difficult to solve if the crustal features (which can
include lateral heterogeneity) are not known beforehand. In any
case, it is important to determine the influence of errors of the
crustal model on the reconstruction of the source location and
moment tensor,  and to do this requires an appropriate theoretical
analysis.

The theoretical-numerical investigation herein is focused on
topics (a) and (b). In contrast to the {\it inverse-scattering}
topics (c) and (d) (to which our analysis could be applied)
wherein the response is known and the propagation medium and/or
the source are to be determined, the problem we are faced with
herein deals with {\it forward-scattering}: given the seismic
source and the characteristics of the propagation medium,
determine the response (displacement in the frequency and/or time
domain) on the ground. More specifically, we shall be concerned
with a (deceivingly-) simple canonical scattering problem: that of
a cylindrical SH pulse wave impinging on a soft homogeneous layer,
the latter being horizontal, of infinite lateral extent, bounded
above by the ground and below by an interface with a half space
filled with hard homogeneous rock. The questions we address, and
that we think can  be answered with the help of such a simple
model, are:

(i) is it possible to obtain anomalous (in the sense mentioned
above in connection with the Michoacan earthquake) response
without any lateral heterogeneity (arising from volumic inclusions
or uneveness of interfaces) in the underground medium?

(ii) what is the relation of 1D to 2D response and how adequate is
it to model the general response of the configuration by its
response to a (nearly) vertically-incident plane wave?

(iii) how does the focal distance of the source affect the
response?

(iv) how does the epicentral distance affect the response?

(v) how does the contrast of mechanical properties between the
layer and the half space affect the ground response?

(vi) how does the thickness of the layer affect the response?

(vii) how do the spectral characteristics of the incident pulse
affect the response?

It will be shown that a source radiating cylindrical waves in a
fully-elastic soft layer/hard half space medium produces a ground
response which is the sum of three terms corresponding  to various
combinations of two types of waves in the soft layer (SL) and hard
half space (HHS):

(1) standing body waves (SBW) in the SL and body waves (BW) in the
HHS,

(2) standing body waves in the SL and surface waves (SW) in the
HHS,

(3) standing surface waves (SSW) in the SL and surface waves in
the HHS.

Only type (2) waves correspond to Love modes (at the resonance
frequencies of these modes) and the conditions for optimal
excitation and maximal contribution of these modes will be
rendered explicit. It will be shown that large-duration (i.e.,
anomalous) response generally requires a preponderant contribution
of at least one (usually the lowest-order) of the Love modes to
the overall response. The type (1) waves dominate in the situation
in which the focal distance is large and do not usually produce
{\it long-duration response}, although they can produce {\it
strong (although normal) response} when the contrast of mechanical
properties between the SL and HHS is large. Beating phenomena will
be shown to be a consequence of interference between type (1) and
type (2) waves which both lead to maxima in response at nearly the
same (low) frequency. Type (3) waves turn out to have negligible
contribution to overall response. Most of these features carry
over to the case in which the layer is lossy. The practical
consequences of these results, in relation with topics (a) and
(b), will be discussed.

\section{Description of the configuration}\label{s2}
\begin{figure}
[ptb]
\begin{center}
  \includegraphics[scale=0.5] {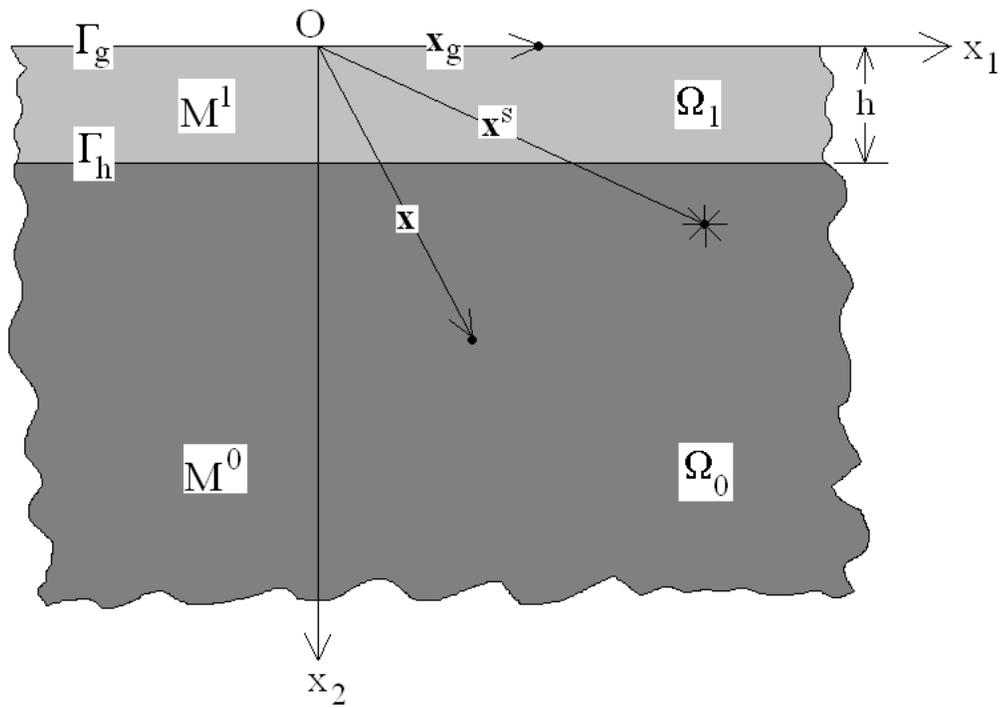}
  \caption{Cross section view of the configuration.}
  \label{wfig1}
  \end{center}
\end{figure}

Fig. \ref{wfig1} represents a cross-section (sagittal plane) view
of the site. $\Gamma_{g}$ is the ground, assumed to be flat and
horizontal, above which is located the air medium, assumed to be
the vacumn. $\Omega_{1}$ is the laterally-infinite domain occupied
by the mechanically-soft layer and $h$ is its thickness.
$\Omega_{0}$ is the semi-infinite domain (substratum) occupied by
a mechanically-hard medium, and $\Gamma_{h}$ the flat, horizontal
interface between the layer and the substratum. A
$Ox_{1}x_{2}x_{3}$ cartesian coordinate system is attached to this
configuration such that $O$ is on the ground, $x_{2}$ increases
with depth and $x_{3}$
is perpendicular to the (sagittal) plane of the figure. With
$\mathbf{i}_{j}$ the unit vector along the positive $x_{j}$ axis,
we note that the unit vectors normal to $\Gamma_{g}$ and
$\Gamma_{h}$  are $\mathbf{i}_{2}$. The media filling
$\Omega_{0}$ and  $\Omega_{1}$ are $M^{0}$ and
$M^{1}$ respectively and the latter are assumed to be initially
stress-free, linear, isotropic and homogeneous. We assume that
$M^{0}$ is non-dissipative and $M^{1}$ is generally (unless
specified otherwise) dissipative.

The seismic disturbance is delivered to the site in the form of a
shear-horizontal (SH) cylindrical pulse wave radiated by a line
source (perpendicular to the sagittal plane) located at
$\mathbf{x}^{s}:=(x_{1}^{s},x_{2}^{s})$, with, by hypothesis,
$x_{2}^{s}>h$ (i.e., $\mathbf{x}^{s}\in \Omega_{0}$). The SH
nature of this wave means that the motion associated with it is
strictly transverse (i.e., in the $x_{3}$ direction  and
independent of the $x_{3}$ coordinate). Both the SH polarization
and the invariance of the incident wave with respect to $x_{3}$
are communicated to the fields that are generated at the site in
response to the incident wave. Thus, our analysis will deal only
with the propagation of 2D SH waves (i.e., waves that depend
exclusively on the two cartesian coordinates $x_{1},~x_{2}$ and
that are associated with motion in the $x_{3}$ direction only).

We shall be concerned with a description of the elastodynamic
wavefield on the ground (i.e., on $\Gamma_{g}$) resulting from the
cylindrical seismic wave sollicitation of the site.
\section{Governing equations}\label{s3}
\subsection{Space-time domain wave equations}
In a generally-inhomogeneous, isotropic elastic or viscoelastic
medium $M$ occupying $\mathbb{R}^{3}$, the time-domain wave
equation for SH waves is:
\begin{equation}\label{w32.7}
 \nabla\cdot (\mu(\mathbf{x},\omega)\nabla u(\mathbf{x},t))-
 \rho(\mathbf{x})\partial_{t}^{2}u(\mathbf{x},t)=-\rho(\mathbf{x})
 f(\mathbf{x},t) ~,
\end{equation}
wherein $u$ is the displacement component in the $\mathbf{i}_{3}$
direction, $f$ the component of applied force density in the
$\mathbf{i}_{3}$ direction, $\mu$ the Lam\'e descriptor of
rigidity, $\rho$ the mass density, $t$ the time variable, $\omega$
the angular frequency, $\partial^{n}_{t}$ the $n-$th partial
derivative with respect to $t$, and $\mathbf{x}=(x_{1},x_{2})$.
Since our configuration involves two homogeneous media and the
applied force is assumed to be non-vanishing only in $\Omega_{0}$,
we have
\begin{equation}\label{w33.1}
 \left ( c^{m}(\omega)\right ) ^{2}\nabla\cdot\nabla u^{m}(\mathbf{x},t)-
 \partial_{t}^{2}u^{m}(\mathbf{x},t)=
 -f(\mathbf{x},t)\delta_{m0}~~;~~\mathbf{x}\in\Omega_{m}~,
\end{equation}
wherein $m$ superscripts designate the medium (0 for $M^{0}$ or 1
for $M^{1}$), $\delta_{m0}=1$ for $m=0$ and equal to zero
otherwise, and $c^{m}$ is the generally-complex velocity of shear
body waves in $M^{m}$, related to the density and rigidity by
\begin{equation}\label{w33.2}
  \left ( c^{m}(\omega)\right ) ^{2}=\frac{\mu^{m}(\omega)}{\rho^{m}}~,
\end{equation}
it being understood that $\rho^{m},\mu^{m}(\omega)~;~m=0,1$ are
constants with respect to $\mathbf{x}$.
\subsection{Space-time domain representation of the impulsive force}
In all that follows we choose the pseudo-Ricker type of impulse
line source function
\begin{multline}\label{w33.2a}
f(\mathbf{x},t)=-\delta(\mathbf{x}-\mathbf{x^{s}})
3\frac{\partial}{\partial t}
\left [ - 2 \alpha^{2} \left ( 1-2\alpha^{2} \left(t-t_{0}
\right) ^{2} \right ) e^{-\alpha^{2} \left(t-t_{0}
\right)^{2}} \right ]=
\\
\delta(\mathbf{x}-\mathbf{x^{s}})12 \alpha^{4}
\left [ -3\left(t-t_{0} \right)+2\alpha^{2} \left(t-t_{0}
\right)^{3} \right ] e^{-\alpha^{2} \left(t-t_{0}
\right)^{2}}~,
\end{multline}
wherein $\alpha=\pi \nu_{0}$ and $t_{0}=1/\nu_{0}$ and $\delta(~~)$ the
Dirac delta distribution.
\subsection{Space-frequency domain wave equations}
The frequency-domain versions of the wave equations are obtained
by expanding the force density and displacement in Fourier
integrals:
\begin{equation}\label{w33.3}
  f(\mathbf{x},t)=\int_{-\infty}^{\infty}f(\mathbf{x},\omega)e^{-i\omega t}
  d\omega~~,~~u^{m}(\mathbf{x},t)=\int_{-\infty}^{\infty}
  u^{m}(\mathbf{x},\omega)e^{-i\omega t}d\omega~,\forall t\in
  \mathbb{R}~,
\end{equation}
so as to give rise to the Helmholtz equations
\begin{multline}\label{w34.1}
 \nabla\cdot\nabla
  u^{m}(\mathbf{x},\omega)+\left ( k^{m}(\omega)\right ) ^{2}
  u^{m}(\mathbf{x},\omega)=
  \\
  -f(\mathbf{x},\omega)\delta_{m0}~~;~~\forall\mathbf{x}\in
  \Omega_{m}~~;~~m=0,1~,
\end{multline}
wherein
\begin{equation}\label{w34.2}
  k^{m}(\omega):=\frac{\omega}{c^{m}(\omega)}=
  \omega\sqrt{\frac{\rho^{m}}{\mu^{m}(\omega)}}~.
\end{equation}
is the generally-complex wavenumber in $M^{m}$. Actually, due to
the assumptions made in sect. \ref{s1}:
\begin{equation}\label{w34.3}
  k^{0}(\omega):=\frac{\omega}{c^{0}}=
  \omega\sqrt{\frac{\rho^{0}}{\mu^{0}}}~,
\end{equation}
(i.e., $k^{0}$ is real),
\begin{equation}\label{w34.3a}
  f^{m}(\mathbf{x},\omega)=S(\omega)
  \delta(\mathbf{x}-\mathbf{x^{s}})~,
\end{equation}
wherein $S(\omega)$ is the spectrum of the incident pulse. In fact, the spectrum
corresponding to the chosen (see (\ref{w33.2a}) pseudo-Ricker
impulsive force is
\begin{equation}\label{w34.3b}
S(\omega)=3\frac{i \omega ^{3}}{2 \sqrt{\pi}\alpha}
e^{i\omega t_{0}-\frac{\omega^{2}}{4\alpha^{2}}}~.
\end{equation}
\subsection{Material constants in a dissipative medium}
A word is now in order about the dissipative nature of the layer.
When a medium $M$ is lossy, the wavenumber therein is complex and
can be written (omitting, for the moment, the $\omega$ dependence)
as
\begin{equation}\label{w34.4a}
 k=k'+ik"~,
\end{equation}
where, by convention,
\begin{equation}\label{w34.4b}
 \Re k=k'\geq 0~~, \Im k=k"\geq 0~.
\end{equation}
We now refer to (\ref{w34.2}) and note that complex $k$ implies
complex $\mu$, due to the fact that it is advisable to consider
the mass density to be a real quantity. Thus, we write
\begin{equation}\label{w34.4c}
 \mu=\mu'-i\mu"~.
\end{equation}
In order to retain the positive real aspect of the rigidity for
elastic materials, we take
\begin{equation}\label{w34.4d}
 \Re\mu=\mu'\geq 0~,
\end{equation}
and inquire as to the sign of the imaginary part of $\mu$.
Introducing (\ref{w34.4c}) into (\ref{w34.2}) gives
\begin{equation}\label{w34.4e}
 k=\omega\rho^{1/2}(\mu'-i\mu")^{-1/2}=\omega
 \left ( \frac{\mu'}{\rho}\right ) ^{-1/2}
 \left [ 1-i\frac{\mu"}{\mu'}\right ] ^{-1/2}~.
\end{equation}
We assume, as is generally the case for moderately-dissipative
media, that $|\mu"/\mu'|<<1$, so that a Taylor series expansion of
$[~~]^{-1/2}$ limited to the first two terms yields
\begin{equation}\label{w34.4f}
 k=k'+ik"\approx\frac{\omega}{c'}
 \left [ 1+i\frac{\mu"}{2\mu'}\right ] ~,
\end{equation}
wherein, by definition,
\begin{equation}\label{w34.4f}
 c'=\left ( \frac{\mu'}{\rho}\right ) ^{1/2}
~.
\end{equation}
Making use of (\ref{w34.4b}) and (\ref{w34.4c}) thus necessarily
leads to
\begin{equation}\label{w34.4g}
 \Im\mu=-\mu"\leq 0.
~.
\end{equation}
We define the positive real quantity  known as the quality factor
$Q$ by the ratio
\begin{equation}\label{w34.4h}
 Q:=\frac{\mu'}{\mu"}~,
\end{equation}
and note that it is infinite for a lossless medium such as $M^{0}$
(because $\mu"=0$ in this case). Furthermore, the complex
wavenumber becomes
\begin{equation}\label{w34.4i}
 k=k'+ik"=\frac{\omega}{c'}\left ( 1+\frac{i}{2Q}\right )~,
\end{equation}
from which we find
\begin{equation}\label{w34.4j}
 Q=\frac{k'}{2k"}~.
\end{equation}
A question arises as to the proper definition of the complex
body wave velocity $c$ in $M$. We write
\begin{equation}\label{w34.4k}
 c=c'-ic"~,
\end{equation}
and require
\begin{equation}\label{w34.4l}
 \Re c=c'\geq 0~,
\end{equation}
due to the fact that the body wave velocity is positive in a
non-lossy medium. We have
\begin{equation}\label{w34.4m}
 k=k'+ik"=\frac{\omega}{c}=\frac{\omega}{c'-ic"}=
 \frac{\omega c'+i\omega c"}{|c|^{2}}~,
\end{equation}
from which we see that in order for $\Im k=k"\geq 0$, we must have
\begin{equation}\label{w34.4n}
 \Im c=-c"\leq 0~.
\end{equation}
The remaining question is that of the $\omega$-dependence of $\mu$
and $Q$ (the $\omega$-dependence of $k$ and $c$ follows from that
of $\mu$ and $Q$). In seismological applications involving
viscoelastic media the quality factor is found to be either
constant or a weakly-varying function of frequency \cite{fasu94}.
We shall assume that $Q^{1}(\omega)=Q^{1}=$const., and it can be
shown \cite{kj79} that this implies
\begin{equation}\label{w34.5a}
 \mu^{1}(\omega)=\mu^{1}_{ref} \left( \frac{-{i} \omega}
 {\omega_{ref}} \right) ^{\frac{2}{\pi}
 \arctan\left ( \frac{1}{Q^{1}}\right ) }~,
\end{equation}
wherein: $\omega_{ref}$ is a reference angular frequency, chosen
herein to be equal to $9\times 10^{-2}$Hz. Hence
\begin{equation}\label{w34.6}
 c^{1}(\omega)=c^{1}_{ref} \left( \frac{-{i} \omega}
 {\omega_{ref}} \right) ^{\frac{1}{\pi}
 \arctan\left ( \frac{1}{Q^{1}}\right ) }~,
\end{equation}
and
\begin{equation}\label{w34.7}
 c^{1}_{ref}:=\sqrt\frac{\mu^{1}_{ref}}{\rho^{1}}~.
\end{equation}
Note should be taken of the fact that even though $Q^{1}$ is
non-dispersive (i.e., does not depend on $\omega$) under the
present assumption, the phase velocity $c^{1}$ {\it is}
dispersive.
\subsection{Boundary and radiation conditions}
We assume the two media to be in {\it welded contact} so that the
displacement and the normal components of stress are continuous
across the interface $\Gamma_h$:
\begin{equation}\label{w35.4}
u^{1}(\mathbf{x},\omega)-u^{0}(\mathbf{x},\omega)~;~\mathbf{x} \in
\Gamma_{h}~,
\end{equation}
\begin{equation}\label{w35.5}
\mu^{1}(\omega)\partial_{n}
u^{1}(\mathbf{x},\omega)-\mu^{0}(\omega)\partial_{n}
u^{0}(\mathbf{x},\omega)~;~\mathbf{x} \in \Gamma_{h}.
\end{equation}
Since the air/layer interface $\Gamma_g$ (i.e., the ground) is
assumed to separate the vacumn from an elastic medium, the normal
component of stress must vanish on this boundary, i.e.,
\begin{equation}\label{w35.6}
\mu^{1}(\omega)\partial_{n} u^{1}(\mathbf{x},\omega)=0~;~
\mathbf{x} \in \Gamma_{g}~,
\end{equation}
wherein $\partial_{n}=\mathbf{i}_{2}\cdot\nabla=\partial_{x_{2}}$.
The uniqueness of the solution to the forward-scattering problem
is assured by the radiation condition in the substratum:
\begin{equation}\label{w35.7}
u^{0}(\mathbf{x},\omega)\sim ~ \text{  outgoing waves}~,~
\|\mathbf{x}\|\rightarrow \infty,~~x_{2}>h~.
\end{equation}
\subsection{Statement of the boundary-value (forward-scattering)
problem} The problem is to determine the time record of the ground
displacement field $u^{1}(\mathbf{x}_{g},t)$ (with
$\mathbf{x}_{g}:=(x_{1},0)$) from the spectrum of the ground
displacement $u^{1}(\mathbf{x}_{g},\omega)$ via the Fourier
transform
\begin{equation}\label{w35.8}
 u^{1}(\mathbf{x}_{g},t)= \int_{-\infty}^{\infty}u^{1}(\mathbf{x}_{g},\omega)
 e^{-i\omega t}
  d\omega~.
\end{equation}
Note that due to the fact that  $u^{1}(\mathbf{x}_{g},t)$ is a
real function, we must have
\begin{equation}\label{w35.9}
 \left [ u^{1}(\mathbf{x}_{g},\omega)\right ] ^{*}=
 u^{1}(\mathbf{x}_{g},-\omega)~,
\end{equation}
(wherein the symbol * designates the complex conjugate operator)
from which it follows that
\begin{equation}\label{w35.10}
 u^{1}(\mathbf{x}_{g},t)= 2\Re\int_{0}^{\infty}u^{1}(\mathbf{x}_{g},\omega)
 e^{-i\omega t}
  d\omega~.
\end{equation}
\section{Exact solutions in the frequency domain by separation of
variables}\label{s4}
\subsection{Preliminaries}
Although the material in this section (4) is classical as regards
the way of obtaining plane wave integral representations of the
fields, the way these integrals are decomposed, analyzed, and
computed  is  different than in previous investigations (e.g.,
\cite{ha64},\cite{beha64},\cite{kn64},\cite{ha70},\cite{sc70},
\cite{he74},\cite{bo82},\cite{aplu83},\cite{luap83},\cite{ke83},
\cite{sc84},\cite{pa85},
\cite{drmo88},\cite{flsu91},\cite{zhch03}; the considerable
quantity and variety of these publications attests to the richness
and importance of the subject, and to the fact that certain
features of the latter certainly remain to be discovered). In a
first subclass of these investigations, the plane wave integrals
(with the horizontal wavenumber as the variable of integration)
are reduced to residue series (so-called modal series) plus branch
cut integrals which are usually neglected if the
source-to-observation point is large compared to the wavelength.
In a second subclass of the aforementioned investigations, various
devices are employed to evaluate in numerically-efficient,
accurate, or asymptotic manner the plane wave integrals. Our
contribution is essentially of the second variety, but numerical
efficiency (more important in the inverse problem context) is of
less interest to us than the physical significance of the terms
entering into our choice of the decomposition of the integrals.
\subsection{Frequency-domain solutions in the absence of the layer and the free
surface}\label{ss41} In the absence of the layer and the free
surface, the problem is that of the radiation of a SH wave from a
line source in  2D free space ($\mathbb{R}^{2}$) occupied by the
homogeneous medium $M^{0}$. We term this radiated wave the
'incident wave' and designate it by $u^{i}$.

By applying separation of variables in the cartesian coordinate
system to the Helmholtz equation and using the radiation
condition, it can be shown that $u^{i}$ takes the form
\cite{mofe53}
\begin{equation}\label{w41.1}
  u^{i}(\mathbf{x},\omega)=\frac{i}{4\pi}S(\omega)\int_{-\infty}^{\infty}
  e^{i\left [
  k_{1}(x_{1}-x_{1}^{s}+k_{2}^{0}(\omega)|x_{2}-x_{2}^{s}|\right
  ]}\frac{dk_{1}}{k_{2}^{0}}~,
\end{equation}
or
\begin{equation}\label{w41.4}
  u^{i}(\mathbf{x},\omega)=\frac{i}{4}S(\omega)
  H_{0}^{(1)}\left ( k^{0}(\omega)\|\mathbf{x}-\mathbf{x^{s}}\|\right ) ~,
\end{equation}
wherein $ H_{0}^{(1)}(~~)$ is the zeroth-order Hankel function of
the first kind and:
\begin{equation}\label{w41.3}
 k_{2}^{j}(\omega):=\sqrt{\left ( k^{j}(\omega)\right )
 ^{2}-k_{1}^{2}}~~,~~\Re k_{2}^{j}(\omega)\geq 0~,~\Im
 k_{2}^{j}(\omega)\geq 0~~,~~j=0,1~.
\end{equation}
We shall make use in sect. \ref{ss44} of the form taken by $u^{i}$
in the region $\Omega_{0}^{-}:=\{x_{2}^{s}>x_{2}>h~;~\forall
x_{1}\in \mathbb{R}\}$:
\begin{equation}\label{w41.1}
  u^{i}(\mathbf{x},\omega)=\int_{-\infty}^{\infty}A^{0}(k_{1},\omega)
  e^{i\left [
  k_{1}x_{1}-k_{2}^{0}(\omega)x_{2}\right ] }dk_{1}~~;~~
  \forall \mathbf{x}\in\Omega_{0}^{-}~,
\end{equation}
wherein
\begin{equation}\label{w41.7}
  A^{0}(k_{1},\omega)=S(\omega)\frac{i}{4\pi k_{2}^{0}(\omega)}e^{-i\left [
  k_{1}x_{1}^{s}-k_{2}^{0}(\omega)x_{2}^{s}\right ] }~.
\end{equation}
\subsection{Field representations in cartesian
coordinates for the configuration including the layer and the free
surface}\label{ss42} When the layer and free surface are present,
the incident field described in the previous section cannot
proceed in unobstructed manner, i.e., it gives rise to a
'diffracted' field (indicated by the superscript '$d$') so that by
re-use of separation of variables in cartesian coordinates and the
radiation condition we are led to represent the total fields in
the substrate and the layer by
\begin{equation}\label{w42.1}
 u^{0}(\mathbf{x},\omega)=u^{i}(\mathbf{x},\omega)+u^{0d}(\mathbf{x},\omega)~,
\end{equation}
\begin{equation}\label{w42.2}
 u^{1}(\mathbf{x},\omega)=u^{1d}(\mathbf{x},\omega)~.
\end{equation}
wherein:
\begin{equation}\label{w42.3}
  u^{0d}(\mathbf{x},\omega)=\int_{-\infty}^{\infty}B^{0}(k_{1},\omega)
  e^{i\left [
  k_{1}x_{1}+k_{2}^{0}(\omega)(x_{2}-h)\right
  ]}dk_{1}~;~~x_{2}>h~,~\forall x_{1}\in \mathbb{R},
\end{equation}
\begin{multline}\label{w42.4}
  u^{1d}(\mathbf{x},\omega)=
  \\
  \int_{-\infty}^{\infty}\left ( A^{1}(k_{1},\omega)
  e^{i\left [
  k_{1}x_{1}-k_{2}^{1}(\omega)x_{2}\right
  ]}+B^{1}(k_{1},\omega)
  e^{i\left [
  k_{1}x_{1}+k_{2}^{1}(\omega)x_{2})\right
  ]}\right ) dk_{1}~;
  \\
  ~~0<x_{2}<h~,~\forall x_{1}\in \mathbb{R}~,
\end{multline}
it being understood that the diffraction coefficients $B^{0}$,
$A^{1}$, $B^{1}$ are, as yet, undetermined.
\subsection{Determination of the diffraction coefficients  and
frequency domain fields by application of the boundary
conditions}\label{ss44} The free-surface boundary condition
entails:
\begin{equation}\label{w44.2}
 A^{1}(k_{1},\omega)=
 B^{1}(k_{1},\omega)~,~\forall k_{1}\in \mathbb{R}~,
\end{equation}
whence
\begin{equation}\label{w44.3}
  u^{1d}(\mathbf{x},\omega)=2\int_{-\infty}^{\infty}A^{1}(k_{1},\omega)
  e^{i
  k_{1}x_{1}}\cos(k_{2}^{1}(\omega)x_{2})dk_{1}~;~~0<x_{2}<h~,
  ~\forall x_{1}\in \mathbb{R}.
\end{equation}
The continuity of displacement condition leads to:
\begin{equation}\label{w44.7}
 B^{0}(k_{1},\omega)-2A^{1}(k_{1},\omega)
  \cos(k_{2}^{1}(\omega)h) = -A^{0}(k_{1},\omega)
  e^{-ik_{2}^{0}(\omega)h}~;~\forall k_{1}\in \mathbb{R}~,
\end{equation}
whereas the continuity of normal stress boundary condition
implies:
\begin{multline}\label{w44.8}
i\mu^{0}(\omega)k_{2}^{0}(\omega)B^{0}(k_{1},\omega) +
  2\mu^{1}k_{2}^{1}(\omega)
  A^{1}(k_{1},\omega)\sin(k_{2}^{1}(\omega)h)=
  \\
  i\mu^{0}(\omega)k_{2}^{0}(\omega)A^{0}(k_{1},\omega)
  e^{-ik_{2}^{0}(\omega)h}~;~\forall k_{1}\in \mathbb{R}~.
\end{multline}
The solution of this linear system of equations is:
\begin{multline}\label{w44.9}
 B^{0}(k_{1},\omega)=A^{0}(k_{1},\omega)e^{-ik_{2}^{0}(\omega)h}\times
 \\
\left (
\frac{-\mu^{1}(\omega)k_{2}^{1}(\omega)\sin(k_{2}^{1}(\omega)h)+
i\mu^{0}(\omega)k_{2}^{0}(\omega)\cos(k_{2}^{1}(\omega)h)}
{\mu^{1}(\omega)k_{2}^{1}(\omega)\sin(k_{2}^{1}(\omega)h)+
i\mu^{0}(\omega)k_{2}^{0}(\omega)\cos(k_{2}^{1}(\omega)h)}\right )
  ~;~\forall k_{1}\in \mathbb{R}~,
\end{multline}
\begin{multline}\label{w44.10}
 A^{1}(k_{1},\omega)=\frac{A^{0}(k_{1},\omega)}{2}
 e^{-ik_{2}^{0}(\omega)h}\times
 \\
\left ( \frac{2i\mu^{0}(\omega)k_{2}^{0}(\omega)}
{\mu^{1}(\omega)k_{2}^{1}(\omega)\sin(k_{2}^{1}(\omega)h)+
i\mu^{0}(\omega)k_{2}^{0}(\omega)\cos(k_{2}^{1}(\omega)h)}\right )
  ~;~\forall k_{1}\in \mathbb{R}~.
\end{multline}
so that the solutions for the fields in the frequency domain are:
\begin{multline}\label{w47.1}
  u^{0}(\mathbf{x},\omega)=S(\omega)\int_{-\infty}^{\infty}
  \frac{i}{4\pi k_{2}^{0}(\omega)} e^{i\left [
  k_{1}\left ( x_{1}-x_{1}^{s}\right ) +k_{2}^{0}(\omega)\left | x_{2}-
  x_{2}^{s}\right | \right
  ]}dk_{1}+
  \\
  S(\omega)\int_{-\infty}^{\infty}
  \frac{i}{4\pi k_{2}^{0}(\omega)}
\left [
\frac{i\mu^{0}(\omega)k_{2}^{0}(\omega)\cos(k_{2}^{1}(\omega)h)-
\mu^{1}(\omega)k_{2}^{1}(\omega)\sin(k_{2}^{1}(\omega)h)} {
i\mu^{0}(\omega)k_{2}^{0}(\omega)\cos(k_{2}^{1}(\omega)h)+
\mu^{1}(\omega)k_{2}^{1}(\omega)\sin(k_{2}^{1}(\omega)h)}\right ]
\times
\\
e^{i\left [ k_{1}\left ( x_{1}-x_{1}^{s}\right )
+k_{2}^{0}(\omega)\left ( x_{2}+x_{2}^{s}-2h\right ) \right
 ] }dk_{1}
  ~;~\forall \mathbf{x}\in \Omega_{0}~,
\end{multline}
\begin{multline}\label{w47.2}
  u^{1}(\mathbf{x},\omega)=u^{1d}(\mathbf{x},\omega)=
  \\
  S(\omega)\int_{-\infty}^{\infty}\frac{i}{4\pi k_{2}^{0}(\omega)}
\left [ \frac{2i\mu^{0}(\omega)k_{2}^{0}(\omega)} {
i\mu^{0}(\omega)k_{2}^{0}(\omega)\cos( k_{2}^{1}(\omega)h) +
\mu^{1}(\omega)k_{2}^{1}(\omega)\sin( k_{2}^{1}(\omega)h) }\right
] \times
\\
 \cos\left ( k_{2}^{1}(\omega)x_{2}\right ) e^{i\left [
k_{1}\left ( x_{1}-x_{1}^{s}\right ) -k_{2}^{0}(\omega)\left (
h-x_{2}^{s}\right ) \right ] }dk_{1}
  ~;~\forall \mathbf{x}\in \Omega_{1}~.
\end{multline}
Finally, the frequency-domain ground response takes the form:
\begin{multline}\label{w47.3}
  u^{1}(\mathbf{x}_{g},\omega)=
  \\
  S(\omega)\int_{-\infty}^{\infty}\frac{i}{4\pi k_{2}^{0}(\omega)}
\left [ \frac{2i\mu^{0}(\omega)k_{2}^{0}(\omega)} {
i\mu^{0}(\omega)k_{2}^{0}(\omega)\cos( k_{2}^{1}(\omega)h) +
\mu^{1}(\omega)k_{2}^{1}(\omega)\sin( k_{2}^{1}(\omega)h) }\right
] \times
\\
 e^{i\left [
k_{1}\left ( x_{1}-x_{1}^{s}\right ) -k_{2}^{0}(\omega)\left (
h-x_{2}^{s}\right ) \right ] }dk_{1}
  ~;~\forall \mathbf{x}\in \Omega_{1}~.
\end{multline}
\section{Structure of the frequency-domain response in the case of
a non-lossy layer}\label{s5}
\subsection{Frequency domain response in the layer}\label{ss51}
When the layer is free of dissipation, i.e., elastic, then
$\mu^{1}$ is real and does not depend on $\omega$, and
$k^{1}(\omega)$ is real (recall that we assumed the substratum to
be elastic, which means that $\mu^{0}$ is real and does not depend
on $\omega$, and $k^{0}(\omega)$ is real also). Consequently, in
the integrals of the previous section we encounter intervals of
$k_{1}$ over which $k_{2}^{0}$ and $k_{2}^{1}$ are either purely
real or purely imaginary:
\begin{equation}\label{w48.1}
  k_{2}^{j}(\omega)=K_{2}^{j}(\omega):=\left | \sqrt{\left (
  k^{j}(\omega)\right ) ^{2}-k_{1}^{2}}\right | ~~;~~|k_{1}|\leq
  k^{j}(\omega)~;~\omega\geq 0~,
\end{equation}
\begin{equation}\label{w48.1a}
  k_{2}^{j}(\omega)=i\kappa_{2}^{j}(\omega):=i\left |
  \sqrt{k_{1}^{2}-\left ( k^{j}(\omega)^{2}\right ) }\right | ~~;~~|k_{1}|\geq
  k^{j}(\omega)~;~\omega\geq 0~,
\end{equation}
It is important to note that the terms 'soft layer' and
(relatively) 'hard substratum' have the following meaning in the
present context:
\begin{equation}\label{w48.2}
  c^{0}(\omega)>c^{1}(\omega)~~\Rightarrow~~k^{0}(\omega)<k^{1}(\omega)~,
\end{equation}
\begin{equation}\label{w48.2a}
  \mu^{0}>\mu^{1}~,
\end{equation}
so that (\ref{w47.2}) can be expressed as:
\begin{equation}\label{w48.3}
  u^{1}(\mathbf{x},\omega)=I^{1}_{1}(\mathbf{x},\omega)+
  I^{1}_{2}(\mathbf{x},\omega)+I^{1}_{3}(\mathbf{x},\omega)
  ~;~\forall \mathbf{x}\in \Omega_{1}~,
\end{equation}
with:
\begin{multline}\label{w48.3a}
  I^{1}_{1}(\mathbf{x},\omega)=\int_{-k^{0}}^{k^{0}}
  du^{1}_{1}(\mathbf{x},\mathbf{x}_{g},k_{1},\omega)=
  \\
  -\frac{S(\omega)}{2\pi}\int_{-k^{0}}^{k^{0}}
  F^{1}_{1}(k_{1},\omega)
  \cos\left ( K_{2}^{1}(\omega)x_{2}\right )
  e^{i\left [
k_{1}\left ( x_{1}-x_{1}^{s}\right ) -K_{2}^{0}(\omega)\left (
h-x_{2}^{s}\right ) \right ] }dk_{1}~,
\end{multline}
\begin{equation}\label{w48.4}
  F^{1}_{1}(k_{1},\omega)=\frac{\mu^{0}} {
i\mu^{0}K_{2}^{0}(\omega)\cos( K_{2}^{1}(\omega)h) +
\mu^{1}K_{2}^{1}(\omega)\sin( K_{2}^{1}(\omega)h)}~,
\end{equation}
\begin{multline}\label{w48.5}
  I^{1}_{2}(\mathbf{x},\omega)=\left [
  \int_{-k^{1}}^{-k^{0}}+\int_{k^{0}}^{k^{1}}\right ]
  du^{1}_{2}(\mathbf{x},\mathbf{x}_{g},k_{1},\omega)=
  \\
  -\frac{S(\omega)}{2\pi}\left [
  \int_{-k^{1}}^{-k^{0}}+\int_{k^{0}}^{k^{1}}\right ]
  F^{1}_{2}(k_{1},\omega)
  \cos\left ( K_{2}^{1}(\omega)x_{2}\right )\times
  \\
  e^{\left [
ik_{1}\left ( x_{1}-x_{1}^{s}\right ) +\kappa_{2}^{0}(\omega)\left
( h-x_{2}^{s}\right ) \right ] }dk_{1}~,
\end{multline}
\begin{equation}\label{w48.6}
  F^{1}_{2}(k_{1},\omega)=\frac{\mu^{0}} {
-\mu^{0}\kappa_{2}^{0}(\omega)\cos( K_{2}^{1}(\omega)h) +
\mu^{1}K_{2}^{1}(\omega)\sin( K_{2}^{1}(\omega)h)}~,
\end{equation}
\begin{multline}\label{w48.7}
  I^{1}_{3}(\mathbf{x},\omega)=\left [ \int_{-\infty}^{-k^{1}}+
  \int_{k^{1}}^{\infty}\right ]
  du^{1}_{3} (\mathbf{x},\mathbf{x}_{g},k_{1},\omega)=
  \\
  -\frac{S(\omega)}{2\pi}\left [
  \int_{-\infty}^{-k^{1}}+\int_{k^{1}}^{\infty}\right ]
  F^{1}_{3}(\mathbf{x}_{g},k_{1},\omega)
  \cosh\left ( \kappa_{2}^{1}
  (\omega)x_{2}\right )\times
  \\
  e^{\left [
ik_{1}\left ( x_{1}-x_{1}^{s}\right ) +\kappa_{2}^{0}(\omega)\left
( h-x_{2}^{s}\right ) \right ] }dk_{1}~,
\end{multline}
\begin{equation}\label{w48.8}
  F^{1}_{3}(k_{1},\omega)=\frac{-\mu^{0}} {
\mu^{0}\kappa_{2}^{0}(\omega)\cosh( \kappa_{2}^{1}(\omega)h) +
\mu^{1}\kappa_{2}^{1}(\omega)\sinh( \kappa_{2}^{1}(\omega)h) }~.
\end{equation}
We write:
\begin{multline}\label{w48.9}
  du^{1}_{1}(\mathbf{x},\mathbf{x}_{g},k_{1},\omega)=
  dG^{1}_{1}(\mathbf{x}_{g},k_{1},\omega)
  e^{i\left [ k_{1}x_{1}+K_{2}^{1}(\omega)x_{2}\right ]}+
  \\
  dG^{1}_{1}(\mathbf{x}_{g},k_{1},\omega)
  e^{i\left [ k_{1}x_{1}-K_{2}^{1}(\omega)x_{2}\right ]}~,
\end{multline}
\begin{equation}\label{w48.10}
dG^{1}_{1}(\mathbf{x}_{g},k_{1},\omega)=-\frac{S(\omega)}{\pi}F^{1}_{1}(k_{1},\omega)
  e^{-i\left [
k_{1}x_{1}^{s}+K_{2}^{0}(\omega)\left ( h-x_{2}^{s}\right ) \right
] }dk_{1} ~,
\end{equation}
which, together with (\ref{w48.3a}), express the fact that a part
(i.e., $I^{1}_{1}$) of the field in the layer is composed of a sum
of standing body waves (SBW), each of which is the sum of two
plane body waves having wavevectors with the same length.

In the same manner, we write:
\begin{multline}\label{w48.11}
  du^{1}_{2}(\mathbf{x},\mathbf{x}_{g},k_{1},\omega)=
  dG^{1}_{2}(\mathbf{x}_{g},k_{1},\omega)
  e^{i\left [ k_{1}x_{1}+K_{2}^{1}(\omega)x_{2}\right ] }+
  \\
  dG^{1}_{2}(\mathbf{x}_{g},k_{1},\omega)
  e^{i\left [ k_{1}x_{1}-K_{2}^{1}(\omega)x_{2}\right ] }~,
\end{multline}
\begin{equation}\label{w48.12}
dG^{1}_{2}(\mathbf{x}_{g},k_{1},\omega)=
-\frac{S(\omega)}{\pi}F^{1}_{2}(k_{1},\omega)
  e^{-ik_{1}x_{1}^{s}+\kappa_{2}^{0}(\omega)
  \left ( h-x_{2}^{s}\right ) }dk_{1} ~,
\end{equation}
which, together with (\ref{w48.5}), express the fact that another
part (i.e., $I^{1}_{2}$) of the field in the layer is again
composed of a sum of standing body waves, each of which is the sum
of two plane body waves with wavevectors having the same length.
Note however that neither the wavevectors nor the amplitudes of
these SBW are the same as those of the SBW (henceforth termed
SBW1) in $I^{1}_{1}$ (because the range of integration in the
latter is different from that in $I^{1}_{2}$). In fact,
(\ref{w48.12}) tells us that the amplitudes $dG^{1}_{2}$ of the
SBW in $I^{1}_{2}$ (henceforth termed SBW2) decrease exponentially
as the focal distance (i.e., $x_{2}^{s}$) increases, so that {\it
the SBW2 make themselves felt all the less the farther the source
is (in the vertical direction) from the ground}. On the other
hand, the amplitudes of the SBW1 are sinusoidal functions of focal
distance, so that {\it the SBW1 can possibly make themselves felt
strongly for a large variety of source locations}.

Finally, we write:
\begin{multline}\label{w48.13}
  du^{1}_{3}(\mathbf{x},\mathbf{x}_{g},k_{1},\omega)=
  dG^{1}_{3}(\mathbf{x}_{g},k_{1},\omega)
  e^{ik_{1}x_{1}+\kappa_{2}^{1}(\omega)x_{2}}+
  \\
  dG^{1}_{3}(\mathbf{x}_{g},k_{1},\omega)
  e^{ik_{1}x_{1}-\kappa_{2}^{1}(\omega)x_{2}}~,
\end{multline}
\begin{equation}\label{w48.14}
dG^{1}_{3}(\mathbf{x}_{g},k_{1},\omega)=
-\frac{S(\omega)}{\pi}F^{1}_{3}(k_{1},\omega)
  e^{-ik_{1}x_{1}^{s}+\kappa_{2}^{0}(\omega)
  \left ( h-x_{2}^{s}\right ) }dk_{1} ~,
\end{equation}
which, together with (\ref{w48.7}), express the fact that the
third part (i.e., $I^{1}_{3}$) of the field in the layer is
composed of a sum of standing surface waves (SSW), each of which
is the sum of two plane  surface waves with wavevectors having the
same length (note that each such plane surface wave is an
inhomogeneous wave (with complex wavevector) whose phase is
constant on $x_{1}=$const. and whose amplitude either increases or
decreases as $x_{2}$ approaches some horizontal surface
$x_{2}=$const.). Eq. (\ref{w48.14}) tells us that the amplitudes
$dG^{1}_{3}$ of the SSW in $I^{1}_{3}$ decrease exponentially as
the focal distance increases, so that {\it the SSW make themselves
felt all the less the farther the source is (in the vertical
direction) from the ground}.

The main conclusion of this discussion is that for focal distances
of the source that are sufficiently large, the field in the layer
is essentially given by $I^{1}_{1}$ and is expressed by a sum of
SBW1. This corresponds more or less to the situation in the
quasi-1D analysis of the forward-scattering problem, but, as we
shall see further on, it is, by no means, a valid picture of the
response of the configuration when the focal distance of the
source is not large.
\subsection{Frequency domain response in the hard half space}\label{ss52}
We shall concentrate our attention exclusively on the diffracted
field in the subdomain $\Omega_{0}^{-}$ although the essence of
what will be written applies to the whole half space $\Omega_{0}$
. Proceeding as in sect.\ref{ss51} we find:
\begin{equation}\label{w52.1}
  u^{0d}(\mathbf{x},\omega)=I^{0}_{1}(\mathbf{x},\omega)+
  I^{0}_{2}(\mathbf{x},\omega)+I^{0}_{3}(\mathbf{x},\omega)
  ~;~\forall \mathbf{x}\in \Omega_{0}~,
\end{equation}
with:
\begin{multline}\label{w52.2}
  I^{0}_{1}(\mathbf{x},\omega)=\int_{-k^{0}}^{k^{0}}
  du^{0}_{1}(\mathbf{x},\mathbf{x}_{g},k_{1},\omega)=
  \\
  \frac{S(\omega)}{4\pi}\int_{-k^{0}}^{k^{0}}
  F^{0}_{1}(k_{1},\omega)
  e^{i\left [
k_{1}\left ( x_{1}-x_{1}^{s}\right ) +K_{2}^{0}(\omega)\left (
x_{2}+x_{2}^{s}-2h\right ) \right ] }dk_{1}~,
\end{multline}
\begin{equation}\label{w52.3}
  F^{0}_{1}(k_{1},\omega)=\frac{i}{K_{2}^{0}(\omega)}
  \frac{i\mu^{0}K_{2}^{0}(\omega)\cos( K_{2}^{1}(\omega)h)
  -
\mu^{1}K_{2}^{1}(\omega)\sin( K_{2}^{1}(\omega)h)} {
i\mu^{0}K_{2}^{0}(\omega)\cos( K_{2}^{1}(\omega)h) +
\mu^{1}K_{2}^{1}(\omega)\sin( K_{2}^{1}(\omega)h)}~,
\end{equation}
\begin{multline}\label{w52.4}
  I^{0}_{2}(\mathbf{x},\omega)=\left [
  \int_{-k^{1}}^{-k^{0}}+\int_{k^{0}}^{k^{1}}\right ]
  du^{0}_{2}(\mathbf{x},\mathbf{x}_{g},k_{1},\omega)=
  \\
  \frac{S(\omega)}{4\pi}\left [
  \int_{-k^{1}}^{-k^{0}}+\int_{k^{0}}^{k^{1}}\right ]
  F^{0}_{2}(k_{1},\omega)
  e^{\left [
ik_{1}\left ( x_{1}-x_{1}^{s}\right ) -\kappa_{2}^{0}(\omega)\left
( x_{2}+x_{2}^{s}-2h\right ) \right ] }dk_{1}~,
\end{multline}
\begin{equation}\label{w52.5}
  F^{0}_{2}(k_{1},\omega)=\frac{1}{\kappa_{2}^{0}(\omega)}
  \frac{-\mu^{0}\kappa_{2}^{0}(\omega)\cos( K_{2}^{1}(\omega)h)
  -
\mu^{1}K_{2}^{1}(\omega)\sin( K_{2}^{1}(\omega)h)} {
-\mu^{0}\kappa_{2}^{0}(\omega)\cos( K_{2}^{1}(\omega)h) +
\mu^{1}K_{2}^{1}(\omega)\sin( K_{2}^{1}(\omega)h)}~,
\end{equation}
\begin{multline}\label{w52.6}
  I^{0}_{3}(\mathbf{x},\omega)=\left [ \int_{-\infty}^{-k^{1}}+
  \int_{k^{1}}^{\infty}\right ]
  du^{0}_{3} (\mathbf{x},\mathbf{x}_{g},k_{1},\omega)=
  \\
\frac{S(\omega)}{4\pi}\left [
  \int_{-k^{1}}^{-k^{0}}+\int_{k^{0}}^{k^{1}}\right ]
  F^{0}_{3}(k_{1},\omega)
  e^{\left [
ik_{1}\left ( x_{1}-x_{1}^{s}\right ) -\kappa_{2}^{0}(\omega)\left
( x_{2}+x_{2}^{s}-2h\right ) \right ] }dk_{1}~,
\end{multline}
\begin{equation}\label{w52.7}
  F^{0}_{3}(k_{1},\omega)=\frac{1}{\kappa_{2}^{0}(\omega)}
  \frac{-\mu^{0}\kappa_{2}^{0}(\omega)\cosh( \kappa_{2}^{1}(\omega)h)
  +
\mu^{1}\kappa_{2}^{1}(\omega)\sinh( \kappa_{2}^{1}(\omega)h)} {
-\mu^{0}\kappa_{2}^{0}(\omega)\cosh( \kappa_{2}^{1}(\omega)h) -
\mu^{1}\kappa_{2}^{1}(\omega)\sinh( \kappa_{2}^{1}(\omega)h)}~.
\end{equation}
We write:
\begin{equation}\label{w52.8}
  u^{0}_{1}(\mathbf{x},\mathbf{x}_{g},k_{1},\omega)=
  dG^{0}_{1}(\mathbf{x}_{g},k_{1},\omega)
  e^{i\left [ k_{1}x_{1}+K_{2}^{0}(\omega)x_{2}\right ]}~,
\end{equation}
\begin{equation}\label{w52.9}
dG^{0}_{1}(\mathbf{x}_{g},k_{1},\omega)=\frac{S(\omega)}
{4\pi}F^{0}_{1}(k_{1},\omega)
  e^{-i\left [
k_{1}x_{1}^{s}-K_{2}^{0}(\omega)\left ( x_{2}^{s}-2h\right )
\right ] }dk_{1} ~,
\end{equation}
which, together with (\ref{w52.2}), express the fact that a part
(i.e., $I^{0}_{1}$) of the diffracted field in the half space is
composed of a sum of plane body waves (BW). Thus, to each
horizontal wavenumber $k_{1}$ in the interval $[-k^{0},k^{0}]$,
correspond a SBW1 in $\Omega_{1}$ and a BW in $\Omega_{0}^{-}$.

In the same manner, we write:
\begin{equation}\label{w52.10}
  du^{0}_{2}(\mathbf{x},\mathbf{x}_{g},k_{1},\omega)=
  dG^{0}_{2}(\mathbf{x}_{g},k_{1},\omega)
  e^{ik_{1}x_{1}-\kappa_{2}^{0}(\omega)x_{2}}~,
\end{equation}
\begin{equation}\label{w52.11}
dG^{0}_{2}(\mathbf{x}_{g},k_{1},\omega)=
\frac{S(\omega)}{4\pi}F^{0}_{2}(k_{1},\omega)
  e^{-ik_{1}x_{1}^{s}-\kappa_{2}^{0}(\omega)
  \left ( x_{2}^{s}-2h\right ) }dk_{1} ~,
\end{equation}
which, together with (\ref{w52.4}), express the fact that another
part (i.e., $I^{0}_{2}$) of the diffracted field in the half space
is composed of a sum of plane surface waves (SW), henceforth
denoted by SW2. Eq. (\ref{w52.11}) tells us that the amplitudes
$dG^{0}_{2}$ of the SW2 in $I^{0}_{2}$ decrease exponentially as
the focal distance  increases, so that the SW2 make themselves
felt all the less the farther the source is (in the vertical
direction) from the ground. On the other hand, the amplitudes of
the BW  in $I^{0}_{1}$ are sinusoidal functions of focal distance,
so that these BW can make themselves felt strongly for a large
variety of source locations. In addition, we note that to each
horizontal wavenumber $k_{1}$ in the intervals $[-k^{1},-k^{0}]$
and $[k^{0},k^{1}]$, correspond a SBW2 in $\Omega_{1}$ and a SW2
in $\Omega_{0}^{-}$.

Finally, we write:
\begin{equation}\label{w52.12}
  du^{0}_{3}(\mathbf{x},\mathbf{x}_{g},k_{1},\omega)=
  dG^{0}_{3}(\mathbf{x}_{g},k_{1},\omega)
  e^{ik_{1}x_{1}-\kappa_{2}^{0}(\omega)x_{2}}~,
\end{equation}
\begin{equation}\label{w52.13}
dG^{0}_{3}(\mathbf{x}_{g},k_{1},\omega)=
\frac{S(\omega)}{4\pi}F^{1}_{3}(k_{1},\omega)
  e^{-ik_{1}x_{1}^{s}-\kappa_{2}^{0}(\omega)
  \left ( x_{2}^{s}-2h\right ) }dk_{1} ~,
\end{equation}
which, together with (\ref{w52.6}), express the fact that the
third part (i.e., $I^{0}_{3}$) of the diffracted field in the
substratum is composed of a sum of plane surface waves (henceforth
denoted by SW3). Eq. (\ref{w52.13}) tells us that the amplitudes
$dG^{0}_{3}$ of the SW3 in $I^{0}_{3}$ decrease exponentially as
the focal distance (i.e., $h+x_{2}^{s}$) increases, so that the
SW3 make themselves felt all the less the farther the source is
(in the vertical direction) from the ground. Note however, that
the wavevectors associated with the SW3 are not identical to those
associated with the SW2 because $k_{1}$ spans an interval in
$I^{0}_{3}$ that is different from the one in $I^{0}_{2}$. In
addition, we note that to each horizontal wavenumber $k_{1}$ in
the intervals $]-\infty,-k^{1}]$ and $[k^{1},\infty[$, correspond
a SSW in $\Omega_{1}$ and a SW3 in $\Omega_{0}^{-}$.

The main conclusion of this discussion is that for focal distances
of the source that are sufficiently large, the diffracted field in
the half space is essentially given by $I^{1}_{1}$ and is
expressed by a sum of BW. This corresponds more or less to the
situation in the quasi-1D analysis of the forward-scattering
problem, but, as we shall see further on, it is, by no means, a
valid picture of the response of the configuration when the focal
distance of the source is not large.
\subsection{Amplitudes of the SBW1}\label{ss53}
Henceforth, we restrict our attention to the field in the soft
layer, and, in particular, to the three individual types of
standing waves (SBW1, SBW2, SSW) of which it is composed. Here, we
focus on a generic SBW1 and note that its amplitude $dG_{1}^{1}$
is the product of three factors: the factor $S(\omega)$ associated
with the spectrum of the incident pulse, a geometric factor
associated with the location of the source (whose influence was
already discussed), and a so-called {\it interference factor}
$F_{1}^{1}dk_{1}$. We first discuss $F_{1}^{1}dk_{1}$ and then
close the discussion with some remarks on $S(\omega)$.

We rewrite $I_{1}^{1}$ as
\begin{multline}\label{w53.1}
 I^{1}_{1}(\mathbf{x},\omega)=
 \\
 \frac{-S(\omega)}{\pi}\int_{0}^{k^{0}}
 F^{1}_{1}(k_{1},\omega)\cos\left ( k_{1}(x_{1}-x_{1}^{s})\right )
 \cos\left ( K_{2}^{1}(\omega)x_{2}\right )
 e^{-iK_{2}^{0}(\omega)(h-x_{2}^{s})}dk_{1}~,
\end{multline}
and are therefore interested in
\begin{multline}\label{w53.1a}
  F^{1}_{1}(k_{1},\omega)dk_{1}=\frac{dk_{1}}
{iK_{2}^{0}(\omega)\cos( K_{2}^{1}(\omega)h) +
\frac{\mu^{1}}{\mu^{0}}K_{2}^{1}(\omega)\sin(
K_{2}^{1}(\omega)h)}~~;
\\
k_{1}\in\left [ 0,k^{0}\right ] ~.
\end{multline}
We make the change of variables
\begin{equation}\label{w53.2}
 \eta=k^{0}h=h\frac{\omega}{c^{0}}~~,~~
 \zeta=\frac{k_{1}}{k^{0}}~,
\end{equation}
and examine  $F^{1}_{1}$ in the interval $\zeta\in [0,1]$:
\begin{equation}\label{w53.3}
  F^{1}_{1}(\zeta,\eta)d\zeta=
\frac{d\zeta}{i\psi\cos(\phi\eta)+\upsilon\phi\sin(\phi\eta)}~,
\end{equation}
wherein:
\begin{equation}\label{w53.4}
 \upsilon=\frac{\mu^{1}}{\mu^{0}}~~,~~\gamma=\frac{k^{1}}{k^{0}}=
 \frac{c^{0}}{c^{1}}~~,~~\psi=\sqrt{1-\zeta^{2}}~~,~~
 \phi=\sqrt{\gamma^{2}-\zeta^{2}}~.
\end{equation}
Note that $\gamma >1$ and $\upsilon<1$ due to previous
assumptions. Since $\eta$ and $\zeta$ are real, the denominator in
$F_{1}^{1}$ cannot vanish; however it does attain minima for
certain values of these parameters.

Let us consider $\zeta$ to be constant and inquire for what values
of $\eta$
\begin{equation}\label{w53.5}
  |F^{1}_{1}(\zeta,\eta)|^{-2}=
\psi^{2}\cos^{2}(\phi\eta)+\upsilon^{2}\phi^{2}\sin^{2}(\phi\eta)~,
\end{equation}
attains its minima. A necessary condition is:
\begin{equation}\label{w53.6}
  \partial_{\eta}\left ( |F^{1}_{1}(\zeta,\eta)|^{-2}\right ) =0=
\phi\left ( \upsilon^{2}\phi^{2}-\psi^{2}\right ) \sin(2\phi\eta)
~.
\end{equation}
There are three possibilities, the first one of which is $\phi=0$,
but this implies $\zeta=\gamma >1$ which is in contradiction with
the fact $\zeta$ must lie in $[0,1]$. The second possibility is
that $\psi=\upsilon\phi$; we will re-consider this case further
on. The third possibility is $\sin(2\phi\eta)=0$ whence
$\phi\eta=n\pi/2~;~n=0,1,...$. To determine for what values of $n$
these roots correspond to actual minima of $
|F^{1}_{1}(\zeta,\eta)|^{-2}$ we must have
\begin{equation}\label{w53.7}
  \partial^{2}_{\eta}\left ( |F^{1}_{1}(\zeta,\eta)|^{-2}\right )
  \big | _{\phi\eta=n\pi}=
2\phi^{2}\left ( \upsilon^{2}\phi^{2}-\psi^{2}\right )
\cos(n\pi)>0 ~.
\end{equation}
This condition gives rise to two types of solutions depending on
the sign of $\upsilon^{2}\phi^{2}-\psi^{2}$. The first type, which
we call {\it even body wave} solutions (designated by the
superscript $Be$) is:
\begin{equation}\label{w53.8}
  \eta=\eta_{m}^{Be}=\frac{m\pi}{\phi}~~;~~m=0,1,2,...~\text{when}~\upsilon\phi>\psi~.
\end{equation}
The second type, which we call {\it odd body wave} solutions
(designated by the superscript $Bo$) is:
\begin{equation}\label{w53.9}
  \eta=\eta_{m}^{Bo}=\frac{(2m+1)\pi}{2\phi}~~;~~m=0,1,2,...~
  \text{when}~\upsilon\phi<\psi~.
\end{equation}
Let $\zeta^{B}$ be the value of $\zeta$ for which
$\upsilon\phi=\psi$. We find
\begin{equation}\label{w53.10}
  \zeta^{B}=\sqrt{\frac{1-\upsilon^{2}\gamma^{2}}{1-\upsilon^{2}}}~,
\end{equation}
or
\begin{equation}\label{w53.11}
  \zeta^{B}=\sqrt{1-\frac{(\gamma^{2}-1)\upsilon^{2}}{1-\upsilon^{2}}}~,
\end{equation}
from which it follows that $\zeta^{B}<1$, this meaning that the
second possibility (i.e., $\upsilon\phi=\psi$) is not
contradictory with the constraint $\zeta\in[0,1]$.

Thus, the three types of solutions leading to minima of $F_{1}^1$
are:
\begin{equation}\label{w53.12}
 \text{For}~\zeta>\zeta^{B}:~~ \eta=\eta_{m}^{Be}=\frac{m\pi}{\phi}~~;~~m=0,1,2,...~.
\end{equation}
\begin{equation}\label{w53.13}
 \text{For}~\zeta=\zeta^{B}:~~ \text{all}~ \eta~,
\end{equation}
\begin{equation}\label{w53.14}
\text{For}~\zeta<\zeta^{B}:~~
\eta=\eta_{m}^{Bo}=\frac{(2m+1)\pi}{2\phi}~~;~~m=0,1,2,..~.
\end{equation}
The meaning of all this is that $\big | F_{1}^{1}\big | $ has
regularly-spaced (in terms of $\eta$) maxima for all values of
$\zeta$, which is another way of saying that $\big | F_{1}^{1}\big
| $ is a periodic function of $\eta$ for all $\zeta$. The period
of this function is $\pi/\phi$ (even when $\zeta=\zeta^{B}$,
because a constant is a periodic function with arbitrary period).
However, the function takes different forms in the three cases
((\ref{w53.12})-(\ref{w53.14})). In fact,

(i) for $\zeta>\zeta^{B}$: $\big | F_{1}^{1}\big | $ has maxima
equal to $\psi^{-1}=$ at $\eta=m\pi/\phi$ and minima equal to
$(\upsilon\phi)^{-1}$ at $\eta=(2m+1)\pi/2\phi$,

(ii) for $\zeta=\zeta^{B}$ :$\big | F_{1}^{1}\big | $ is a
constant equal to $\psi^{-1}=(\upsilon\phi)^{-1}$ at all $\eta$,

(iii) for $\zeta<\zeta^{B}$ :$\big | F_{1}^{1}\big | $ has minima
equal to $\psi^{-1}$ at $\eta=m\pi/\phi$ and maxima equal to
$(\upsilon\phi)^{-1}$ at $\eta=(2m+1)\pi/2\phi$.

A numerical example will help to give a measure of the relative
importance of these three types of solutions. Recall that:
\begin{equation}\label{w53.15}
 \upsilon=\frac{\mu^{1}}{\mu^{0}}=\frac{(c^{1})^{2}\rho^{1}}{(c^{0})^{2}\rho^{0}}~,
\end{equation}
so that
\begin{equation}\label{w53.16}
 \upsilon\gamma=\upsilon\frac{c^{0}}{c^{1}}=\frac{c^{1}\rho^{1}}{c^{0}\rho^{0}}~.
\end{equation}
Let us choose parameters that might be pertinent in the context of
topics (a) and (b): $c^{0}=$1000m/s, $\rho^{0}=$1500kg/$m^{3}$,
$c^{1}=$100m/s, $\rho^{0}=$1000kg/$m^{3}$, for which
$\upsilon=0.67\times 10^{-2}$ and $\upsilon\gamma=0.67\times
10^{-1}$, whence $\zeta^{B}=0.995$. Thus, $\big | F_{1}^{1}\big |
$ takes the form of the type (iii) function in most of the
interval $[0,1]$, in fact in $0\leq\zeta<0.995$. In particular,
for body waves whose wavevectors are nearly-vertical (i.e.,
$0\leq\zeta<<1$), the maximum of $\big | F_{1}^{1}\big | $ is
\begin{equation}\label{w53.17}
 (\upsilon\phi)^{-1}=\frac{1}{\upsilon\sqrt{\gamma^{2}-\zeta^{2}}}
 \approx\frac{1}{\upsilon\gamma}=\frac{c^{0}\rho^{0}}{c^{1}\rho^{1}}~.
\end{equation}
which, in the present numerical example, is equal to 15.

The lowest frequency ($\nu=\omega/2\pi$) for which this value is
attained (obtained from $\eta=\pi/2\phi\approx\pi/2\gamma$) is
\begin{equation}\label{w53.19}
 \nu=\frac{c^{1}}{4h}~,
\end{equation}
and is often called either the 'fundamental Haskell resonance
frequency' \cite{babo80} or the 'one-dimensional resonance
frequency' \cite{babo85}, \cite{sedu00} of the soft soil layer/hard substratum
configuration. However, a sinusoidal response function of the type
$F_{1}^{1}$ is not consistent with resonant response (which is
infinite at the resonance frequencies in the absence of a
dissipation mechanism) that would arise, for instance, in the
context of excitation of some sort of structural mode; in fact,
this sinuoidal response results from interference of waves, which
is the reason why we termed $F_{1}^{1}$ the 'interference factor'.
Thus, it is improper to employ the term 'resonances'
\cite{babo80}, \cite{ba85}, \cite{babo85} in connection with body
wave response (embodied in $I^{1}$) of the configuration.

To conclude this discussion, we now consider the spectral factor
$S(\omega)$. It is obvious that if $S(\omega)=S(\eta c^{0}/h)$ is
significantly large near the frequencies $\eta=(2m+1)\pi/2
\phi~;~m=0,1,...$ at which $F^{1}_{1}$ is large, then the product
of these two functions, embodied in $I_{1}^{1}$ will be large at
these frequencies. In particular, if $S(\omega)=S(\eta c^{0}/h)$
is maximal near the low frequency $\eta=\pi/2\phi$, then the
response will be large over a large range of horizontal
wavenumbers due to the contribution of the $m=0$ maximum of the
interference factor $F^{1}_{1}$. This has been noted repeatedly in
the past \cite{babo80}, \cite{babo85}, \cite{ba85}, and termed a
'resonant response' (as mentioned above), which it is not because
$\eta=(2m+1)\pi/2 \phi~;~m=0,1,...$ are not resonance frequencies.
\subsection{Amplitudes of the SBW2}\label{ss54}
We again restrict our attention to the field in the soft layer,
and, in particular, to the SBW2 component. We note that the
amplitude $dG_{2}^{1}$ of the generic SBW2 is the product of three
factors: the factor $S(\omega)$ associated with the spectrum of
the incident pulse, a geometric factor associated with the
location of the source (whose influence was already discussed),
and a so-called interference factor $F_{2}^{1}dk_{1}$. Here we
discuss the product $F^{1}_{2}dk_{1}$ with $S(\omega)$ in order to
evaluate the contribution of generic SBW2 to the overall response
in the layer and on the ground.

We rewrite $I_{2}^{1}$ as
\begin{multline}\label{w54.1}
 I^{1}_{2}(\mathbf{x},\omega)=
 \\
 \frac{-S(\omega)}{\pi}\int_{k^{0}}^{k^{1}}
 F^{1}_{2}(k_{1},\omega)\cos\left ( k_{1}(x_{1}-x_{1}^{s})\right )
 \cos\left ( K_{2}^{1}(\omega)x_{2}\right )
 e^{\kappa_{2}^{0}(\omega)(h-x_{2}^{s})}dk_{1}~,
\end{multline}
wherein
\begin{multline}\label{w54.1a}
  F^{1}_{2}(k_{1},\omega)dk_{1}=\frac{dk_{1}}
{-\kappa_{2}^{0}(\omega)\cos( K_{2}^{1}(\omega)h) +
\frac{\mu^{1}}{\mu^{0}}K_{2}^{1}(\omega)\sin(
K_{2}^{1}(\omega)h)}~~;
\\
k_{1}\in\left [ k^{0},k^{1}\right ] ~.
\end{multline}
We make the same change of variables as in the previous section,
with the additional definition
\begin{equation}\label{w54.2}
 \theta:=\sqrt{\zeta^{2}-1}~,
\end{equation}
and examine  $F^{1}_{2}$ in the interval $\zeta\in [1,\gamma]$:
\begin{equation}\label{w54.3}
  F^{1}_{2}(\zeta,\eta)d\zeta=
\frac{d\zeta}{-\theta\cos(\phi\eta)+\upsilon\phi\sin(\phi\eta)}~.
\end{equation}
Contrary to the previous case, here the denominator in $F_{2}^{1}$
can vanish for real $\eta$ and $\zeta$, i.e.,
\begin{equation}\label{w54.3b}
-\theta\cos(\phi\eta)+\upsilon\phi\sin(\phi\eta)=0~,
\end{equation}
this being none other than the {\it dispersion relation of Love
modes}. The roots of this relation are:
\begin{equation}\label{w54.4}
 \eta=\frac{1}{\phi}\left [ \arctan \left ( \frac{\theta}
 {\upsilon\phi}\right ) +m\pi\right ]~~;~~m=0,1,2~,
\end{equation}
wherein the arctan function is defined in $[-\pi/2,\pi/2]$ and can
be expressed either by the series
\begin{equation}\label{w54.5}
 \arctan y=y+\sum_{l=1}^{\infty}(-1)^{l}\frac{y^{2l+1}}{2l+1}~~;~~y^{2}<1~,
\end{equation}
or by the series
\begin{equation}\label{w54.6}
 \arctan y=\frac{\pi}{2}-\sum_{l=0}^{\infty}(-1)^{l}\frac{y^{-(2l+1)}}{2l+1}
 ~~;~~y^{2}>1~,
\end{equation}
It is easily shown that $\theta=\upsilon\phi$ when
\begin{equation}\label{w54.7}
\zeta=\zeta^{L}:=\sqrt{1+(\gamma^{2}-1)\frac{\upsilon^{2}}{1+\upsilon^{2}}}~,
\end{equation}
so that $\zeta^{L}>1$, as it should be for the constraint
$\zeta\in[1,\gamma]$ to be satisfied.

Thus, three types of solutions lead to a zero in the denominator
of $F_{2}^{1}$:
\begin{equation}\label{w54.8}
\text{For}~\zeta<\zeta^{L}~:~\eta=\eta_{m}^{Le}=\frac{m\pi}{\phi}+\frac{1}{\phi}
\left [ \frac{\theta}{\upsilon\phi}-\frac{1}{3}\left (
\frac{\theta}{\upsilon\phi}\right ) ^{3} +....\right ] ~,
\end{equation}
\begin{equation}\label{w54.9}
\text{For}~\zeta=\zeta^{L}~:~\eta=\eta_{m}^{L}=\frac{(4m+1)\pi}{4\phi}~,
\end{equation}
\begin{equation}\label{w54.10}
\text{For}~\zeta>\zeta^{L}~:~\eta=\eta_{m}^{Lo}=\frac{(2m+1)\pi}{\phi}+\frac{1}{\phi}
\left [ -\frac{\upsilon\phi}{\theta}+\frac{1}{3}\left (
\frac{\upsilon\phi}{\theta}\right ) ^{3} +....\right ] ~,
\end{equation}
and correspond to the existence of three types (even, neutral,
odd) of Love modes whose eigenfrequencies are $\eta_{m}^{Le},~
\eta_{m}^{L},~ \eta_{m}^{Lo}$ respectively.

This means that $\big | F_{2}^{1}\big | $ has regularly-spaced (in
terms of $\eta$) maxima for all values of $\zeta$, which is
another way of saying that $\big | F_{1}^{1}\big | $ is a periodic
function of $\eta$ for all $\zeta$. The period of this function is
$\pi/\phi$ (even when $\zeta=\zeta^{L}$ because a constant is a
periodic function with arbitrary period). However, the function
takes different forms in the three cases (\ref{w54.8})-(\ref{w54.10}).
In fact,
\\\\
(i) for $\zeta<\zeta^{L}$: $\big | F_{2}^{1}\big | $ has maxima
equal to $\infty$ at $\eta=\eta_{m}^{Le}$,
\\
(ii) for $\zeta=\zeta^{L}$ :$\big | F_{2}^{1}\big | $ has maxima
equal to $\infty$ at $\eta=\eta_{m}^{L}$,
\\
(iii) for $\zeta>\zeta^{L}$ :$\big | F_{2}^{1}\big | $ has minima
equal to $\infty$ at $\eta=\eta_{m}^{Lo}$.
\\\\
A numerical example will help give a measure of the relative
importance of these three types of solutions. Let us again choose:
$c^{0}=$1000m/s, $\rho^{0}=$ 1500kg/$m^{3}$, $c^{1}=$100m/s,
$\rho^{0}=$1000kg/$m^{3}$, for which $\gamma=10$,
$\upsilon=0.67\times 10^{-2}$ and $\upsilon\gamma=0.67\times
10^{-1}$, whence $\zeta^{L}=1.0044$. Thus, $\big | F_{2}^{1}\big |
$ takes the form of the type (iii) function for most of the
interval $[1,\gamma]$, in fact in $1.0044\leq\zeta<10$.

A few remarks are in order.

(1) contrary to what may be inferred from works such as
\cite{babo80}, \cite{babo85}, \cite{ba85}, \cite{chba94}, the
individual Love modes do not have the structure of surface waves
in the layer (and, therefore, on the ground) since the SBW2 are
actually standing body waves; the only feature they share with
surface waves (i.e., the SW that coexist in the hard substratum
when Love modes are excited) is their phase velocity
\begin{equation}\label{w54.11}
c^{L}=\frac{c^{0}}{\zeta}~,
\end{equation}
wherein it can be noted that due to the fact that
$\zeta\in[1,\gamma]$,
\begin{equation}\label{w54.12}
c^{L}<c^{0}~,
\end{equation}
which means that the phase velocity of Love modes (shared by the
SBW2 in the layer and the SW in the hard substratum) is less than
the phase velocity of body waves in the the hard substratum,

(2) contrary to the what occurs in connection with the SBW1, the excitation of
Love modes is indeed a resonant process, because Love modes are
actually structural modes of the soft layer/hard substratum
configuration and because the response associated with each of
these modes is infinite at resonance in the absence of dissipation
in both of the media of the configuration (this response can be
large, but finite, when dissipation is present),

(3) the resonant frequencies of the Love modes are not identical
to the frequencies at which the SBW1 attain their maxima; for
instance, the difference of these frequencies, for the $m$-th
prevalent odd-type SBW1 and SBW2, is:
\begin{equation}\label{w54.13}
\eta_{m}^{Bo}-\eta_{m}^{Lo}=\frac{1}{\phi} \left [
\frac{\upsilon\phi}{\theta}-\frac{1}{3}\left (
\frac{\upsilon\phi}{\theta}\right ) ^{3} +....\right ] ~,
\end{equation}
which means that the frequency of occurrence of the maxima of the
$m$-th order SBW1 is higher than (although it can be close to)
that of the corresponding SBW2
(note that the difference in (\ref{w54.13}) does not depend on
$m$).

To conclude this discussion, we again consider the spectral factor
$S(\omega)$. It is obvious that if $S(\omega)=S(\eta c^{0}/h)$ is
significantly large near the frequencies $\eta_{m}^{Lo}$ at which
 $F^{1}_{2}$ is large (infinite if no dissipation is present),
 then the product
of these two functions, embodied in $I_{2}^{1}$, will be large at
these frequencies. In particular, if $S(\eta c^{0}/h)$ is maximal
near the low frequency $\eta_{0}^{Lo}$, then the response will be
large over a large range of horizontal wavenumbers. If
$S(\omega)=S(\eta c^{0}/h)$ is maximal near the low frequency
$\eta_{0}^{Lo}$, and $\eta_{0}^{Lo}$ is not too far from
$\eta_{0}^{Bo}$, then the global response can be even larger due
to the cumulative contribution of both the SBW1 and SBW2.
\subsection{Amplitudes of the SSW}\label{ss55}
We are again concerned with the field in the soft layer, and, in
particular, with its SSW component. We note that the amplitude
$dG_{3}^{1}$ of the generic SSW is the product of three factors:
$S(\omega)$ which is associated with the spectrum of the incident
pulse, a geometric factor associated with the location of the
source (whose influence was already discussed), and the
interference factor $F_{3}^{1}dk_{1}$. Here we discuss the product
of $F_{3}^{1}dk_{1}$ with $S(\omega)$ in order to evaluate the
contribution of generic SSW to the overall response in the layer
and on the ground.

We rewrite $I_{3}^{1}$ as
\begin{multline}\label{w55.1}
 I^{1}_{3}(\mathbf{x},\omega)=
 \\
 \frac{-S(\omega)}{\pi}\int_{k^{1}}^{\infty}
 F^{1}_{3}(k_{1},\omega)\cos\left ( k_{1}(x_{1}-x_{1}^{s})\right )
 \cosh\left ( \kappa_{2}^{1}(\omega)x_{2}\right )
 e^{\kappa_{2}^{0}(\omega)(h-x_{2}^{s})}dk_{1}~,
\end{multline}
wherein
\begin{multline}\label{w55.1a}
  F^{1}_{3}(k_{1},\omega)dk_{1}=-\frac{dk_{1}}
{\kappa_{2}^{0}(\omega)\cosh( \kappa_{2}^{1}(\omega)h) +
\frac{\mu^{1}}{\mu^{0}}\kappa_{2}^{1}(\omega)\sinh(
\kappa_{2}^{1}(\omega)h)}~~;
\\
k_{1}\in\left [ k^{1},\infty\right [ ~.
\end{multline}
We make the same change of variables as in the previous two
sections, with the additional definition
\begin{equation}\label{w55.2}
  \chi:=\sqrt{\zeta^{2}-\gamma^{2}}~,
\end{equation}
and examine $F_{3}^{1}$ for $\zeta$ in the interval $[\gamma,\infty]$:
\begin{equation}\label{w55.3}
  F_{3}^{1}(\zeta,\eta)d\zeta=\frac{-d\zeta}{\theta\cosh(\chi\eta)+
  \upsilon\chi\sinh(\chi\eta)}~.
\end{equation}
Since $\chi\geq 0$ for $\zeta\in [\gamma,\infty]$, and $\eta>0$,
$\sinh(\chi\eta)\geq 0$ and $\cosh(\chi\eta)> 0$ for $\zeta\in
[\gamma,\infty]$, which means that the denominator in the previous
formula cannot vanish for real $\eta$ and $\zeta$. It can however
exhibit minima for  $\zeta\in[\gamma,\infty]$.

Let us consider $\zeta$ to be constant and inquire for what values
of $\eta$ the denominator $ \left ( F_{3}^{1}\right )
^{-1}=\theta\cosh(\chi\eta)+
  \upsilon\chi\sinh(\chi\eta)$ has
minima. This requires that
\begin{equation}\label{w55.4}
  \partial_{\eta}\left ( F_{3}^{1}(\zeta,\eta)\right ) ^{-1}=
  \chi\left [ \theta\sinh(\chi\eta)+
  \upsilon\chi\cosh(\chi\eta)\right ] =0~.
\end{equation}
But $[~~]\neq 0$ except for $\chi=0$, i.e., for $\zeta=\gamma$ and
$\forall \eta$. When $\chi=0$ we find  $\left ( F_{3}^{1}\right )
^{-1}=\theta$, and from the fact that $\sinh(\chi\eta)\geq 0$ and
$\theta\cosh(\chi\eta)\geq\theta$ for $\zeta\in [\gamma,\infty]$,
we conclude that $\left ( F_{3}^{1}\right ) ^{-1}\geq \theta$.
This means that $\zeta=\gamma$ corresponds to the location of a
minimum of $\left ( F_{3}^{1}\right ) ^{-1}$ and this holds for
all $\eta$.

Thus, $ F_{3}^{1}$ is a monotonically-decreasing function of
$\zeta$ for all $\zeta\in ]\gamma,\infty [$ and attains its
maximum equal to $\theta^{-1}=1/\sqrt{\gamma^{2}-1}$ at
$\zeta=\gamma$ for all $\eta$.

To get an idea of the magnitude of this function, notably in
relation to $ F_{1}^{1}$, we again consider the numerical example:
$c^{0}=$1000m/s, $\rho^{0}=$1500kg/$m^{3}$, $c^{1}=$100m/s,
$\rho^{0}=$1000kg/$m^{3}$, for which $\gamma=10$, whence $\max
\big | F_{3}^{1}\big | \leq 0.1005$ which is much less than $\max
| F_{1}^{1}\big |=15$ for the same set of parameters.

Since the maximum of $ F_{3}^{1}$ is attained at all frequencies
(i.e., for all $\eta$), the spectrum function $S(\omega)$ does not
influence the relative contribution of $I_{3}^{1}$ to the ground
response. Thus, to conclude this discussion, we can say that the
SSW contribute relatively little to the ground response in
comparison to the SBW1 and SBW2, except perhaps at frequencies
close to the minima of the functions $F_{1}^{1}$ and $F_{2}^{1}$.
\section{Total frequency domain contributions of the SBW1, SBW2, SSW
as embodied in the cumulative frequency response functions
$I_{1}^{1}$, $I_{2}^{1}$ and $I_{3}^{1}$ for elastic and viscoelastic layers}
Although the theoretical analysis carried out in the sect. 5 may
be useful for underlining the role played by the different types
of body and surface waves that appear in the fields in the layer
and substratum, it does not resolve the practical problem of the
actual
evaluation of the  integrals $I_{1}^{1}$, $I_{2}^{1}$ and
$I_{3}^{1}$. Another drawback of this analysis is that it is
restricted to the case in which the layer is elastic, but the
conclusions that were drawn for the elastic layer case should not be
radically different for the case of a weakly- or
moderately-viscoelastic layer.

Consequently, we resorted to a purely numerical (i.e., Simpson
integration) approach for the evaluation of $I_{1}^{1}$,
$I_{2}^{1}$ and $I_{3}^{1}$ and of their sum to determine the
frequency-domain seismic response of the layer/substratum
configuration. Since physically-realistic configurations involve
viscoelastic layers, we evaluated these integrals and the total
frequency response $u(\mathbf{x}_{g},\omega)$ under the assumption
of viscoelastic layers.  Once $u(\mathbf{x}_{g},\omega)$ was
computed, we determined the temporal signal $u(\mathbf{x}_{g},t)$,
again by purely numerical means, via (\ref{w35.10}).

The weakness of the numerical approach is that it makes it
difficult to discern the mechanisms underlying the observed
response. To overcome this, we will give in sect. 8 a
phenomenological analysis of the frequency-domain and time-domain
responses which should facilitate the comprehension of the
particular features of the temporal signals.

\section{Computational results}
\subsection{Preliminaries}
In all except sect. 7.9  we take the density of the hard half
space $\rho^{0}$ to be 2000kg/$m^{3}$. Contrary to what was
assumed in the preceding theoretical analysis, we henceforth take
into account the lossy nature of the soft layer. The quality
factor $Q^{1}$ is chosen equal to 30 in all the computations
except in sect. 7.9 (recall that the hard half space is non-lossy,
i.e., $Q^{1}=\infty$). The seismic source is associated with the
pseudo-Ricker impulse function given in (\ref{w33.2a}) whose
spectrum is given in (\ref{w34.3b}).
\begin{figure}
[ptb]
\includegraphics[width=4.0cm]{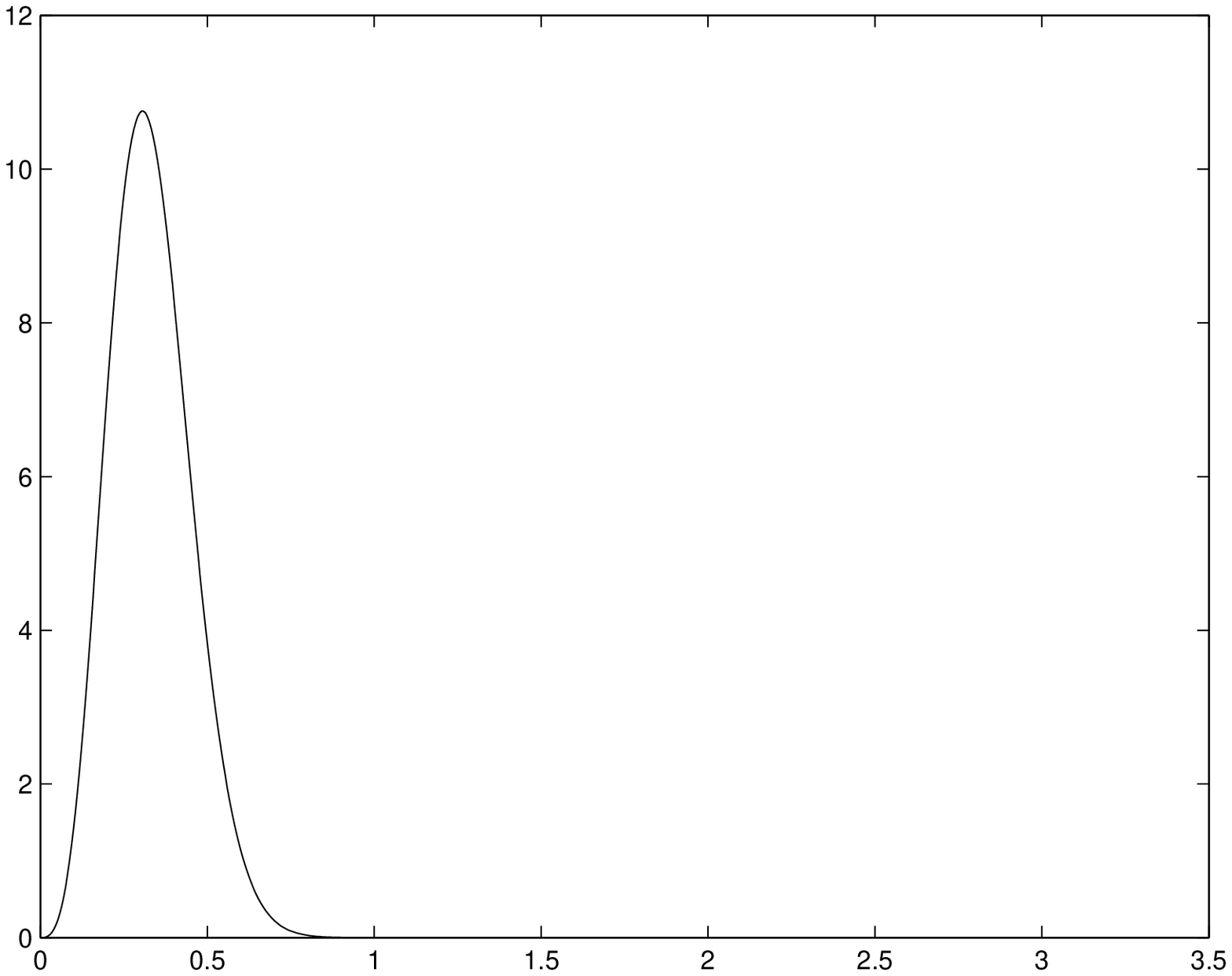}\hfill
\includegraphics[width=4.0cm]{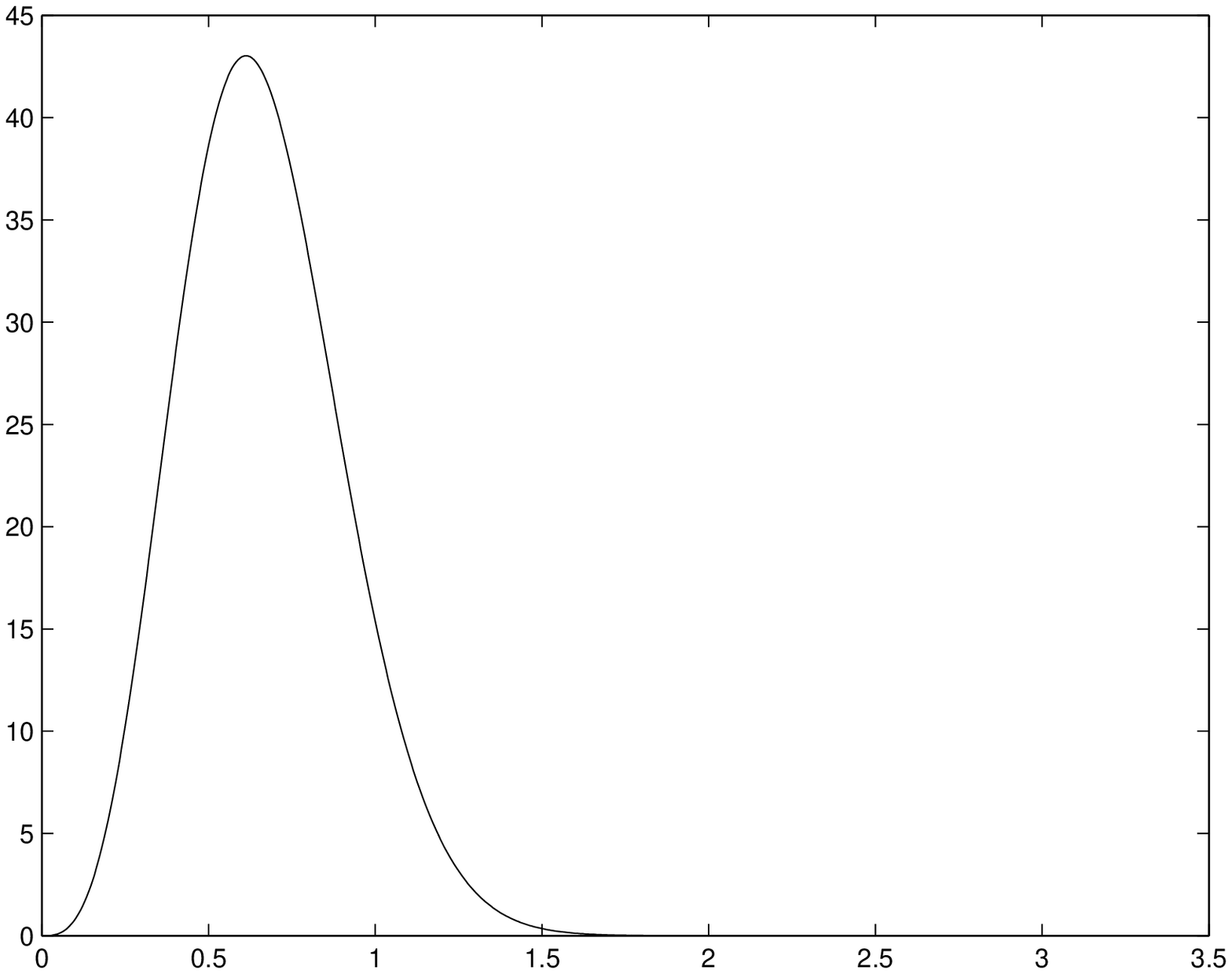}\hfill
\includegraphics[width=4.0cm]{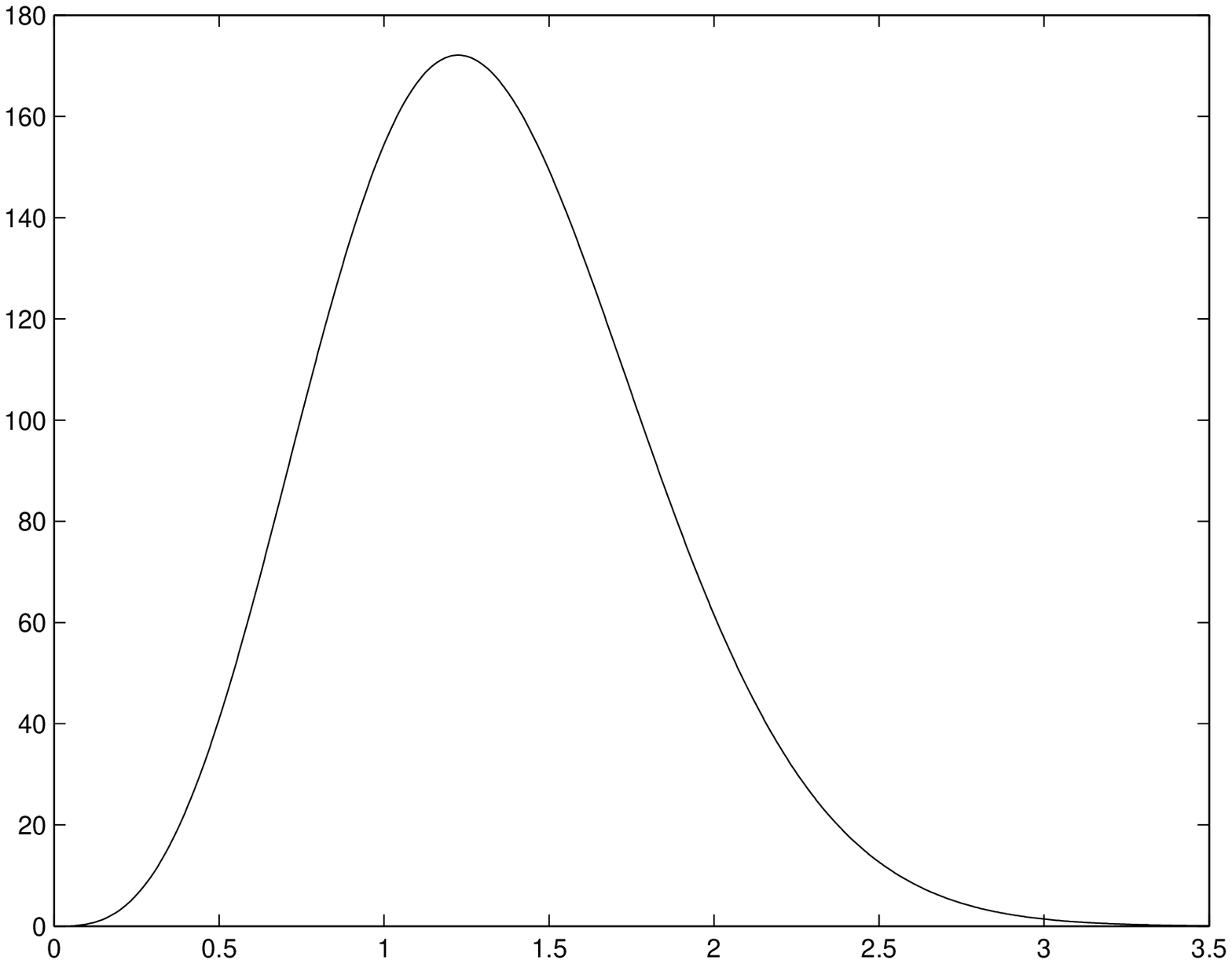}\hfill
\caption{Moduli of source spectrum functions, i.e.
$|S(\nu=\omega/2\pi)|$ versus $\nu$(Hz), for $\nu_{0}=0.25$Hz
(left), 0.50Hz (middle), 1.0Hz (right).} \label{f1a}
\end{figure}
Examples of these spectra (i.e., their moduli) are displayed in
fig. \ref{f1a}.

Unless stated otherwise, the thickness of the layer $h$ is taken
to be 50m. This figure could just as well be 10km provided the
wavelength $\Lambda^{1}$ and/or the wavespeed $c^{1}$ are adjusted
so as to keep the ratio $h/\Lambda^{1}=h\nu/c^{1}$ constant. This
issue is discussed in more depth in the sect. 7.9.
\subsection{Comparison of the results of two methods for determining  the
frequency domain response on the ground} In order to be reasonably
sure that the separation of variables technique employed herein
gives valid results for a viscoelastic layer, we compared these
results to those obtained by a finite element time domain
viscoelastic code developed by one of the present authors (JPG)
with C. Tsogka \cite{grts03}, \cite{grts04}. The time domain
responses obtained by this code were Fourier-transformed to get
the corresponding frequency domain responses.
\begin{figure}[ptb]
\label{f1b}
\includegraphics[width=12.0cm]{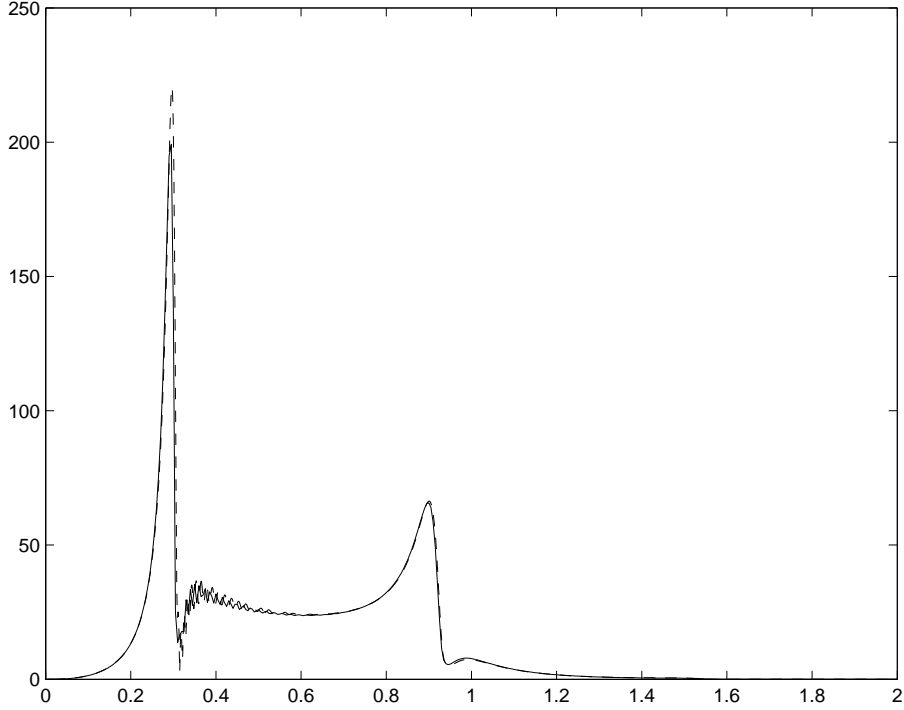}\hfill
\caption{Comparison of the frequency domain ground  response
(i.e., $|u^{1}(\mathbf{x}_{g},\nu=\omega/2\pi)|$ versus $\nu$(Hz))
at $\mathbf{x}=$(3000m,0m), for a shallow $\nu_{0}=$0.5Hz source
at $\mathbf{x}^{s}=$(0m,100m), in a Mexico City-like environment,
i.e., $c^{0}=600$m/s, $\rho^{1}=1300$kg/m$^{3}$, $c^{1}=60$m/s.
The full curve was obtained by the separation of variables
technique described herein whereas the dashed curve was obtained
by the finite element time domain technique described in
\cite{grts03}, \cite{grts04}.}
\end{figure}
An example of these results is given in fig.
\ref{f1b}.
\subsection{The cumulative contributions of the SBW1, SBW2 and
SSW to the overall frequency domain ground response} The
discussion here centers on the transfer functions of frequency
domain ground response. In all that follows, the graph of the
modulus of
$I_{1}^{1}(\mathbf{x}_{g},\nu)/u^{i}(\mathbf{x}_{g},\nu)$ versus
frequency $\nu$ is designated by dots, the graph of the modulus of
$I_{2}^{1}(\mathbf{x}_{g},\nu)/u^{i}(\mathbf{x}_{g},\nu)$ versus
$\nu$ is designated by dashes, the graph of the modulus of
$(I_{2}(\mathbf{x}_{g},\nu)^{1}+I_{3}^{1}(\mathbf{x}_{g},\nu))/
u^{i}(\mathbf{x}_{g},\nu)$ versus frequency $\nu$ is designated by
dot-dashes, and the graph of the modulus of the ground
displacement $u(\mathbf{x}_{g},\nu)/u^{i}(\mathbf{x}_{g},\nu)$
versus frequency by a continuous line.

To begin, consider a configuration thought to be representative of
that in the central portion of the city of Nice (France)
wherein $c^{0}=1000$m/s, $\rho^{1}=1800$kg/m$^{3}$,
$c^{1}=200$m/s.
\begin{figure}
[ptb]
\includegraphics[width=4.0cm]{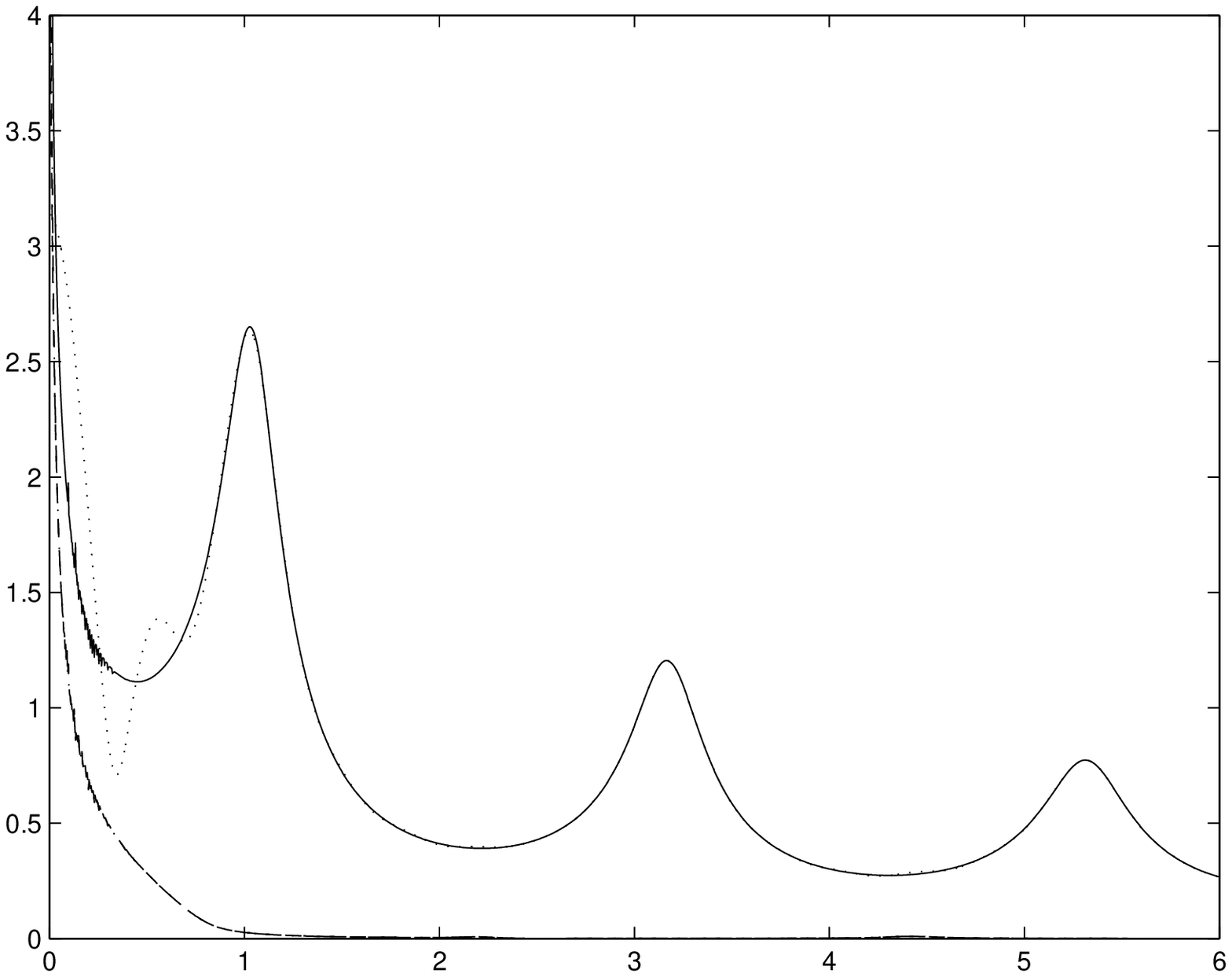}\hfill
\includegraphics[width=4.0cm]{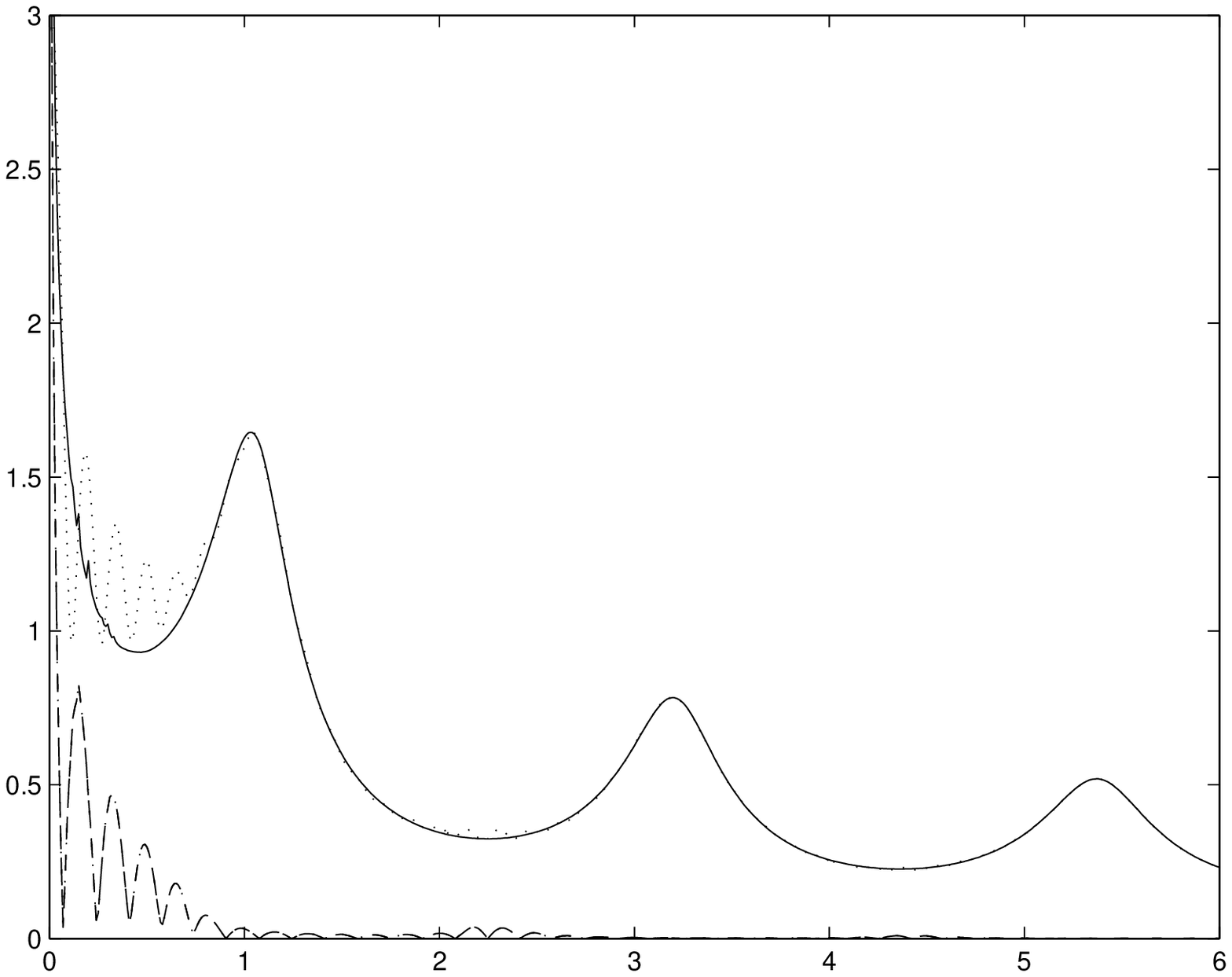}\hfill
\includegraphics[width=4.0cm]{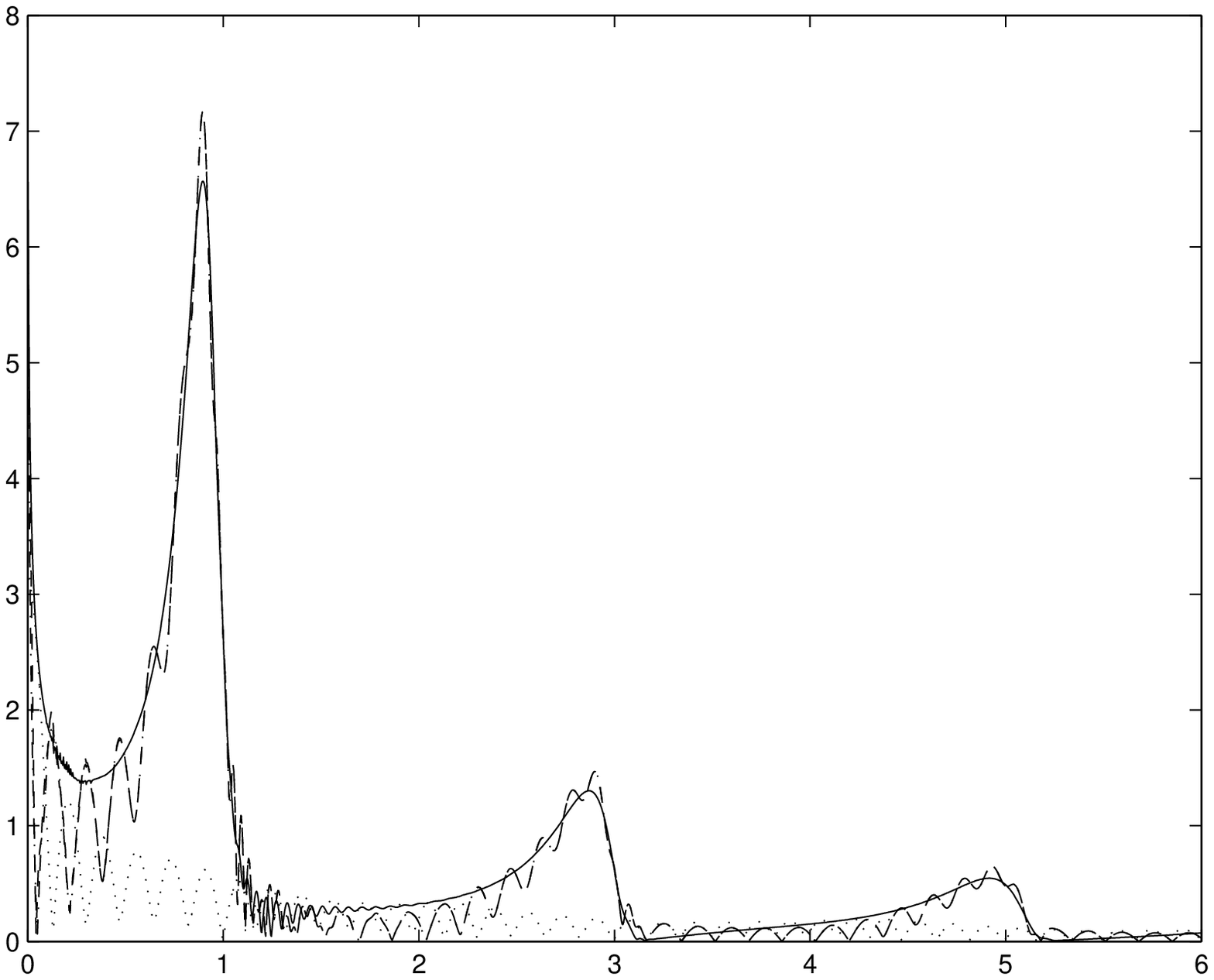}\hfill
\caption{Transfer functions of ground response in Nice-like
environment for various source locations and observation
locations. Left: $\mathbf{x}^{s}=(0\text{m},3000\text{m})$,
$\mathbf{x}=(100\text{m},0\text{m})$. Middle:
$\mathbf{x}^{s}=(0\text{m},3000\text{m})$,
$\mathbf{x}=(3000\text{m},0\text{m})$. Right:
$\mathbf{x}^{s}=(0\text{m},100\text{m})$,
$\mathbf{x}=(3000\text{m},0\text{m})$.}
\label{f2}
\end{figure}
We first place the source at a relatively-large depth of 3km on
the $x_{2}$ axis, i.e., $\mathbf{x}^{s}=(0\text{m},3000\text{m})$
and evaluate the moduli of the ground transfer functions
relatively near the epicenter, i.e.,
$\mathbf{x}=(100\text{m},0\text{m})$ (left subfigure in fig.
\ref{f2}) as well as relatively far from the epicenter, i.e.,
$\mathbf{x}=(3000\text{m},0\text{m})$ (middle subfigure in fig.
\ref{f2}), and then place the source at a relatively-small depth
of 100m on the $x_{2}$ axis, i.e.,
$\mathbf{x}^{s}=(0\text{m},100\text{m})$ and evaluate the ground
transfer functions relatively far from the epicenter, i.e.,
$\mathbf{x}=(3000\text{m},0\text{m})$ (right subfigure in fig.
\ref{f2}).

It will be noticed that, in this and practically all subsequent
results, the curve relative to $\big
|(I_{2}^{1}(\mathbf{x}_{g},\nu)+I_{3}^{1}(\mathbf{x}_{g},\nu))
/u^{i}(\mathbf{x}_{g},\nu)\big | $ is coincident with that
relative to  $\big |
I_{2}^{1}(\mathbf{x}_{g},\nu)/u^{i}(\mathbf{x}_{g},\nu)\big | $
which means, as predicted by the analysis of the preceding
section, that the contribution to overall ground response of the
standing surface waves in the layer is negligible. Thus, we
restrict the following discussion to the sole contribution of the
standing bulk waves of the first (SBW1) and second kinds (SBW2).
The left and middle panels in fig. \ref{f2} show that when the the
focal depth is large the ground response is largely dominated by
the contribution of the SBW1 (i.e., by
$|I_{1}^{1}(\mathbf{x}_{g},\nu)/u^{i}(\mathbf{x}_{g},\nu)|$), and,
in fact, the SBW2 have no influence on the response beyond $\sim
1$Hz. However, the right panel in fig. \ref{f2} gives just the
opposite result when the focal depth is small, since the ground
response is largely dominated by the SBW2 (i.e., by
$|I_{2}^{1}/S(\omega)|$) and the SBW1 have little influence
beyond $\sim 1$Hz. Another interesting feature of these results is
that the total response curves have noticeably-different
appearance when the source is deep  or shallow (notice that this
appearance is qualitatively the same for small and large
epicentral distances, assuming the same, large focal depths in the
two cases).

Next consider a somewhat softer environment
than in Nice
\begin{figure}
[ptb]
\includegraphics[width=6.0cm]{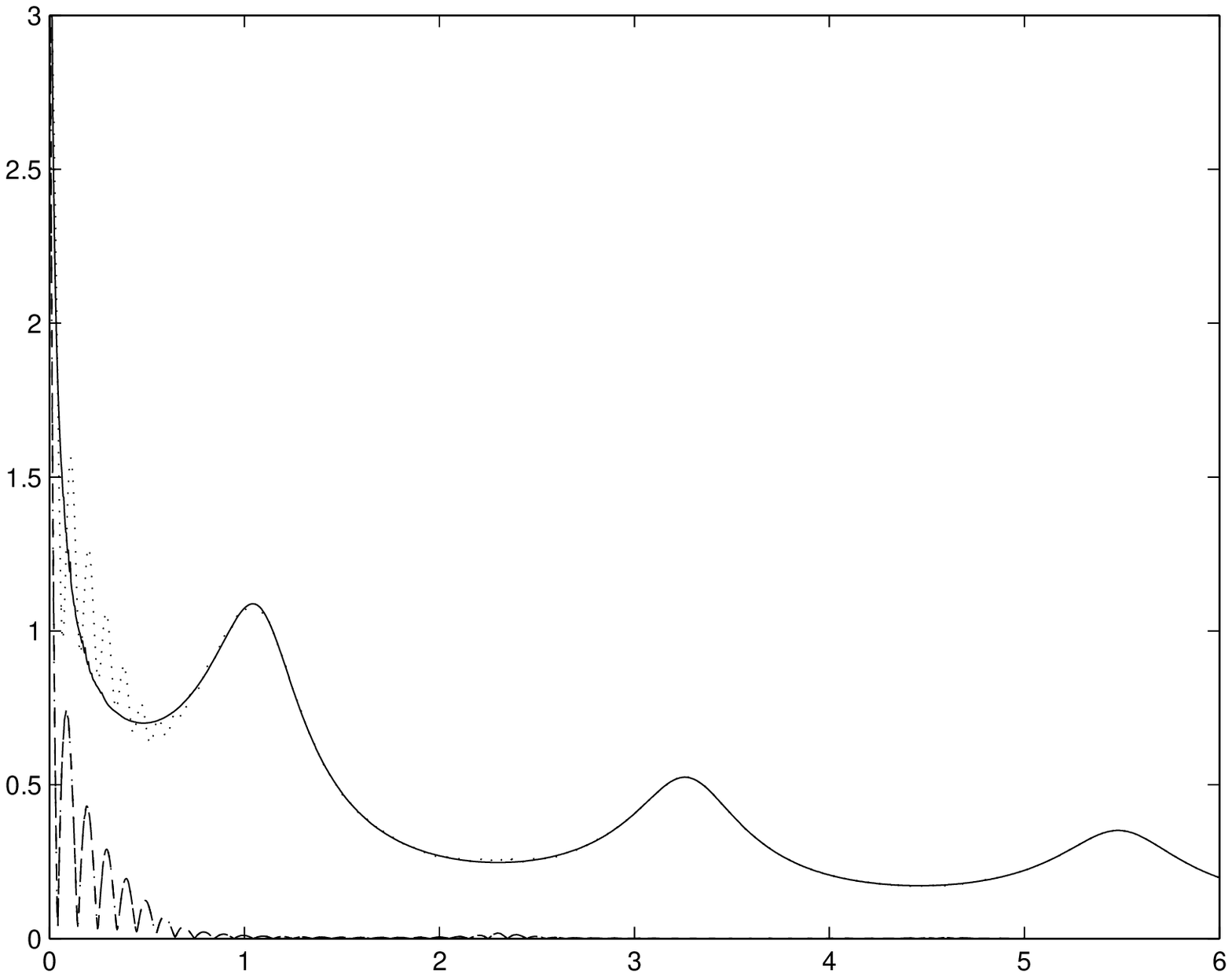}\hfill
\includegraphics[width=6.0cm]{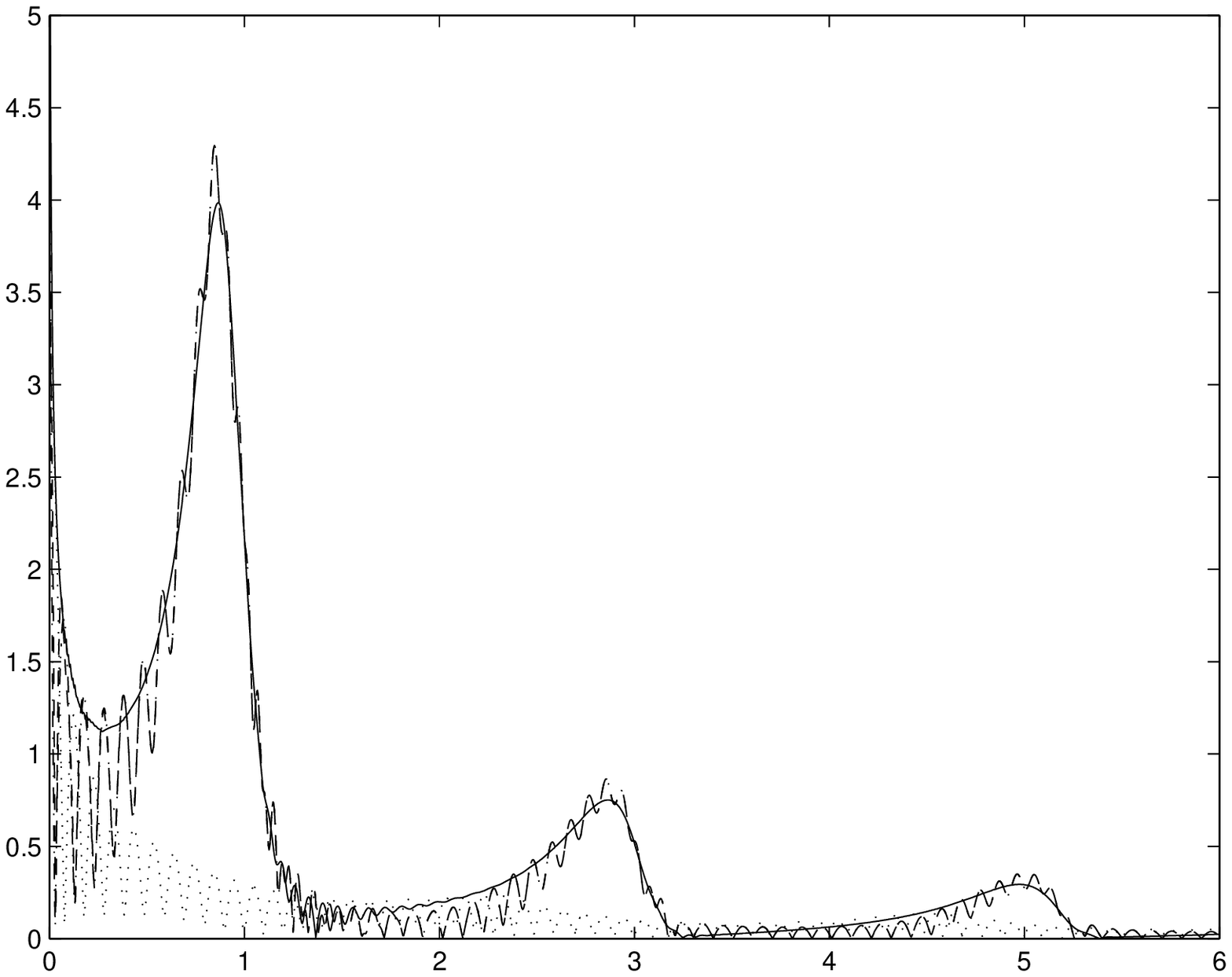}\hfill
\caption{Transfer functions of ground response in softer-than-Nice
environment for various source locations and at the  fixed
observation point $\mathbf{x}=(3000\text{m},0\text{m})$ . Left:
$\mathbf{x}^{s}=(0\text{m},3000\text{m})$. Right:
$\mathbf{x}^{s}=(0\text{m},100\text{m})$.}
\label{f3}
\end{figure}
wherein $c^{0}=600$m/s, $\rho^{1}=1300$kg/m$^{3}$, $c^{1}=200$m/s
(fig. \ref{f3}). We first place the source (rather deep) at
$\mathbf{x}^{s}=(0\text{m},3000\text{m})$ and evaluate the ground
transfer functions relatively far from the epicenter, i.e.,
$\mathbf{x}=(3000\text{m},0\text{m})$ (left subfigure in fig.
\ref{f3}), and then place the source at a relatively-small depth,
i.e., $\mathbf{x}^{s}=(0\text{m},100\text{m})$, and again evaluate
the ground transfer functions relatively far from the epicenter,
i.e., $\mathbf{x}=(3000\text{m},0\text{m})$ (right subfigure in
fig. \ref{f3}). We again observe that the ground response is
dominated by the SBW1 when the source is deep and by the SBW2 when
the source is shallow. Also we notice that the appearance of the
total response curve for a deep source is different from than of a
shallow source.

We next consider  a Mexico-city like site (of course without the
buildings, contrary to the case in \cite{tswi03})
\begin{figure}
[ptb]
\includegraphics[width=6.0cm]{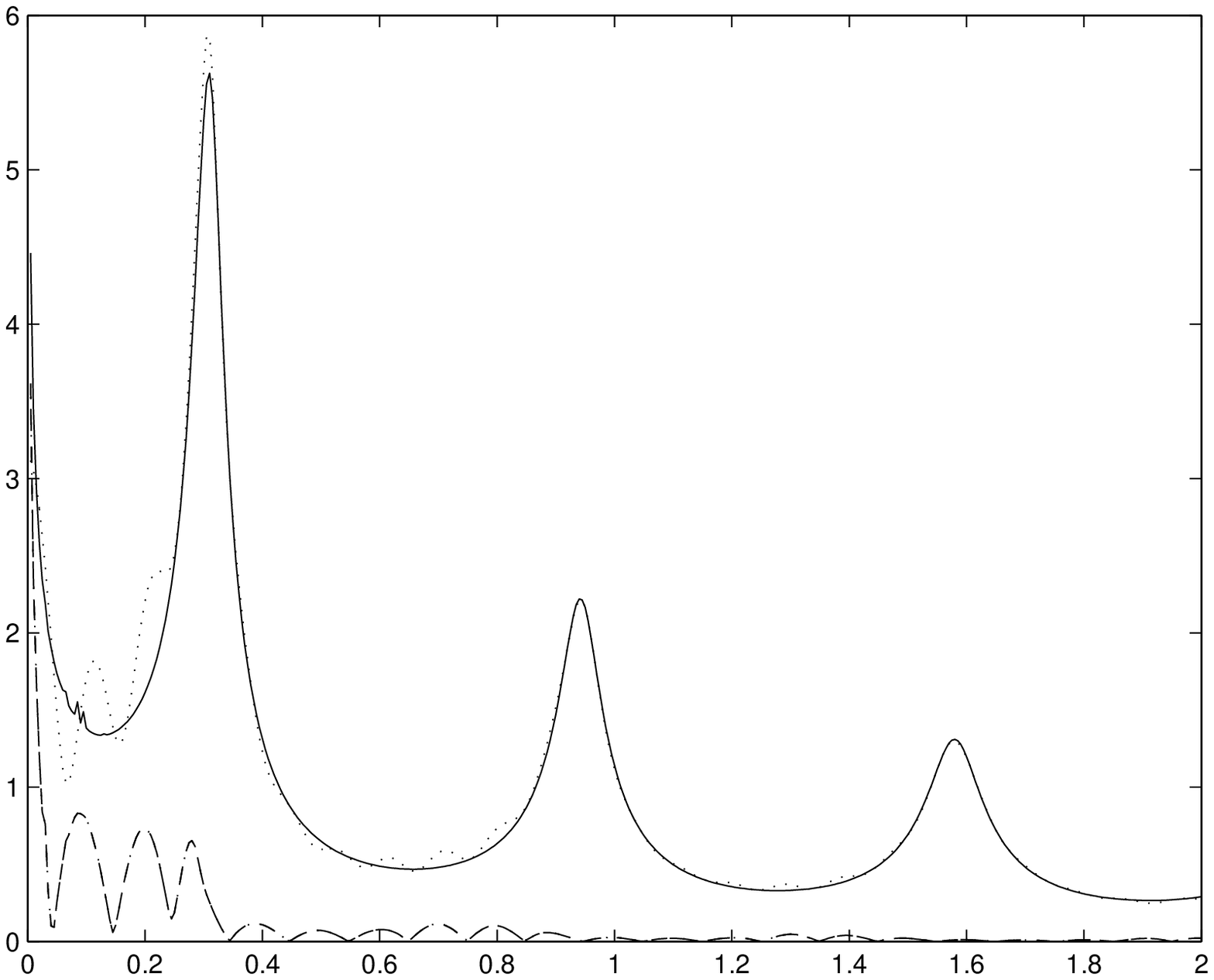}\hfill
\includegraphics[width=6.0cm]{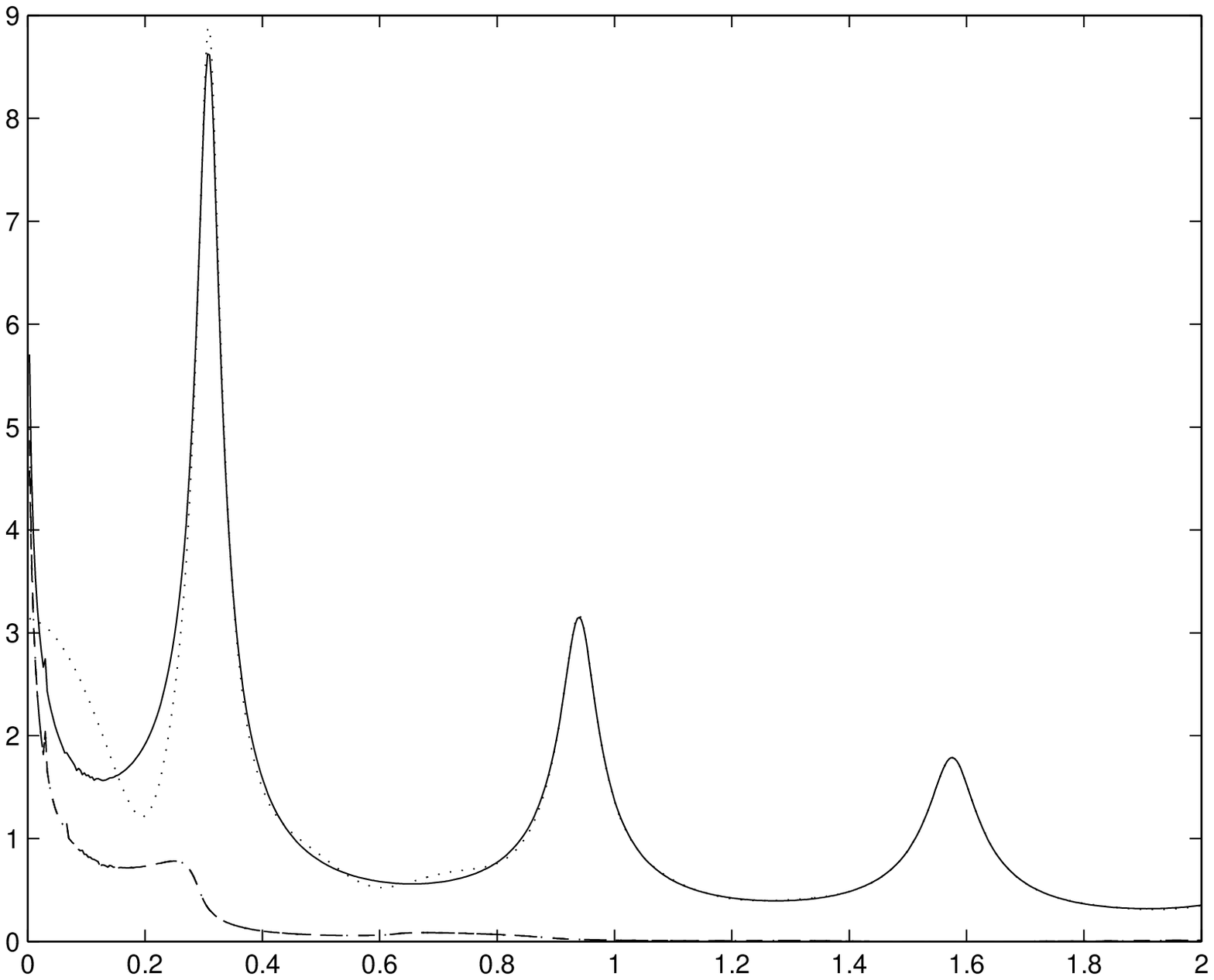}\hfill
\includegraphics[width=6.0cm]{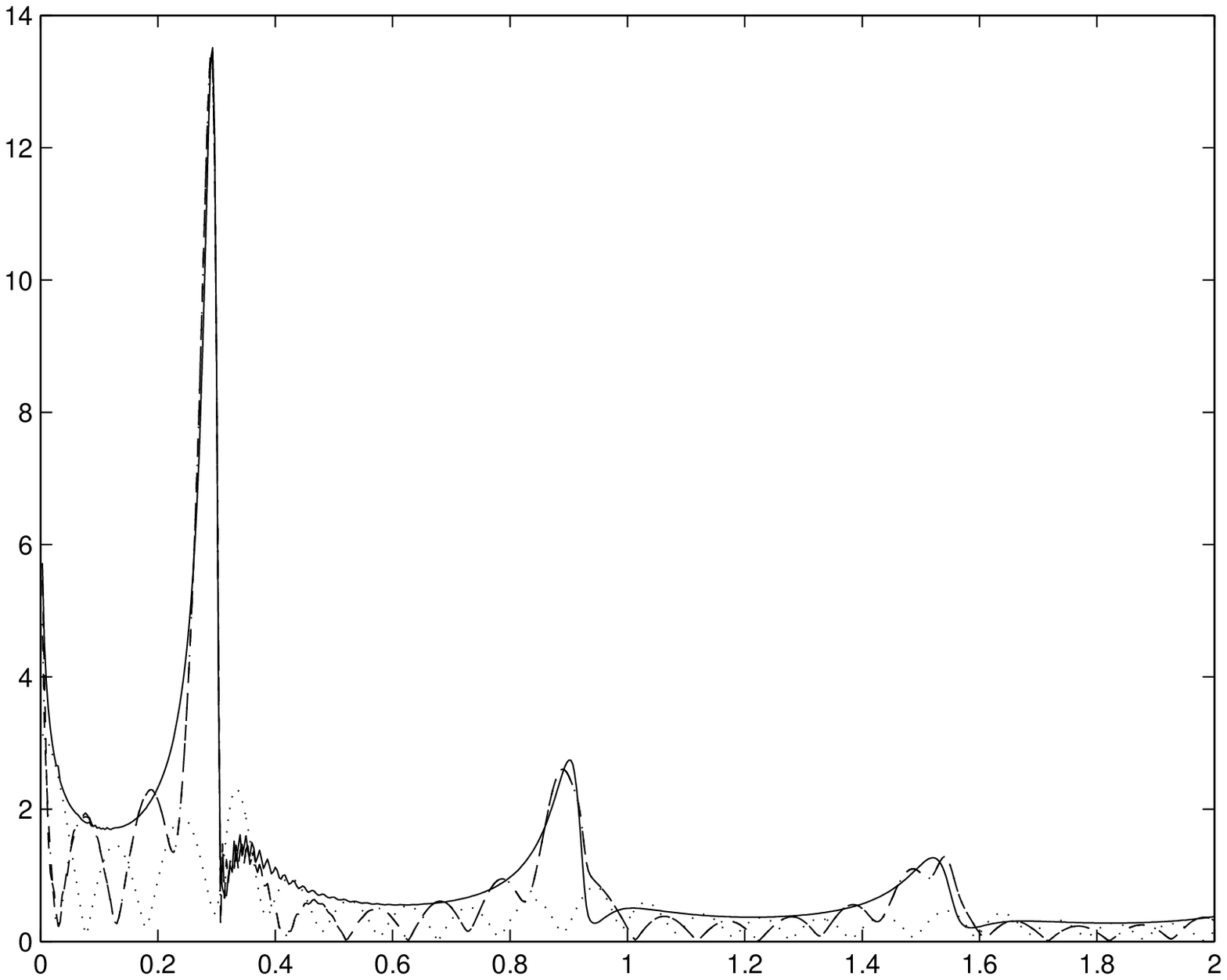}\hfill
\includegraphics[width=6.0cm]{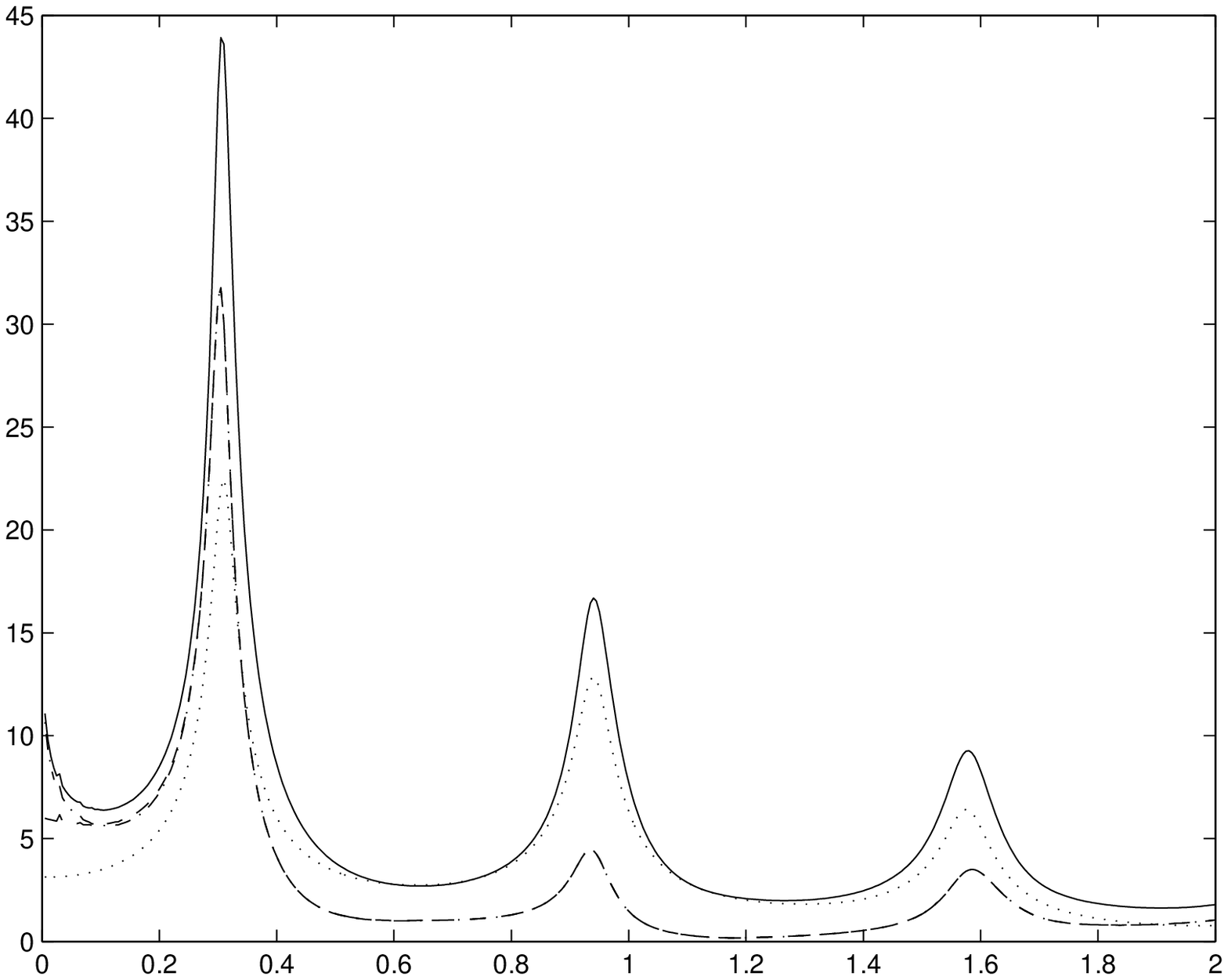}\hfill
\caption{Transfer functions of ground response in Mexico City-like
environment for various source locations and observation points.
Upper-left: $\mathbf{x}^{s}=(0\text{m},3000\text{m})$,
$\mathbf{x}=(3000\text{m},0\text{m})$. Upper-right:
$\mathbf{x}^{s}=(0\text{m},3000\text{m})$,
$\mathbf{x}=(100\text{m},0\text{m})$. Lower-left:
$\mathbf{x}^{s}=(0\text{m},100\text{m})$,
$\mathbf{x}=(3000\text{m},0\text{m})$. Lower-right:
$\mathbf{x}^{s}=(0\text{m},100\text{m})$,
$\mathbf{x}=(100\text{m},0\text{m})$.}
\label{f4}
\end{figure}
in which $c^{0}=600$m/s, $\rho^{1}=1300$kg/m$^{3}$, $c^{1}=60$m/s
(fig. \ref{f4}). In all except the lower right hand panel we again
observe that the ground response is dominated by the SBW1 when the
source is deep and by the SBW2 when the source is shallow. The
exceptional case is that of a shallow source and small epicentral
distance, for which the contributions of the SBW2 and SBW1 to the
overall response are of comparable magnitude, especially near the
first low frequency peak. A plausible cause of this behavior is
the rather large contrast of body-wave velocities between the the
layer and substratum, thus giving rise to a large contribution of
the individual SBW1 at the fundamental Haskell frequency (recall
that this contribution is all the greater the greater the body
wave velocity contrast).

\begin{figure}
[ptb]
\includegraphics[width=6.0cm]{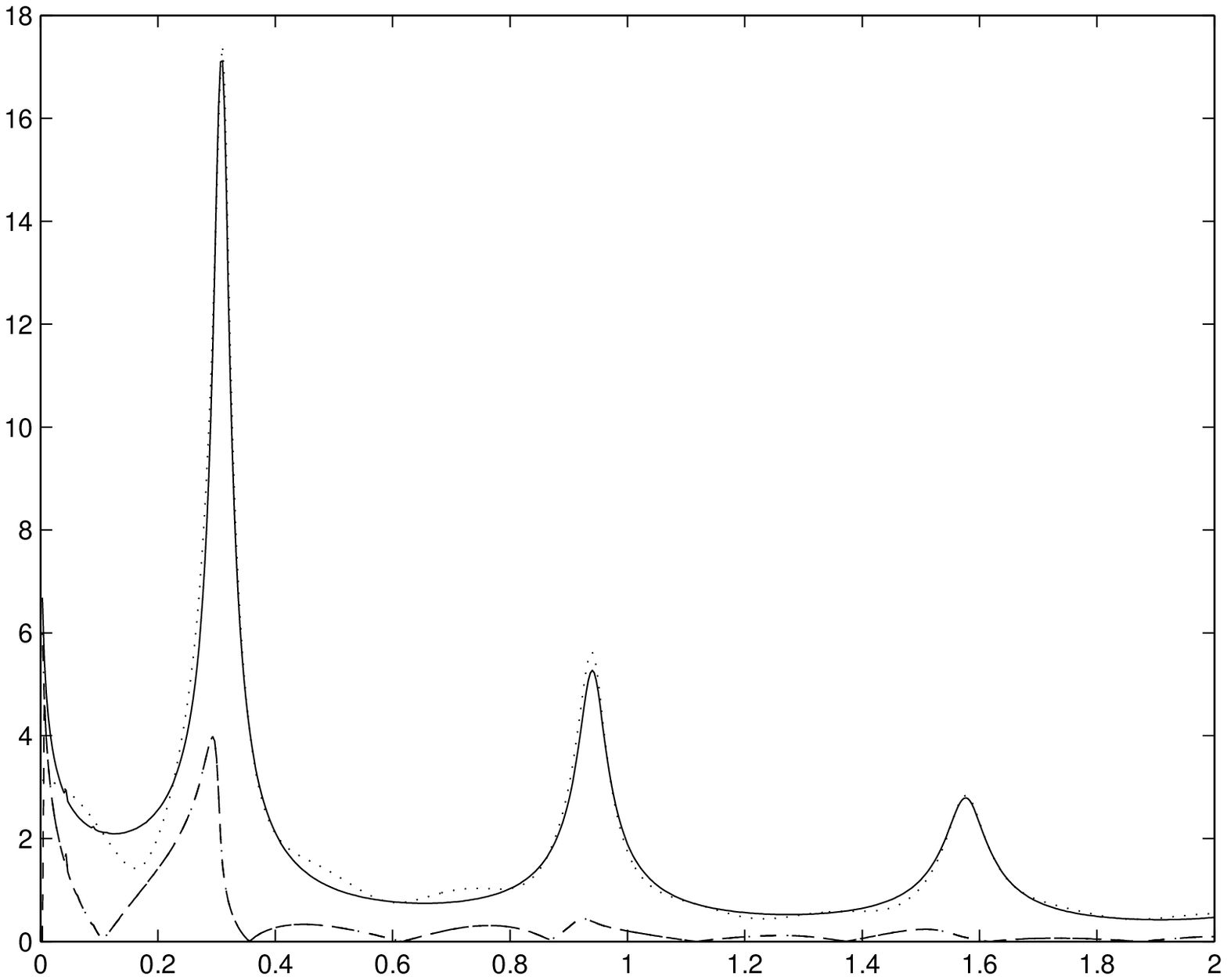}\hfill
\includegraphics[width=6.0cm]{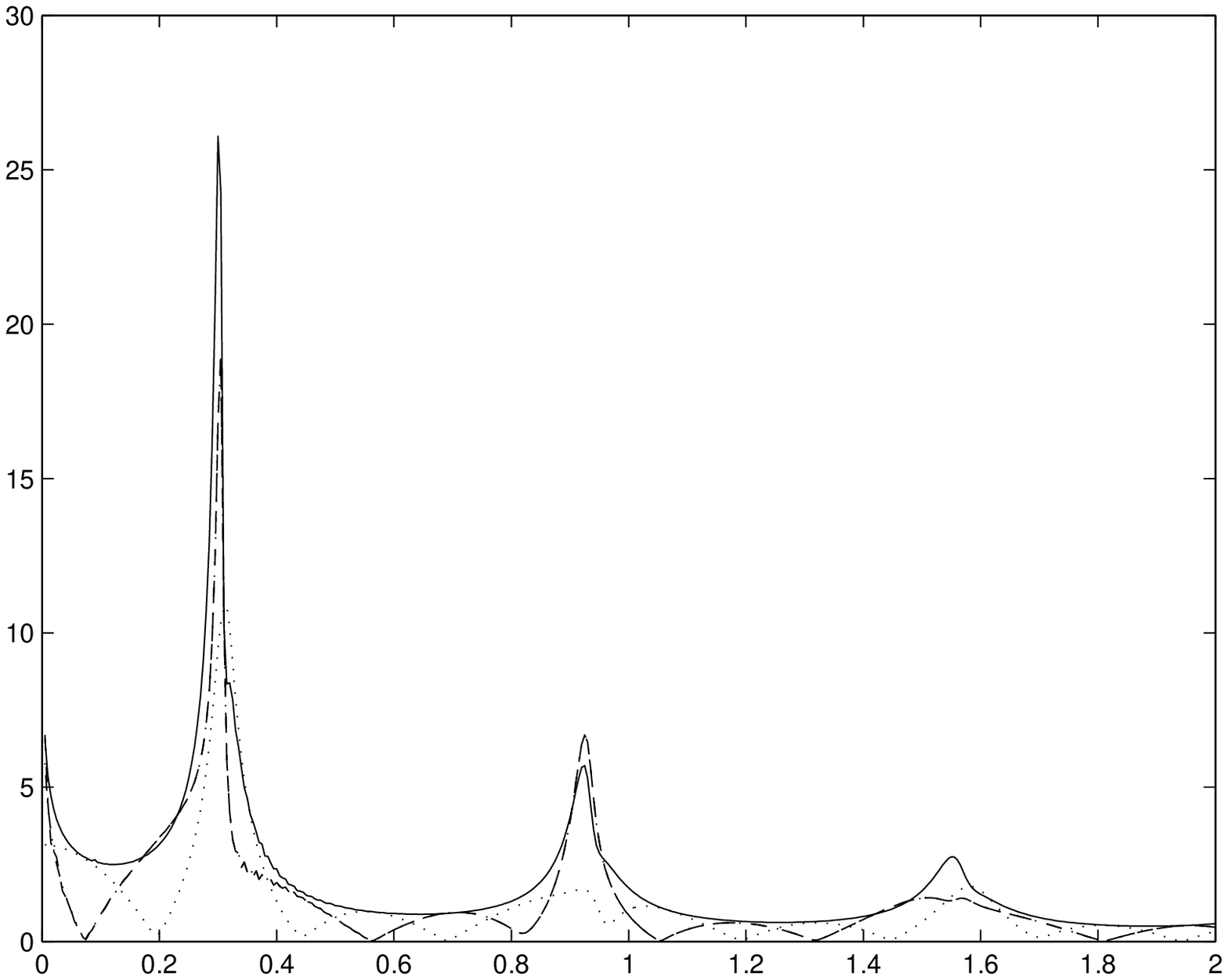}\hfill
\caption{Transfer functions of ground response in Mexico City-like
with harder substratum environment for various source locations
and at the fixed observation point
$\mathbf{x}=(3000\text{m},0\text{m})$ . Left:
$\mathbf{x}^{s}=(0\text{m},3000\text{m})$. Right:
$\mathbf{x}^{s}=(0\text{m},100\text{m})$.}
\label{f5}
\end{figure}

Next we consider a Mexico City-like environment with a somewhat
harder substratum for which $c^{0}=1500$m/s,
$\rho^{1}=1300$kg/m$^{3}$, $c^{1}=60$m/s (fig. \ref{f5}). For a
deep source and large epicentral distance (left panel of the
figure), the response is dominated, as usual, by the SBW1. When
the source is shallow and the epicentral distance is {\it large}
(right panel of the figure) we encounter a new kind of response
characterized by contributions of the SBW1 and SBW2 that are of
comparable magnitude (this was obtained in the previous figure for
a shallow source and {\it small} epicentral distance. That this
should occur even for a large epicentral distance is probably
attributable to the fact that the body wave velocity contrast is
very large (it was smaller in the configuration of the previous
figure) which fact favorizes a substantial contribution of the
SBW1 (notably near the fundamental Haskell frequency), even when
the distance between the source and observation point is large.

\begin{figure}
[ptb]
\includegraphics[width=6.0cm]{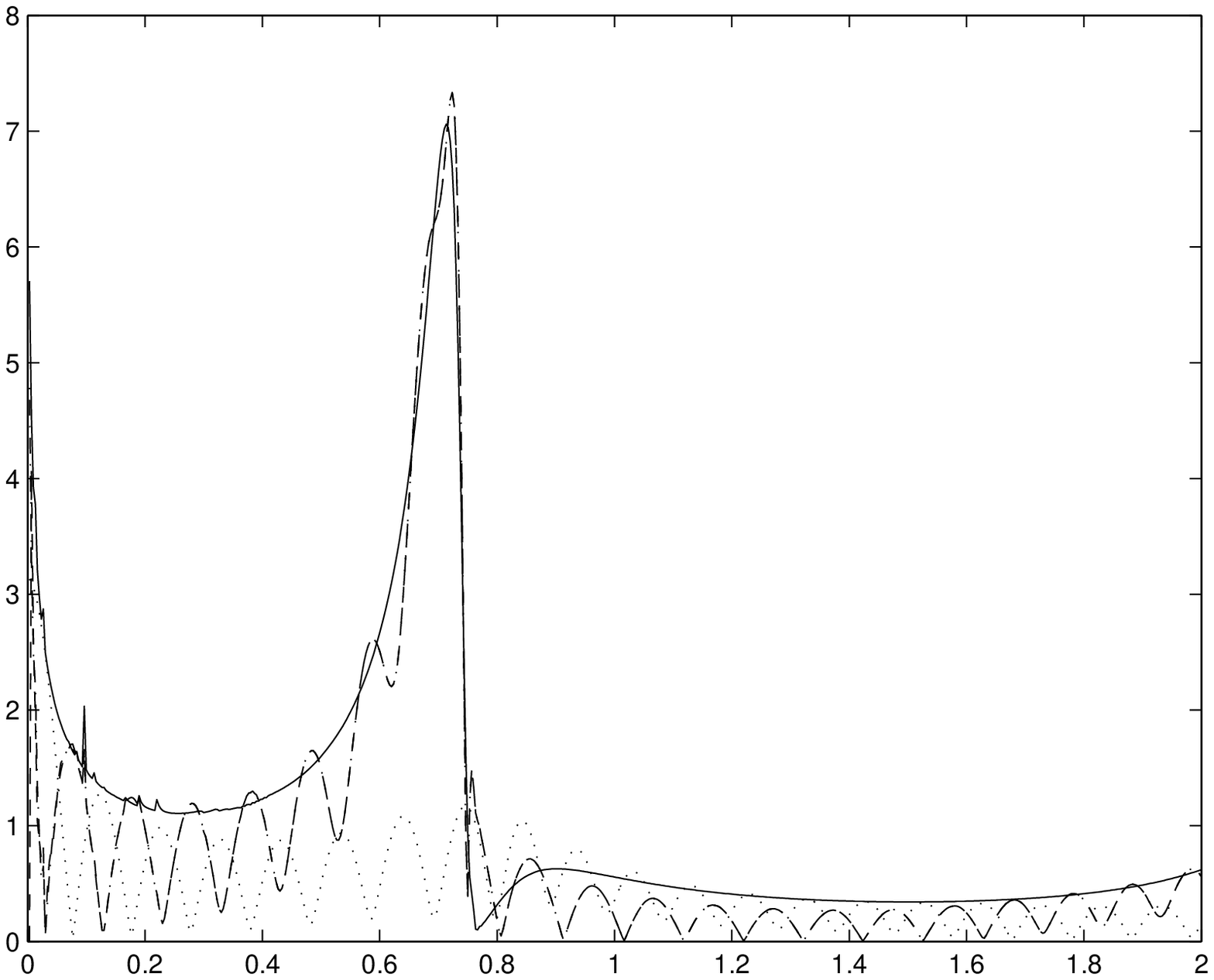}\hfill
\includegraphics[width=6.0cm]{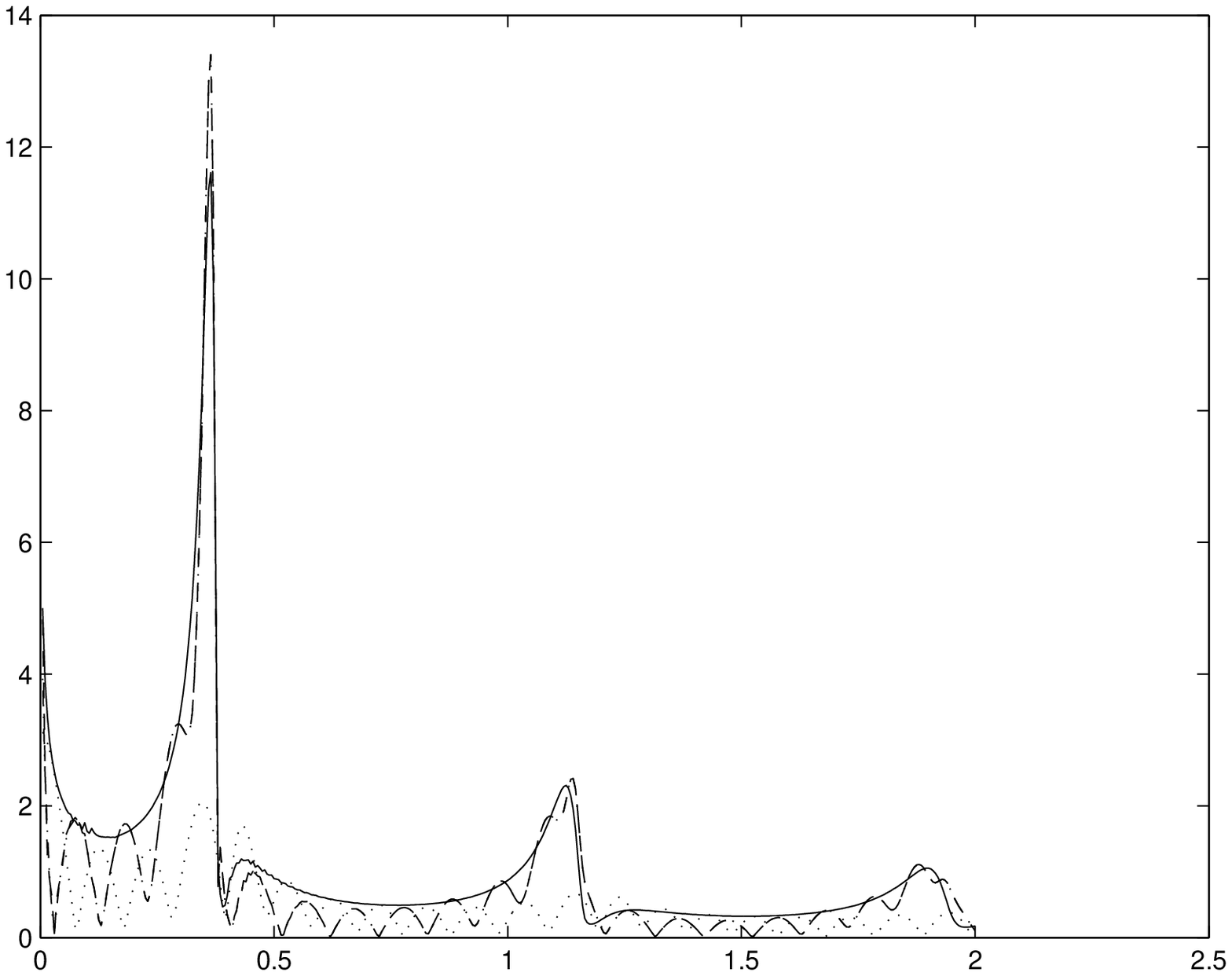}\hfill
\includegraphics[width=6.0cm]{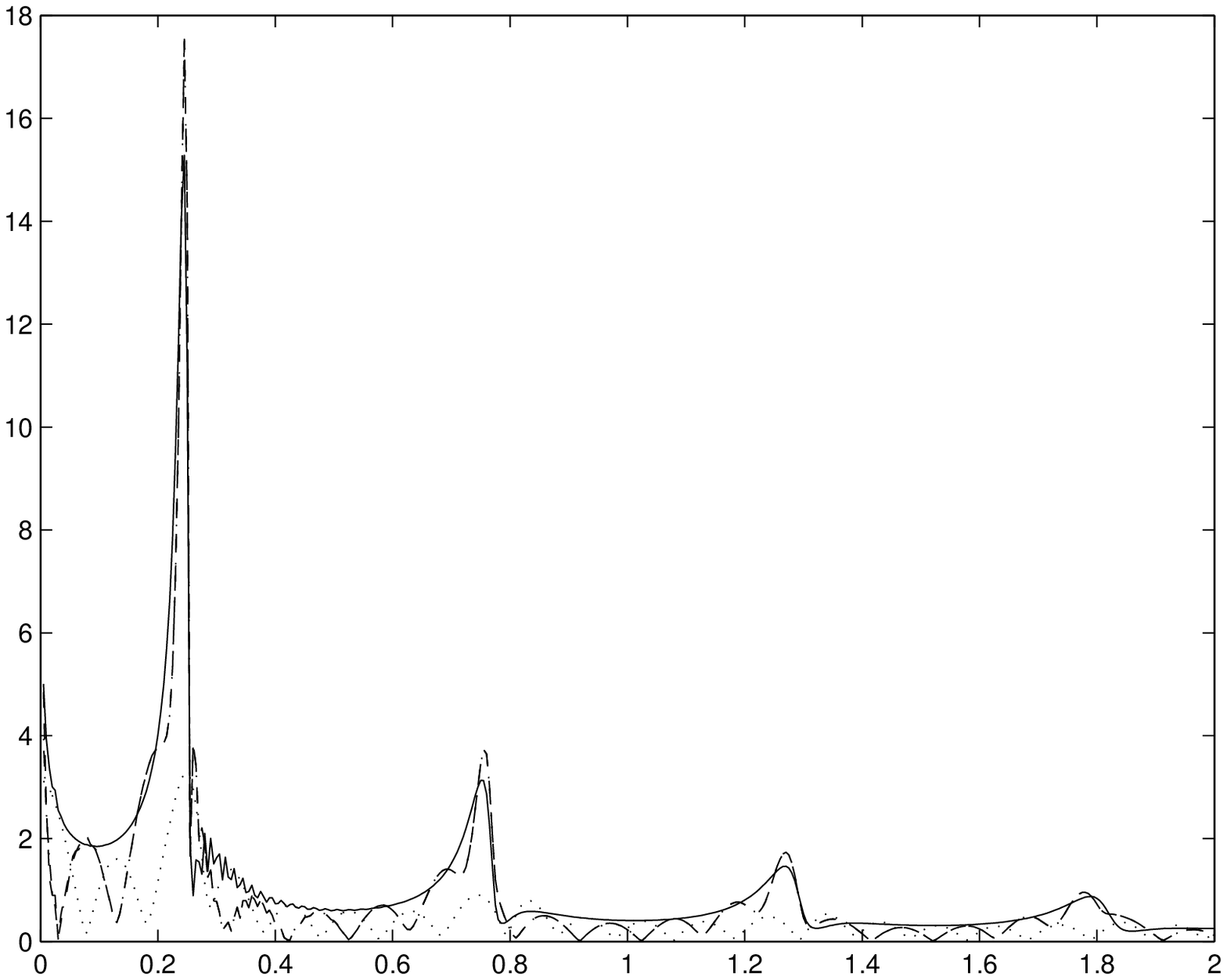}\hfill
\includegraphics[width=6.0cm]{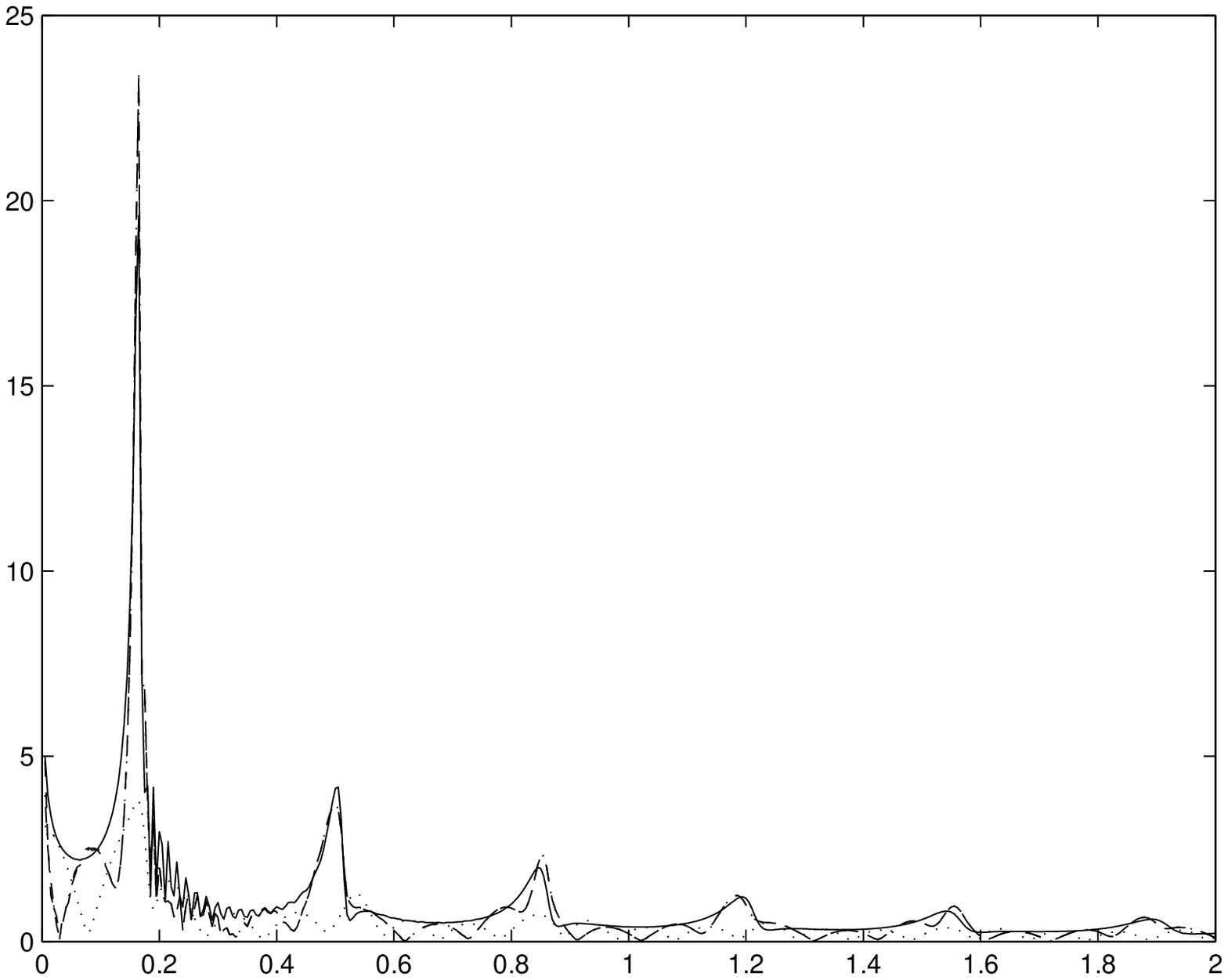}\hfill
\caption{Transfer functions of ground response in Mexico City-like
environment for $\mathbf{x}^{s}=(0\text{m},100\text{m})$,
$\mathbf{x}=(3000\text{m},0\text{m})$ and various layer
thicknesses $h$. Upper-left: $h=20$m. Upper-right: $h=40$m.
Lower-left: $h=60$m. Lower-right: $h=90$m.}
\label{f6}
\end{figure}

The last result in this series concerns once again the Mexico
City-like environment in which $c^{0}=600$m/s,
$\rho^{1}=1300$kg/m$^{3}$, $c^{1}=60$m/s (fig. \ref{f6}). We are
now interested in evaluating the effect of changes in the layer
thickness $h$ for a shallow source and large epicentral distance.
We observe in the figure that  the
response is dominated by the cumulative contribution of the SBW2
for all the layer thicknesses.
Furthermore, the number and finesse of the resonance peaks in the
interval $[0,2{\text Hz}]$ increases with $h$, the dominant peak
always being the one associated with the resonant excitation of the
first (lowest-frequency) Love mode and being located at a
frequency that is all the lower the larger is $h$.
%
\subsection{Time records of various input pulses for different
combinations of source and observation point coordinates} 
\begin{figure}
[ptb]
\includegraphics[width=6.0cm]{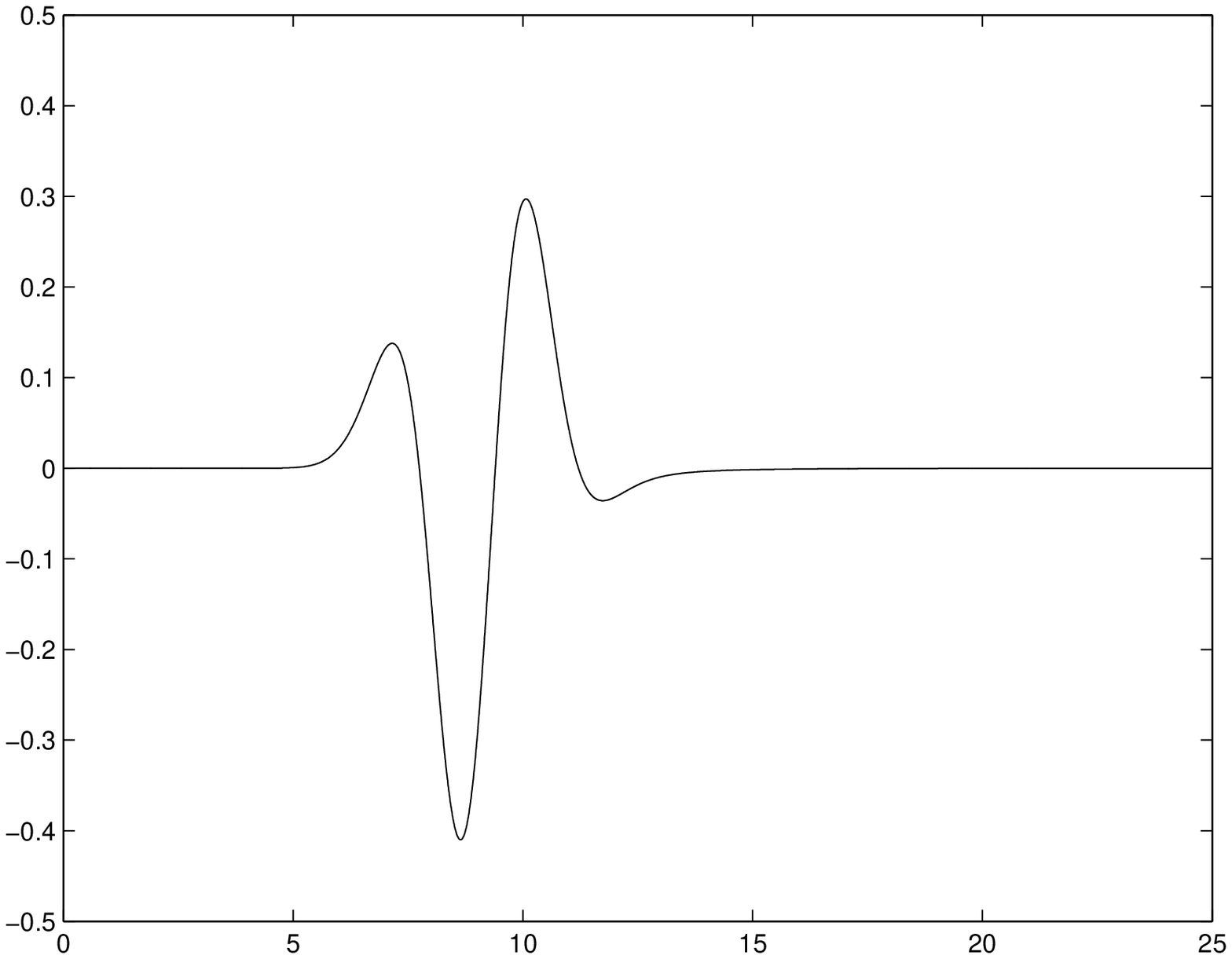}\hfill
\includegraphics[width=6.0cm]{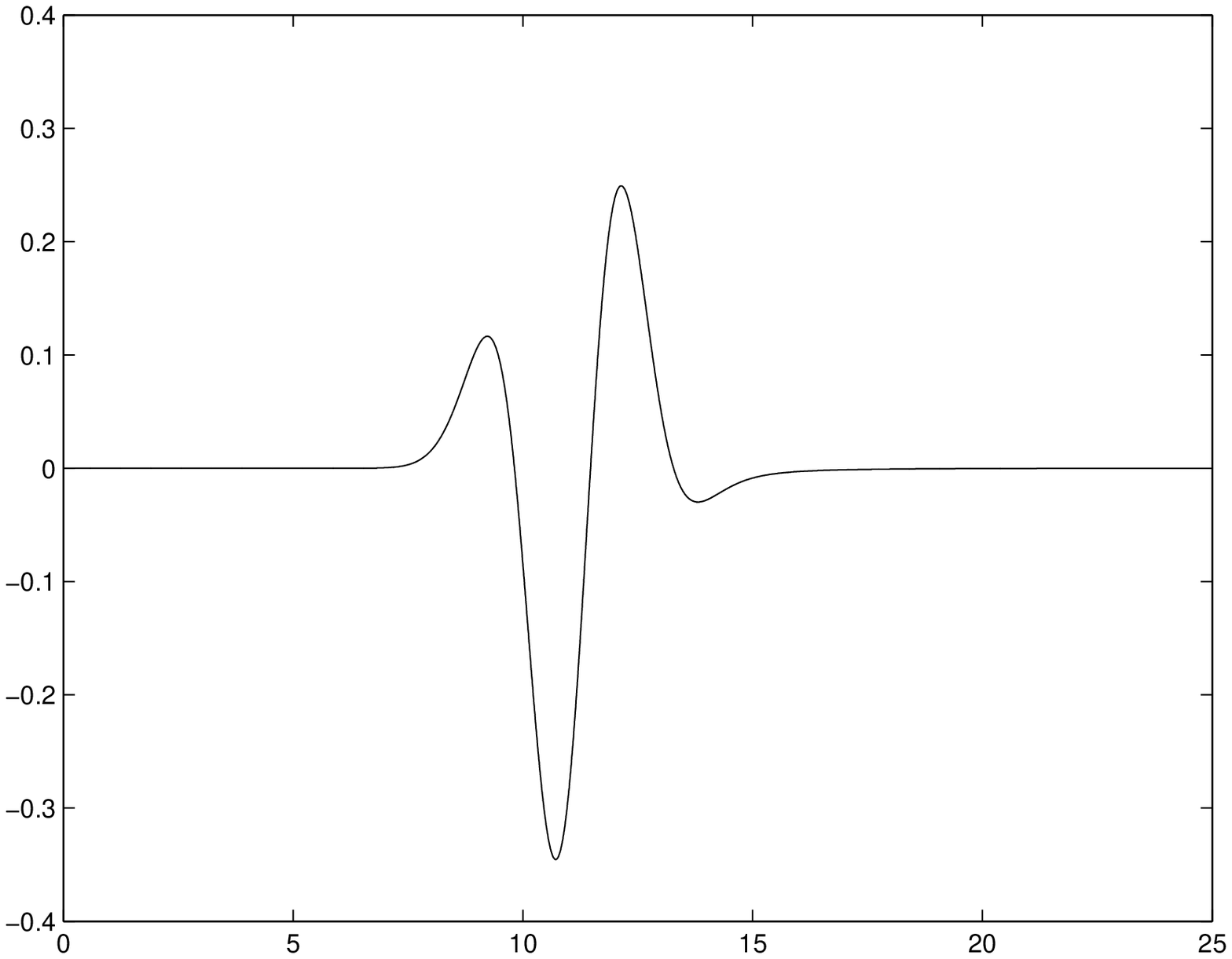}\hfill
\includegraphics[width=6.0cm]{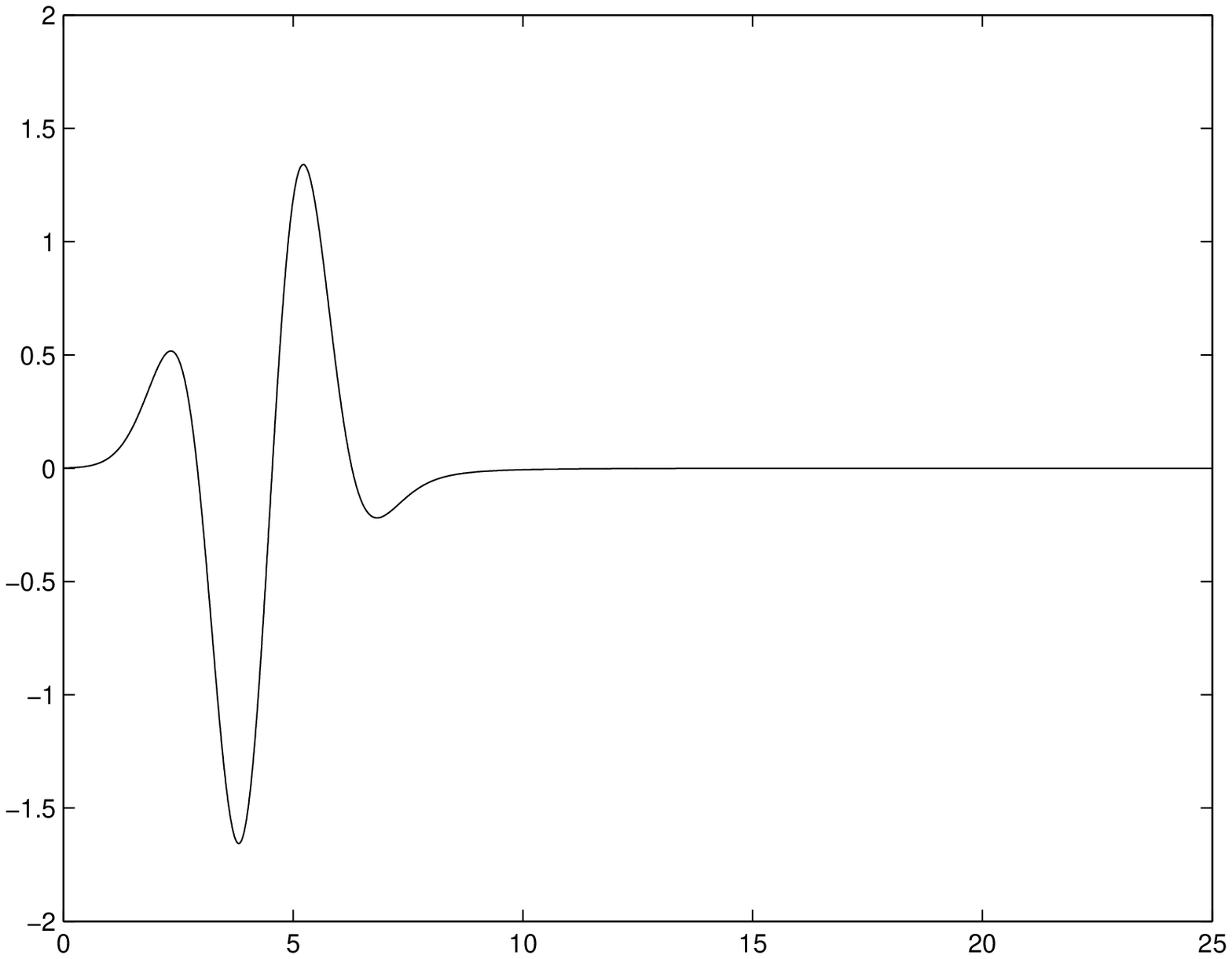}\hfill
\includegraphics[width=6.0cm]{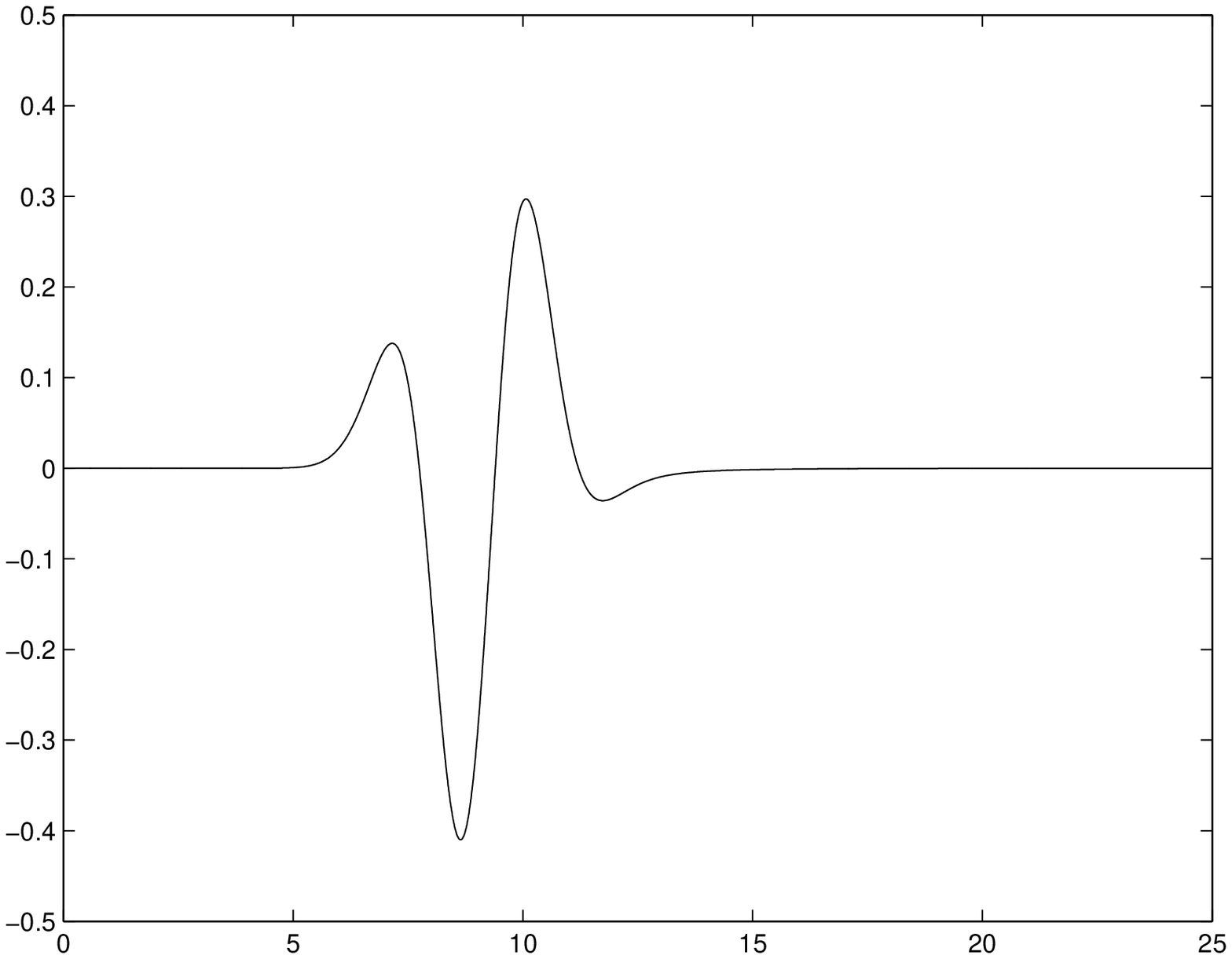}\hfill
\caption{Time records of the incident field (i.e.,
$u^{i}(\mathbf{x},t)$ versus $t$(s)) in the substratum (considered
to fill all space and wherein $\beta^{0}=600$m/s) corresponding to
a $\nu_{0}=0.25$Hz pulse for:
$\mathbf{x}^{s}=(0\text{m},3000\text{m})$,
$\mathbf{x}=(100\text{m},0\text{m})$ (upper left panel),
$\mathbf{x}^{s}=(0\text{m},3000\text{m})$,
$\mathbf{x}=(3000\text{m},0\text{m})$ (upper right panel),
$\mathbf{x}^{s}=(0\text{m},100\text{m})$,
$\mathbf{x}=(100\text{m},0\text{m})$ (lower left panel), and
$\mathbf{x}^{s}=(0\text{m},100\text{m})$,
$\mathbf{x}=(3000\text{m},0\text{m})$ (lower right panel).}
\label{f6a}
\end{figure}

\begin{figure}
[ptb]
\includegraphics[width=6.0cm]{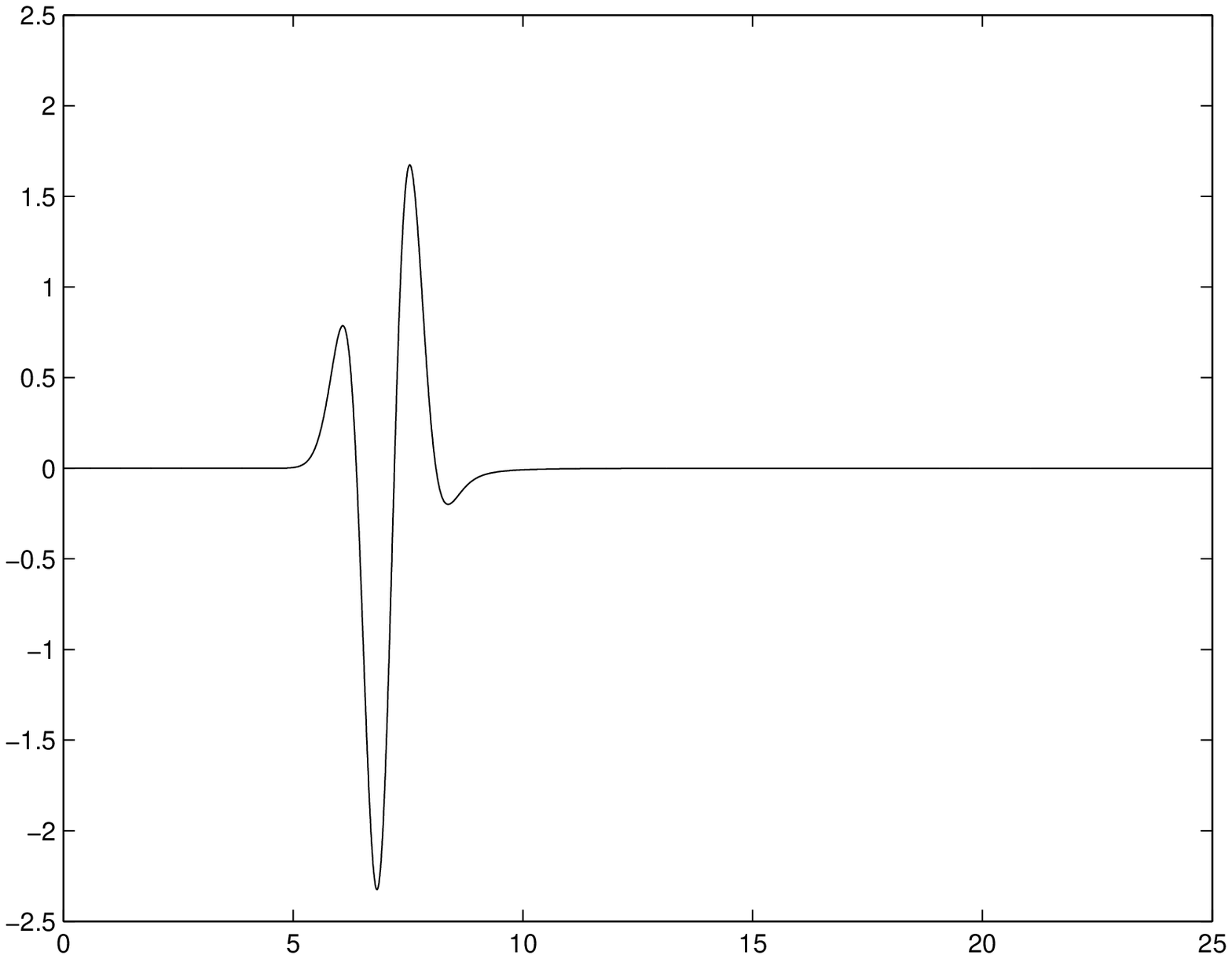}\hfill
\includegraphics[width=6.0cm]{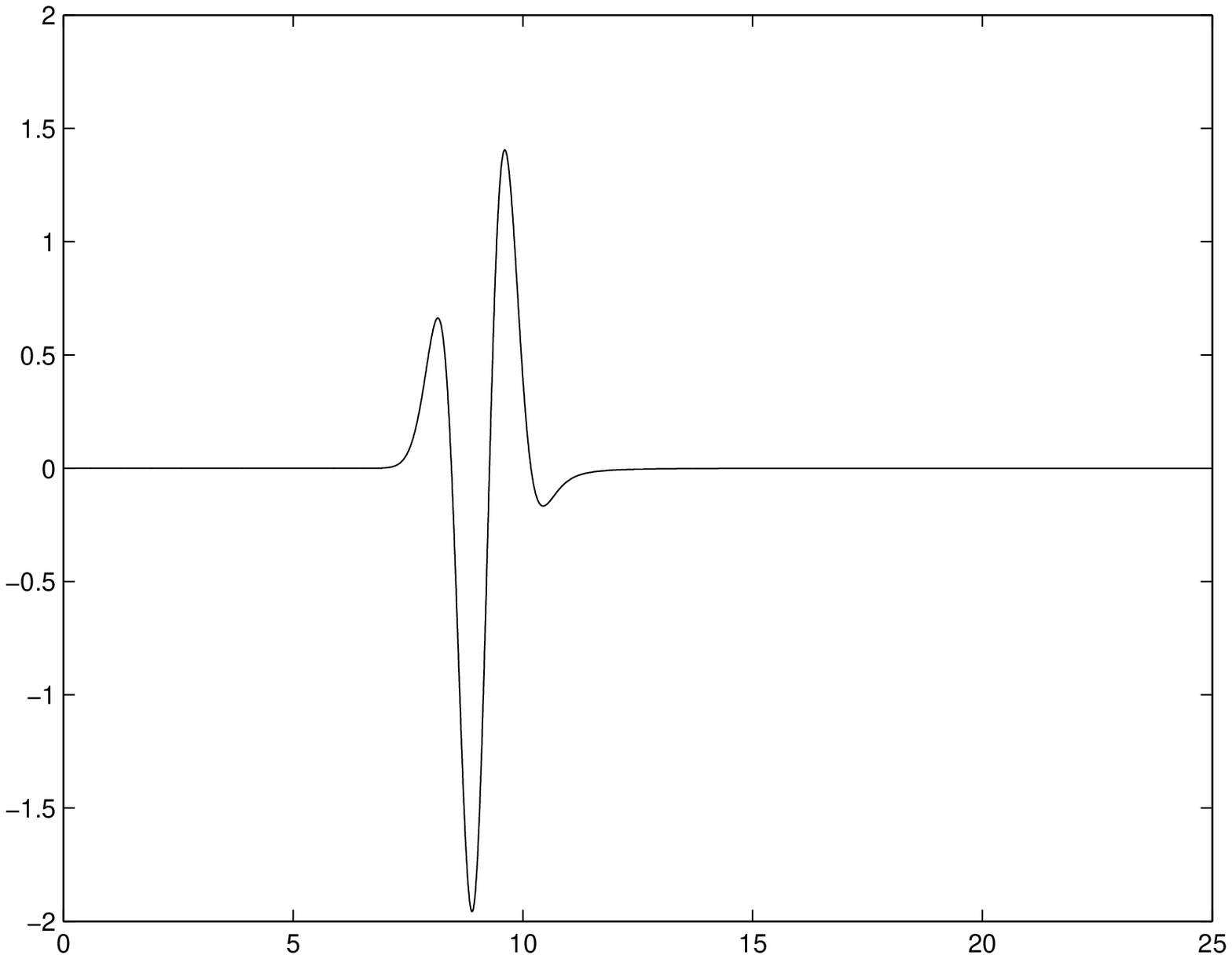}\hfill
\includegraphics[width=6.0cm]{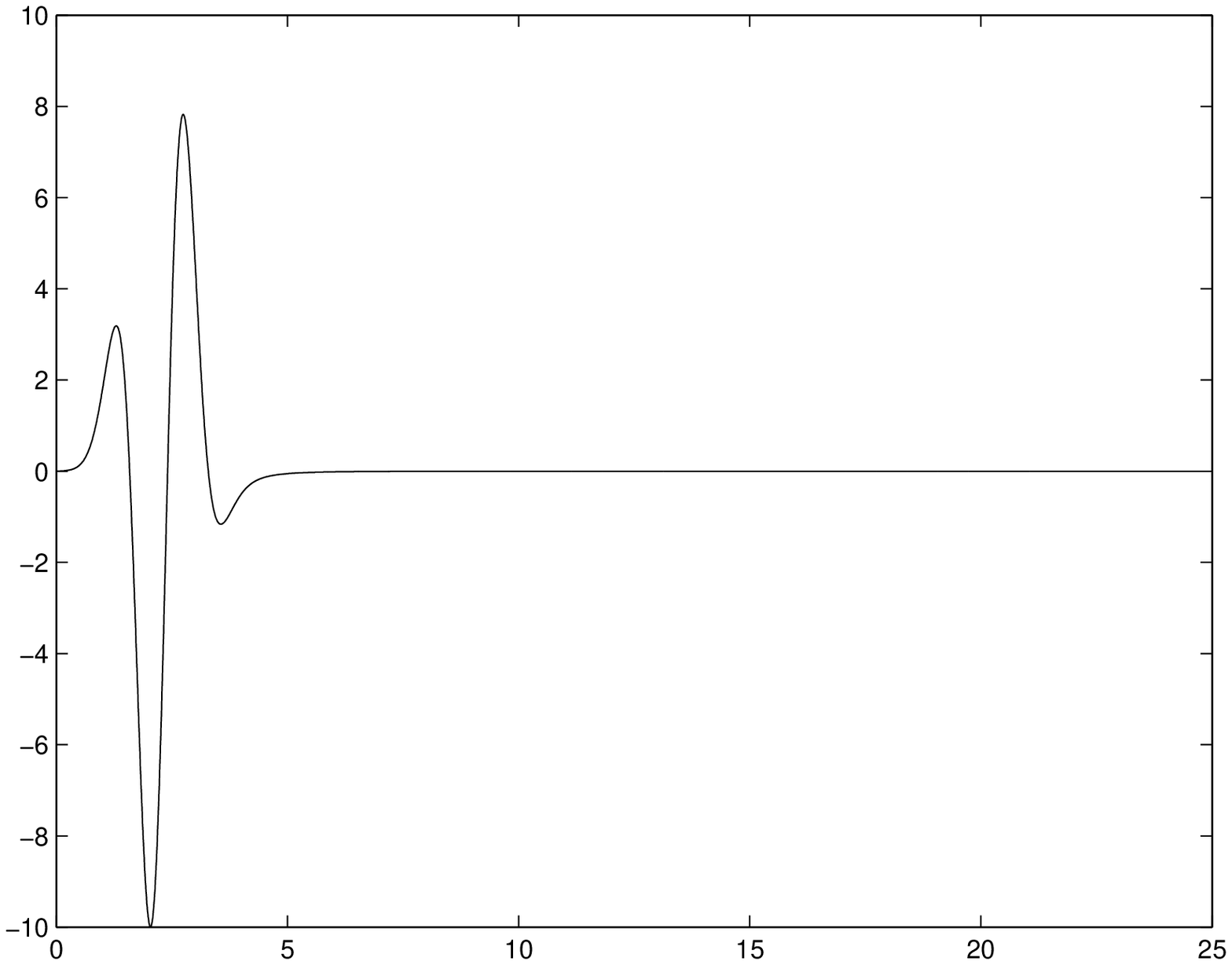}\hfill
\includegraphics[width=6.0cm]{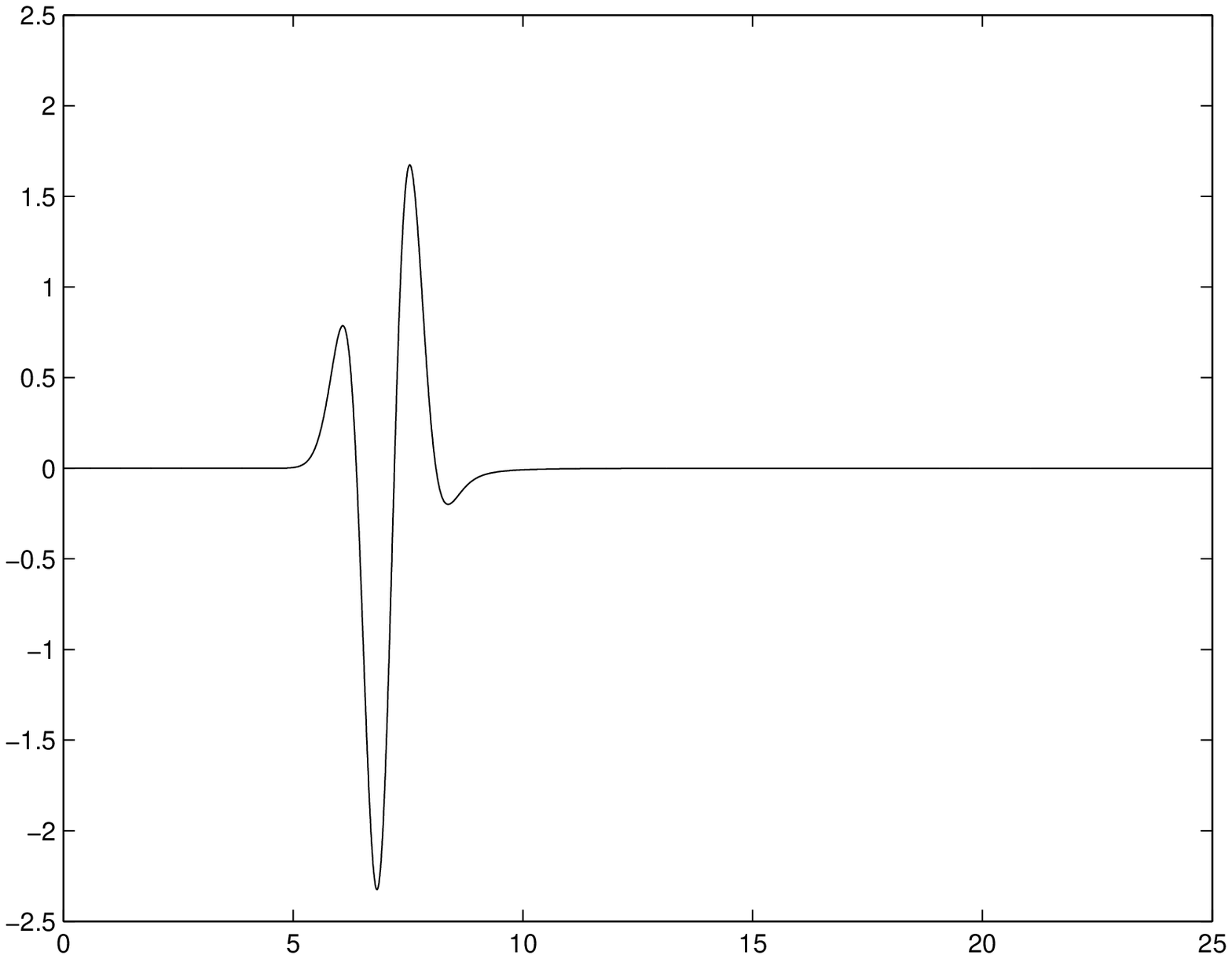}\hfill
\caption{Time records of the incident field in the substratum
(considered to fill all space and wherein $\beta^{0}=600$m/s)
corresponding to a $\nu_{0}=0.5$Hz pulse for:
$\mathbf{x}^{s}=(0\text{m},3000\text{m})$,
$\mathbf{x}=(100\text{m},0\text{m})$ (upper left panel),
$\mathbf{x}^{s}=(0\text{m},3000\text{m})$,
$\mathbf{x}=(3000\text{m},0\text{m})$ (upper right panel),
$\mathbf{x}^{s}=(0\text{m},100\text{m})$,
$\mathbf{x}=(100\text{m},0\text{m})$ (lower left panel), and
$\mathbf{x}^{s}=(0\text{m},100\text{m})$,
$\mathbf{x}=(3000\text{m},0\text{m})$ (lower right panel).}
\label{f6b}
\end{figure}

\begin{figure}
[ptb]
\includegraphics[width=6.0cm]{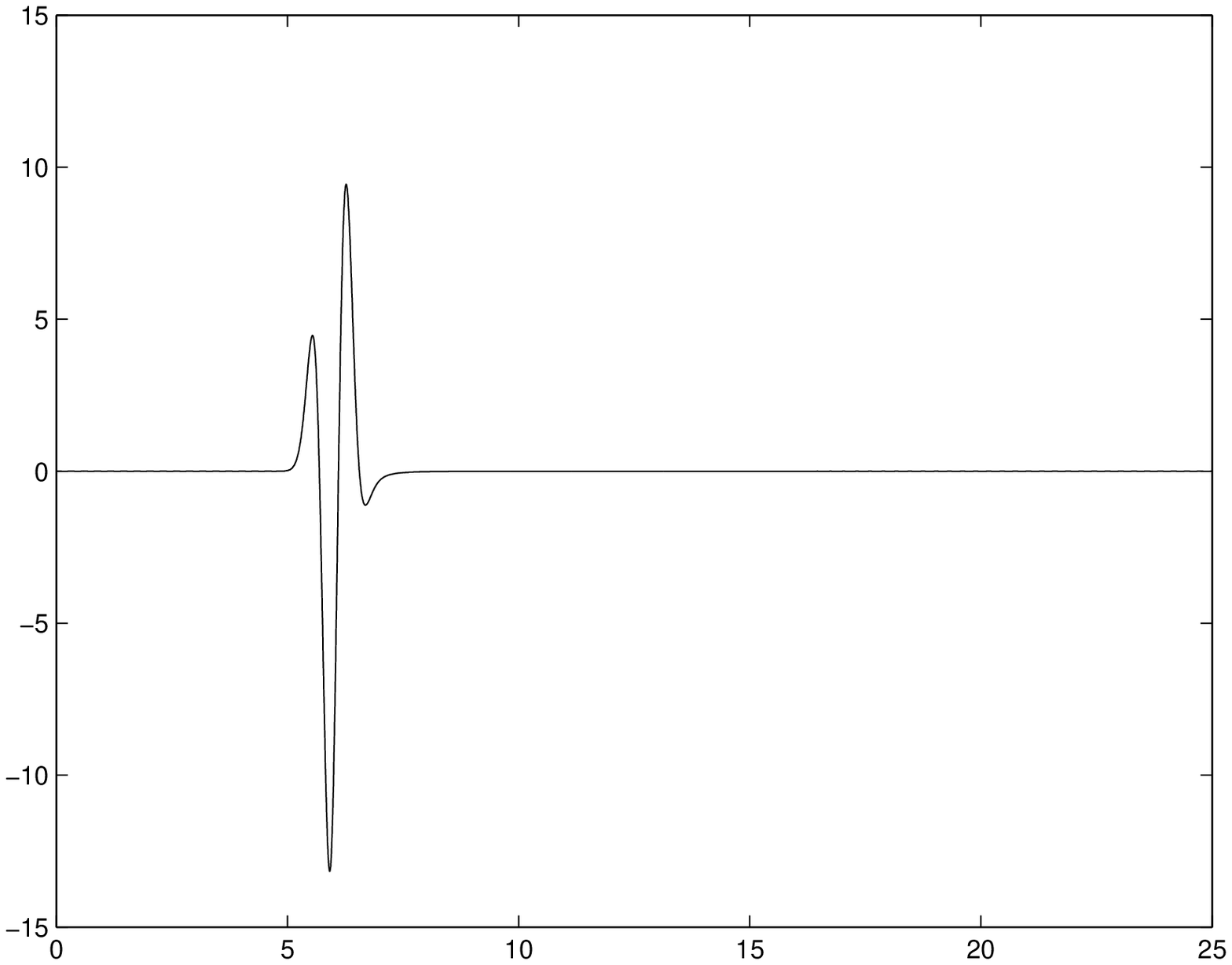}\hfill
\includegraphics[width=6.0cm]{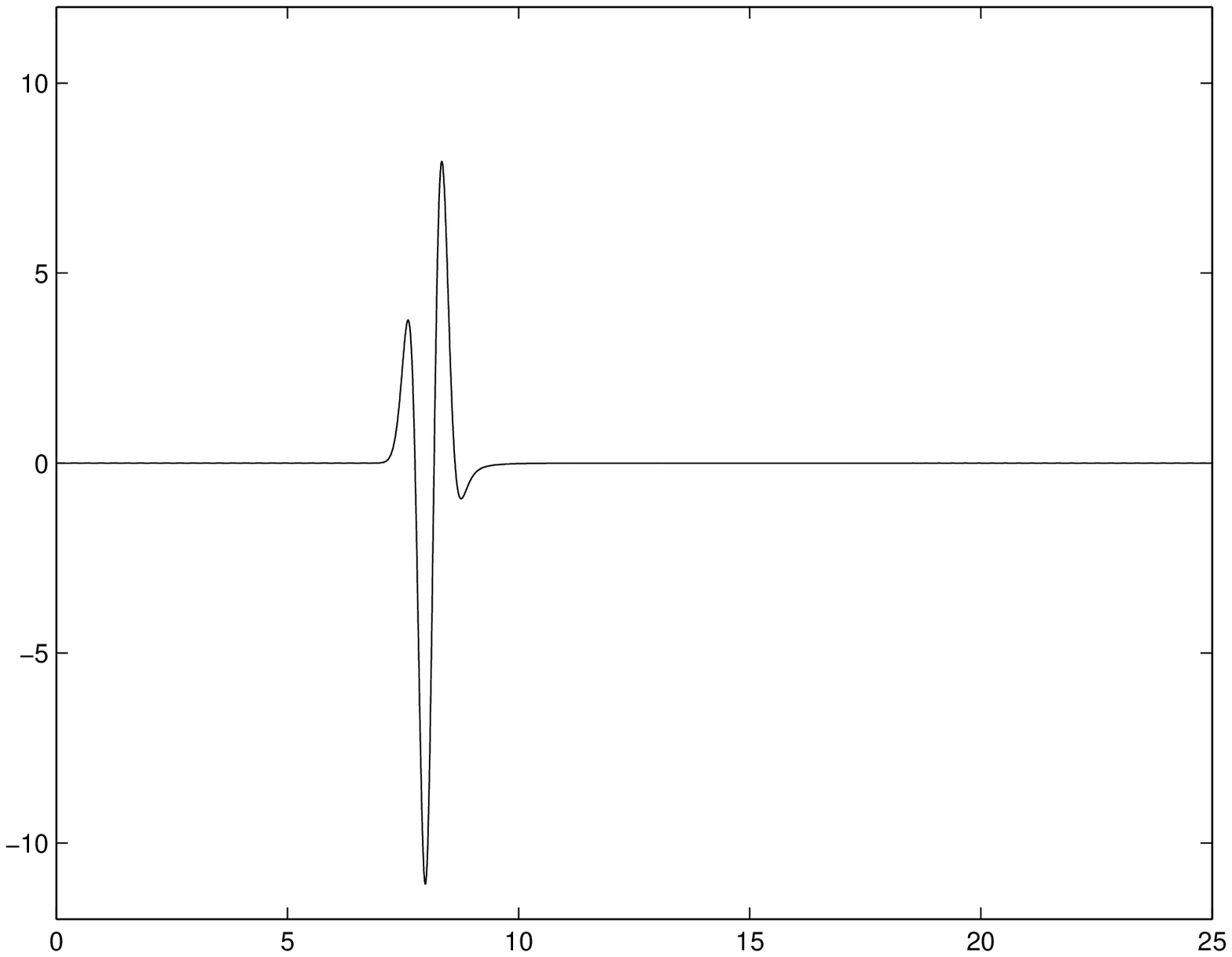}\hfill
\includegraphics[width=6.0cm]{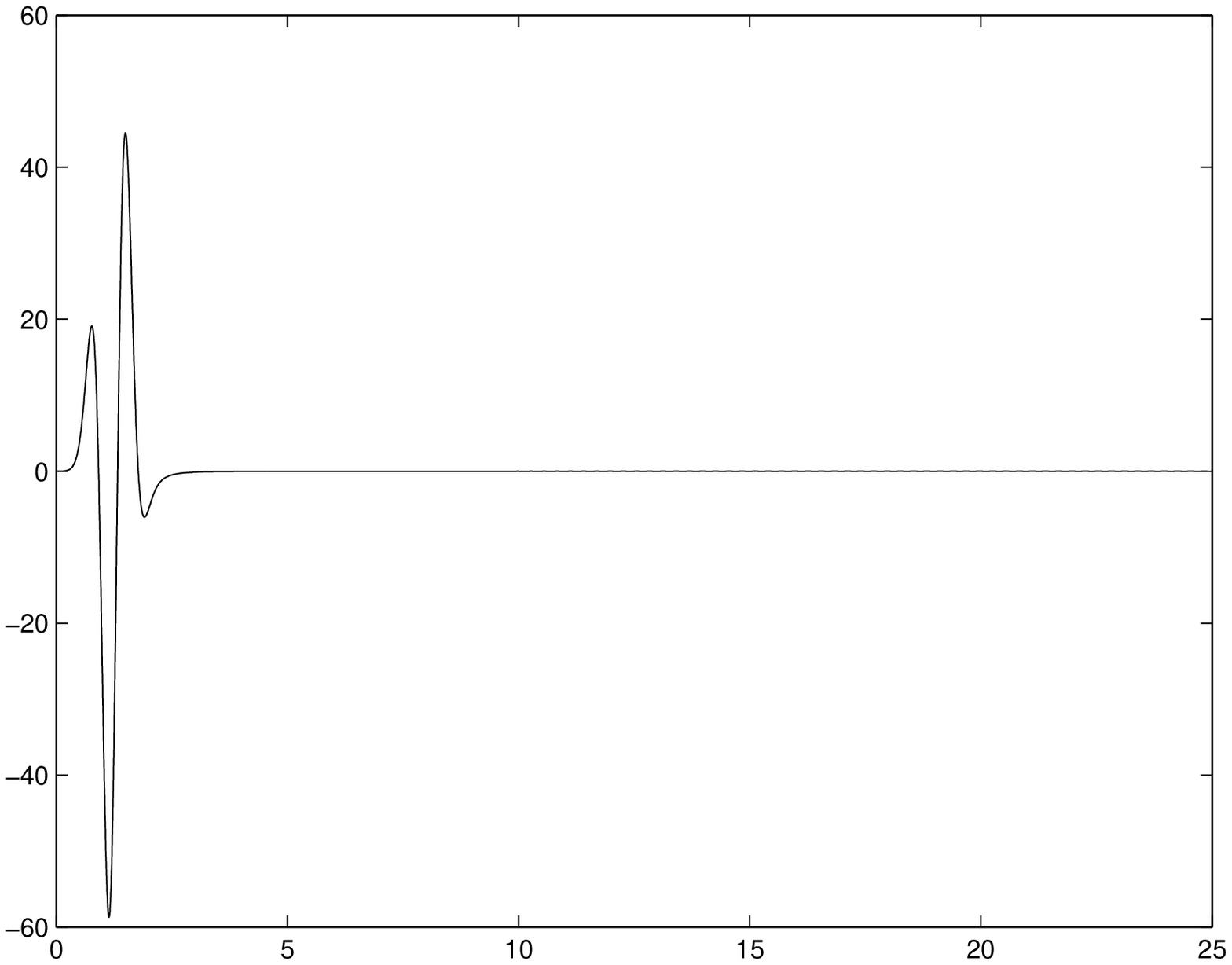}\hfill
\includegraphics[width=6.0cm]{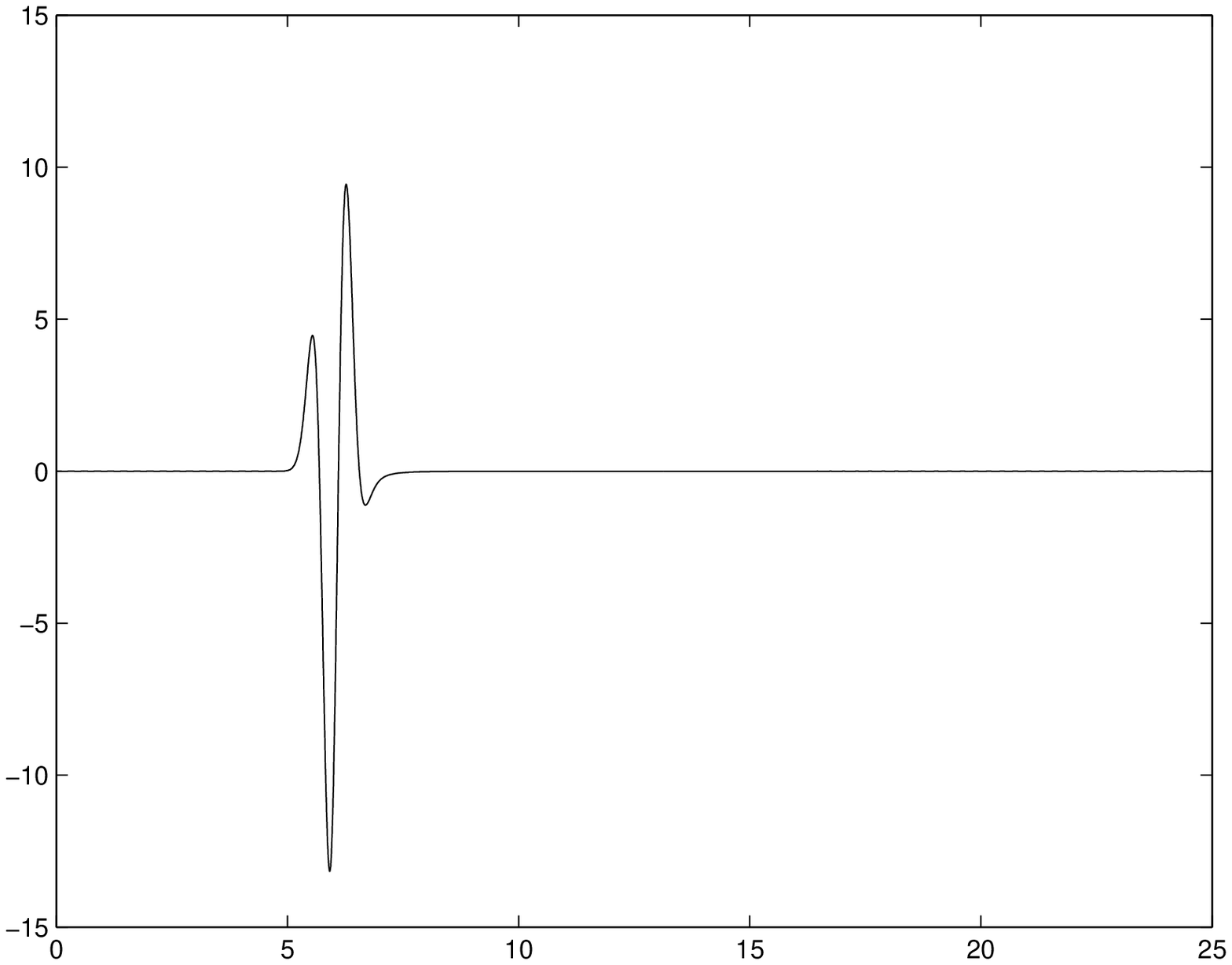}\hfill
\caption{Time records of the incident field in the substratum
(considered to fill all space and wherein $\beta^{0}=600$m/s)
corresponding to a $\nu_{0}=1.0$Hz pulse for:
$\mathbf{x}^{s}=(0\text{m},3000\text{m})$,
$\mathbf{x}=(100\text{m},0\text{m})$ (upper left panel),
$\mathbf{x}^{s}=(0\text{m},3000\text{m})$,
$\mathbf{x}=(3000\text{m},0\text{m})$ (upper right panel),
$\mathbf{x}^{s}=(0\text{m},100\text{m})$,
$\mathbf{x}=(100\text{m},0\text{m})$ (lower left panel), and
$\mathbf{x}^{s}=(0\text{m},100\text{m})$,
$\mathbf{x}=(3000\text{m},0\text{m})$ (lower right panel).}
\label{f6c}
\end{figure}

In the
following, we exhibit (figs. \ref{f6a}-\ref{f6c}) time records of
the three pseudo-Ricker pulses having frequencies:
$\nu_{0}=0.25$Hz, 0.5Hz, 1.0Hz (whose spectra were shown
previously in fig. \ref{f1a}). This is done for all combinations
of the four coordinates (assuming $x_{1}^{s}=0$m, $x_{2}=0$m):
$x_{2}^{s}=100$m, $x_{2}^{s}=3000$m, $x_{1}=100$m, $x_{1}=3000$m.
As expected, the pulses have the same shape for all source and
observation point locations since the substratum was assumed to be
an elastic (i.e., non-dispersive) medium; however their maxima
change as a function of these locations. Not unexpectedly, the
largest pulses are those for which the source to observation point
distances are the smallest. Of particular interest is the fact
that the input pulse duration is approximately $2/\nu_{0}$, which
corresponds to $\sim$8s for the 0.25Hz pulse, $\sim$4s for the
0.5Hz pulse, and $\sim$2s for the 1.0Hz pulse. As will be seen
hereafter, the response to these pulses in the layered
configuration is generally of much longer duration.
\subsection{Comparison of frequency and time domain responses for
constant source-to-observation point distances} 
\begin{figure}
[ptb]
\includegraphics[width=6.0cm]{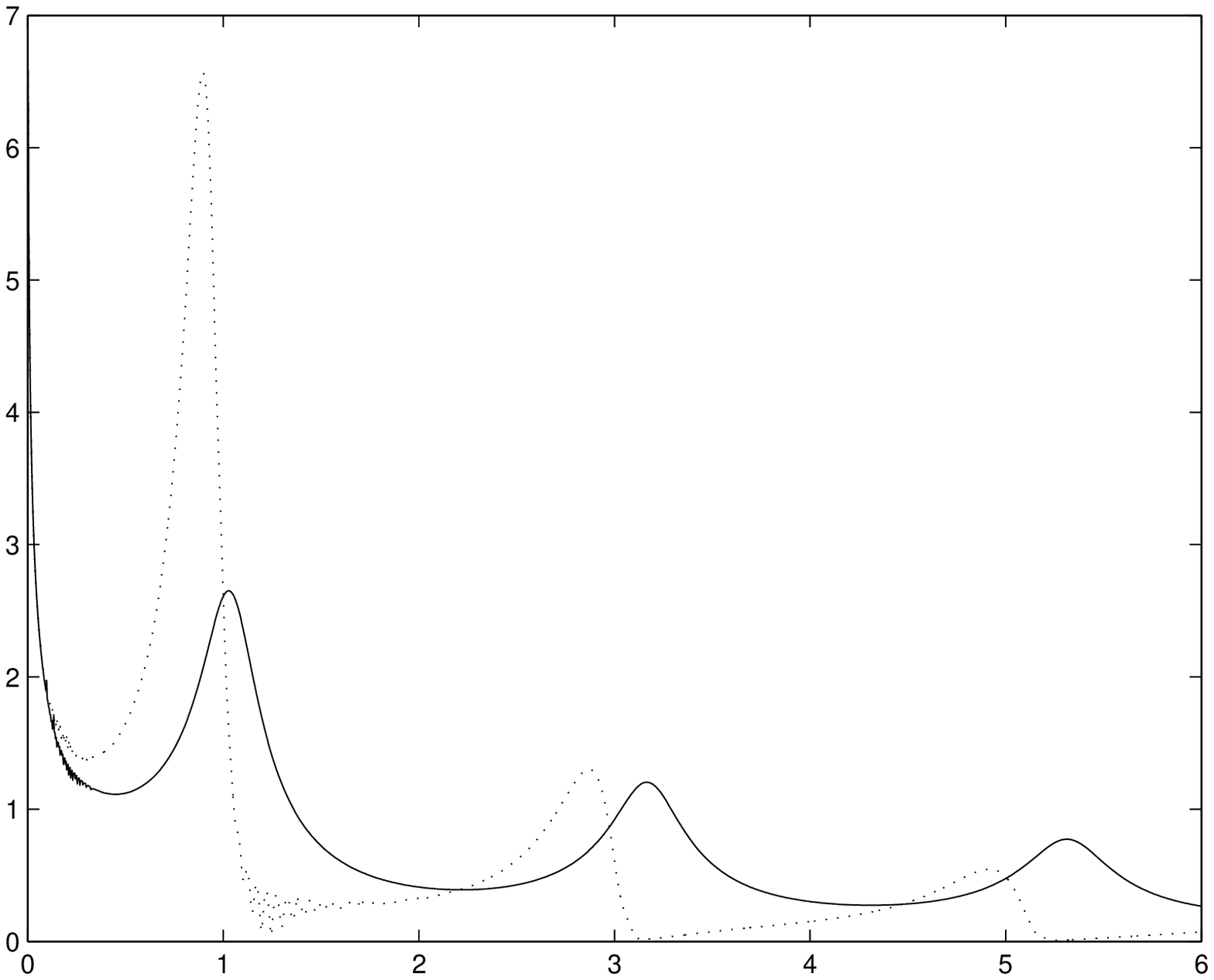}\hfill
\includegraphics[width=6.0cm]{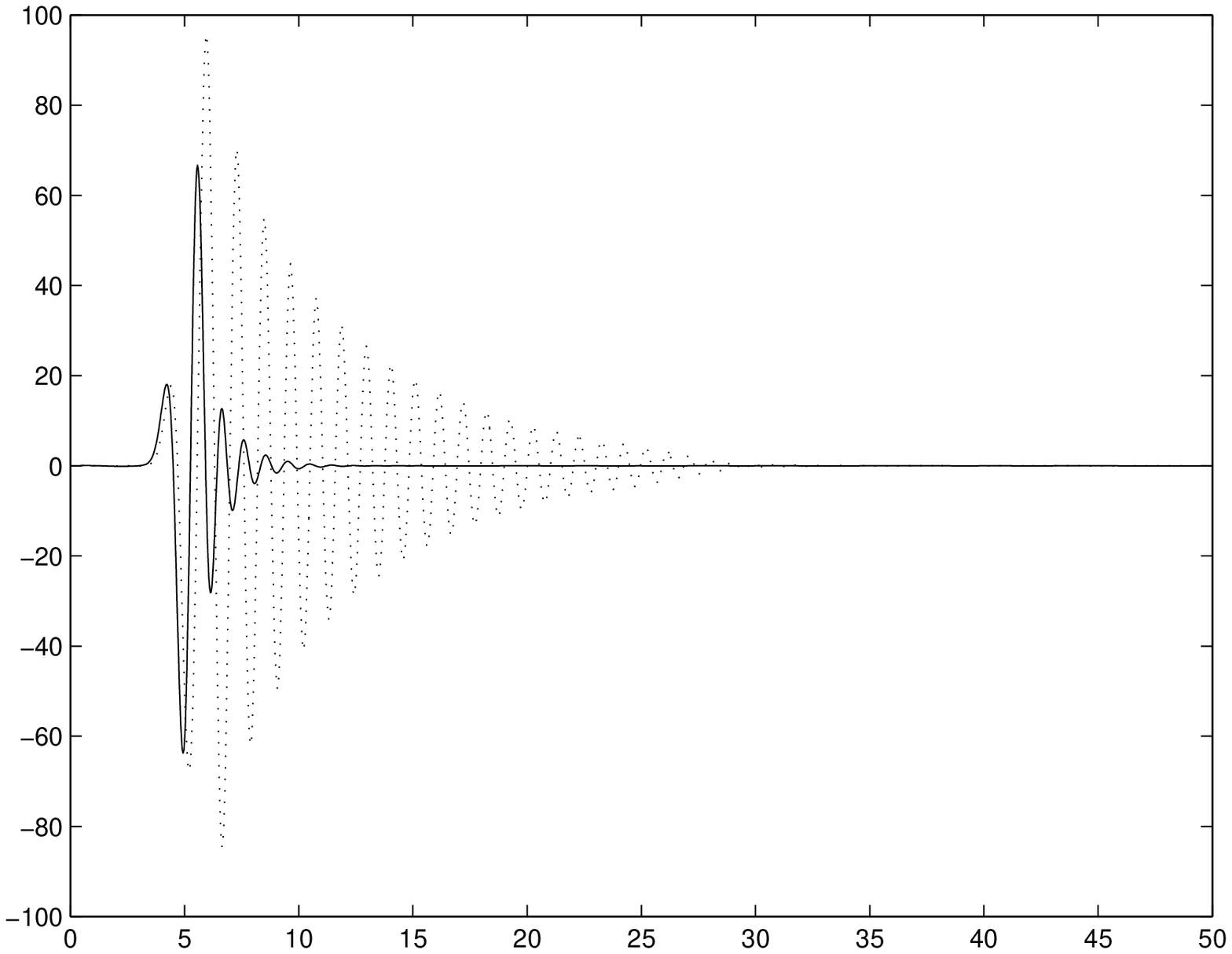}\hfill
\caption{Frequency and time domain representations of ground
response in Nice-like environment for constant
source-to-observation point distances;
$\mathbf{x}^{s}=(0\text{m},3000\text{m})$,
$\mathbf{x}=(100\text{m},0\text{m})$ (solid line curves) and
$\mathbf{x}^{s}=(0\text{m},100\text{m})$,
$\mathbf{x}=(3000\text{m},0\text{m})$ (dotted line curves).  Left:
transfer functions. Right: time-domain responses (i.e.,
$u^{1}(\mathbf{x},t)$ versus $t$(s)) to a $\nu_{0}=0.5$Hz pulse.}
\label{f7}
\end{figure}

Again, we begin by
a configuration thought to be representative of that in the
central portion of the city of Nice (France) wherein
$c^{0}=1000$m/s, $\rho^{1}=1800$kg/m$^{3}$, $c^{1}=200$m/s. Two
constant source-to-observation point distance situations are
considered: a) $\mathbf{x}^{s}=(0\text{m},3000\text{m})$,
$\mathbf{x}=(100\text{m},0\text{m})$ (solid line curves in fig.
\ref{f7}), and b) $\mathbf{x}^{s}=(0\text{m},100\text{m})$,
$\mathbf{x}=(3000\text{m},0\text{m})$ (dotted line curves in fig.
\ref{f7}).
We notice in the left panel of fig. \ref{f7} that the first bump
of the transfer function occurs at a lower frequency  when the
source is near to the layer than when it is far from the layer,
which fact suggests that the lower-frequency bump is due to the
(resonant) excitation of the fundamental Love mode (SBW2) whereas
the higher-frequency peak is associated with the first
(non-resonant) interference (SBW1) maximum. The same remarks apply
to the higher-order bumps. Moreover, the value of the transfer
function at the first couple of bumps is much larger due to Love
mode excitation than to constructive interference effects, and
since the widths of these two bumps are approximately the same,
the finesse (height/width ratio) of the Love mode peak is larger
than that of the interference peak. The translation of this into
the time domain is that the signal associated mainly with the
fundamental Love mode resonance is more intense and of longer
duration than the signal associated mainly with the fundamental
interference bump.
\begin{figure}
[ptb]
\includegraphics[width=6.0cm]{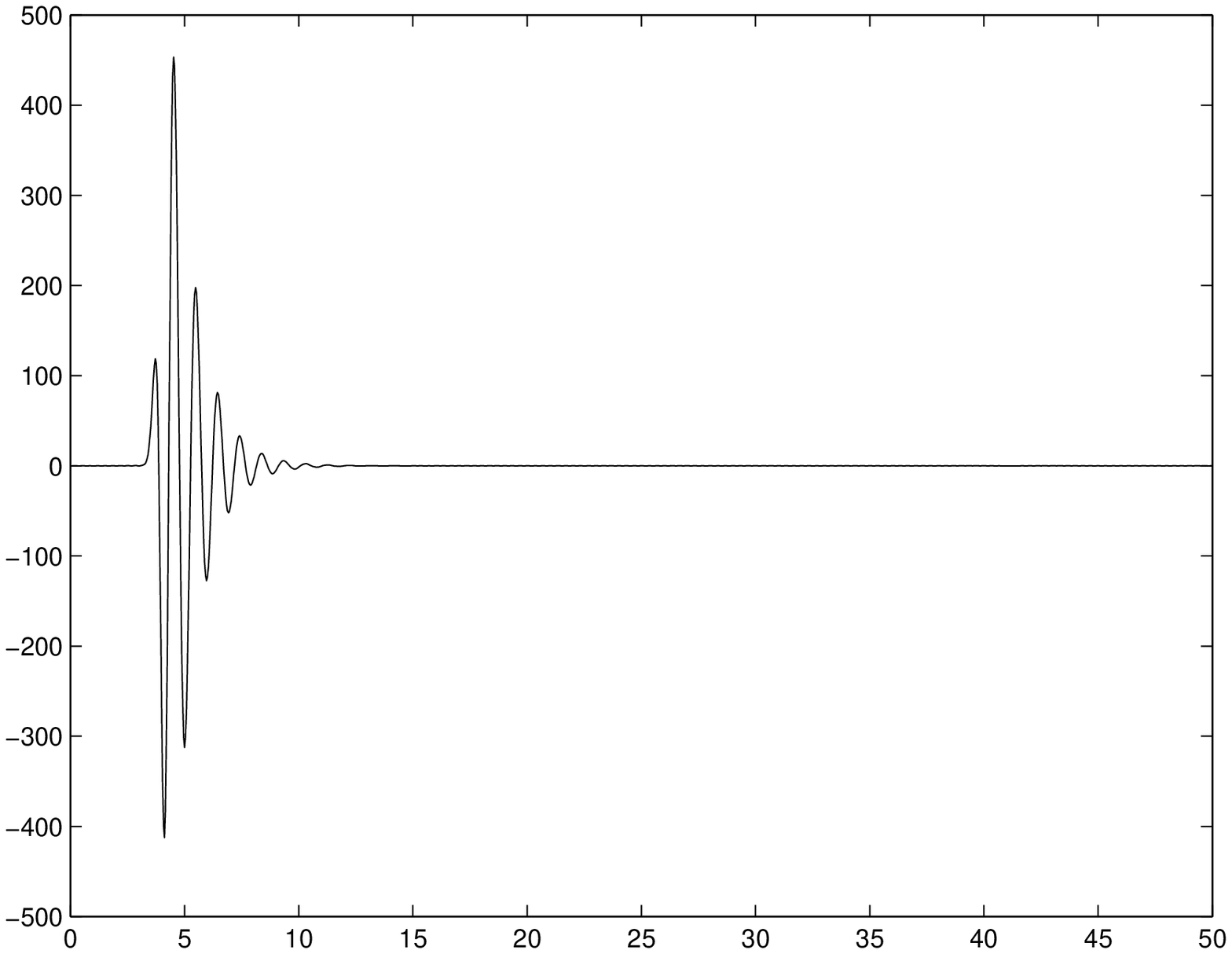}\hfill
\includegraphics[width=6.0cm]{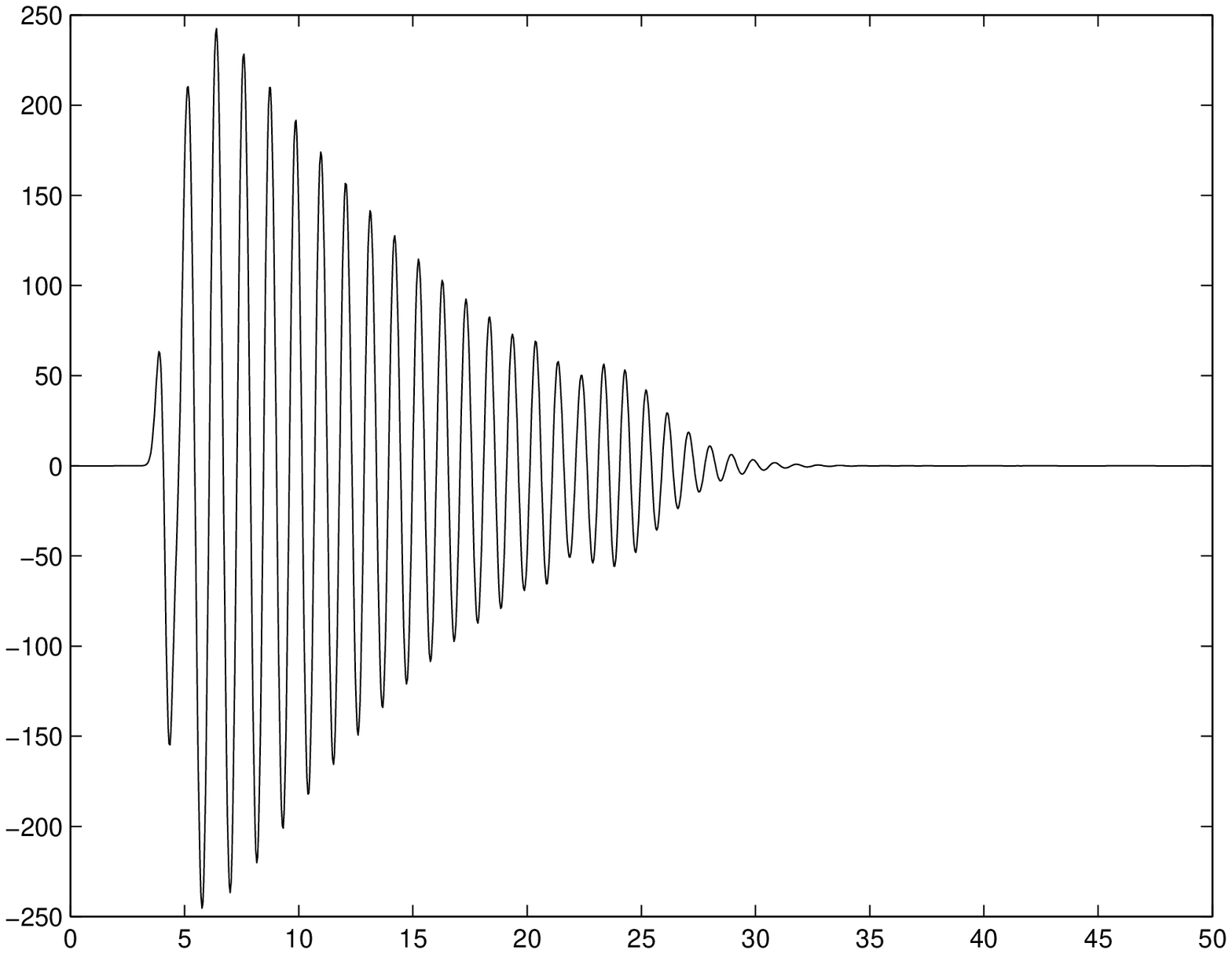}\hfill
\caption{Time domain ground response in Nice-like environment for
constant source-to-observation point distances and input pulse
with $\nu_{0}$ near lowest-frequency maximum of transfer function.
Left: $\mathbf{x}^{s}=(0\text{m},3000\text{m})$,
$\mathbf{x}=(100\text{m},0\text{m})$, $\nu_{0}=1.0$Hz. Right:
$\mathbf{x}^{s}=(0\text{m},100\text{m})$,
$\mathbf{x}=(3000\text{m},0\text{m})$, $\nu_{0}=0.9$Hz.}
\label{f8}
\end{figure}

This last remark should be tempered by consideration of the
spectrum of the input pulse, since the transfer functions do not
take this spectrum into account whereas the temporal signals do.
Thus, when the location of the maximum of the spectrum of the
input pulse is closer to the location of the maximum of the
transfer function, the time-domain response is larger, as seen in
fig. \ref{f8}, this being true for signals that are essentially
due both to Love resonances or to constructive interference
effects (note that the location of the pulse maxima were adjusted
so as to be close to the locations of the transfer function
fundamental peaks). Actually, this figure reveals the existence of
a beating phenomenon in the ground response temporal signal for a
source near the layer, which is probably due to the combined
(amplitude modulation) effects of the fundamental Love mode peak
and the fundamental interference peak. This issue will be
discussed in more depth in the next section.

Next consider the somewhat softer-than-in-Nice environment
\begin{figure}
[ptb]
\includegraphics[width=6.0cm]{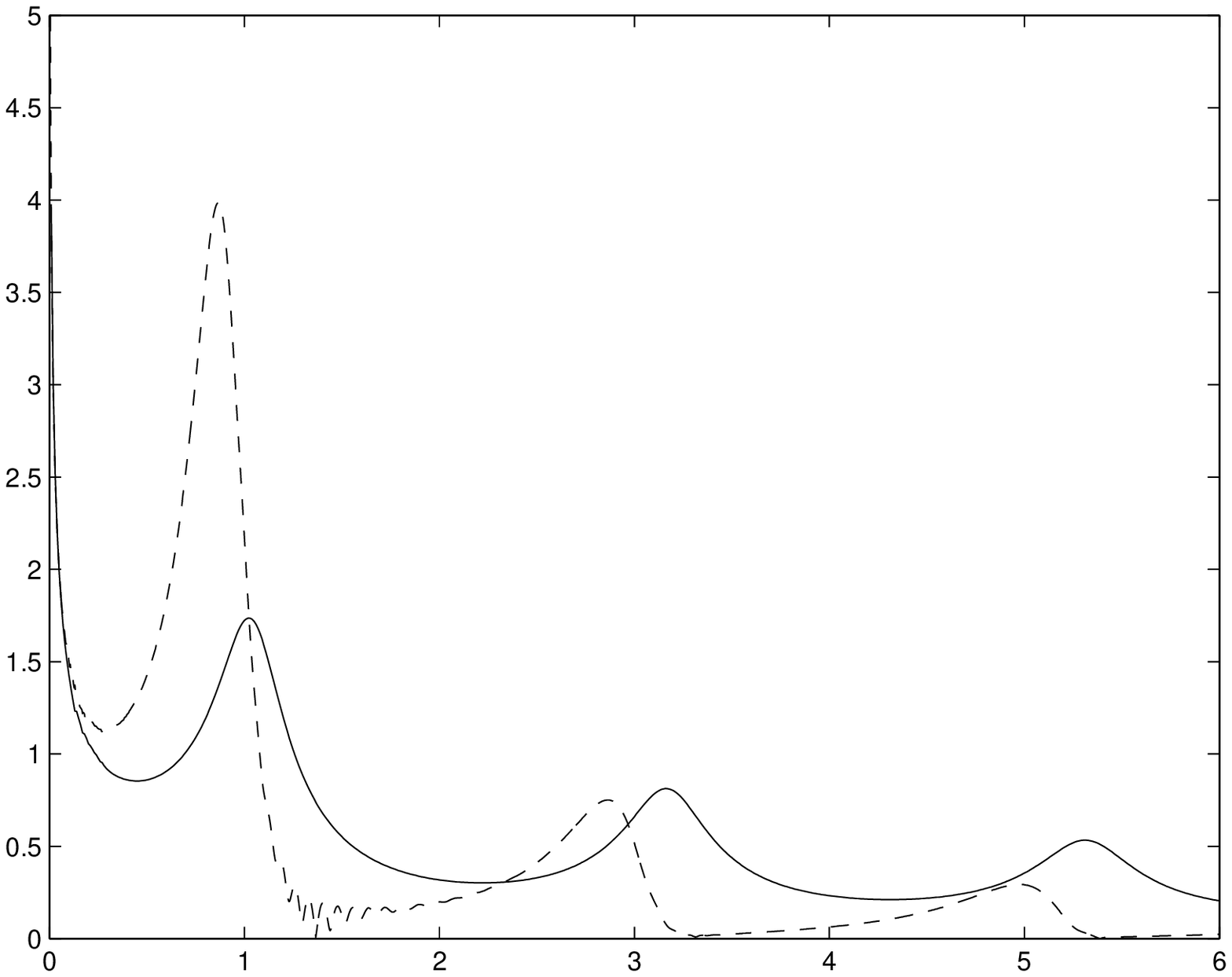}\hfill
\includegraphics[width=6.0cm]{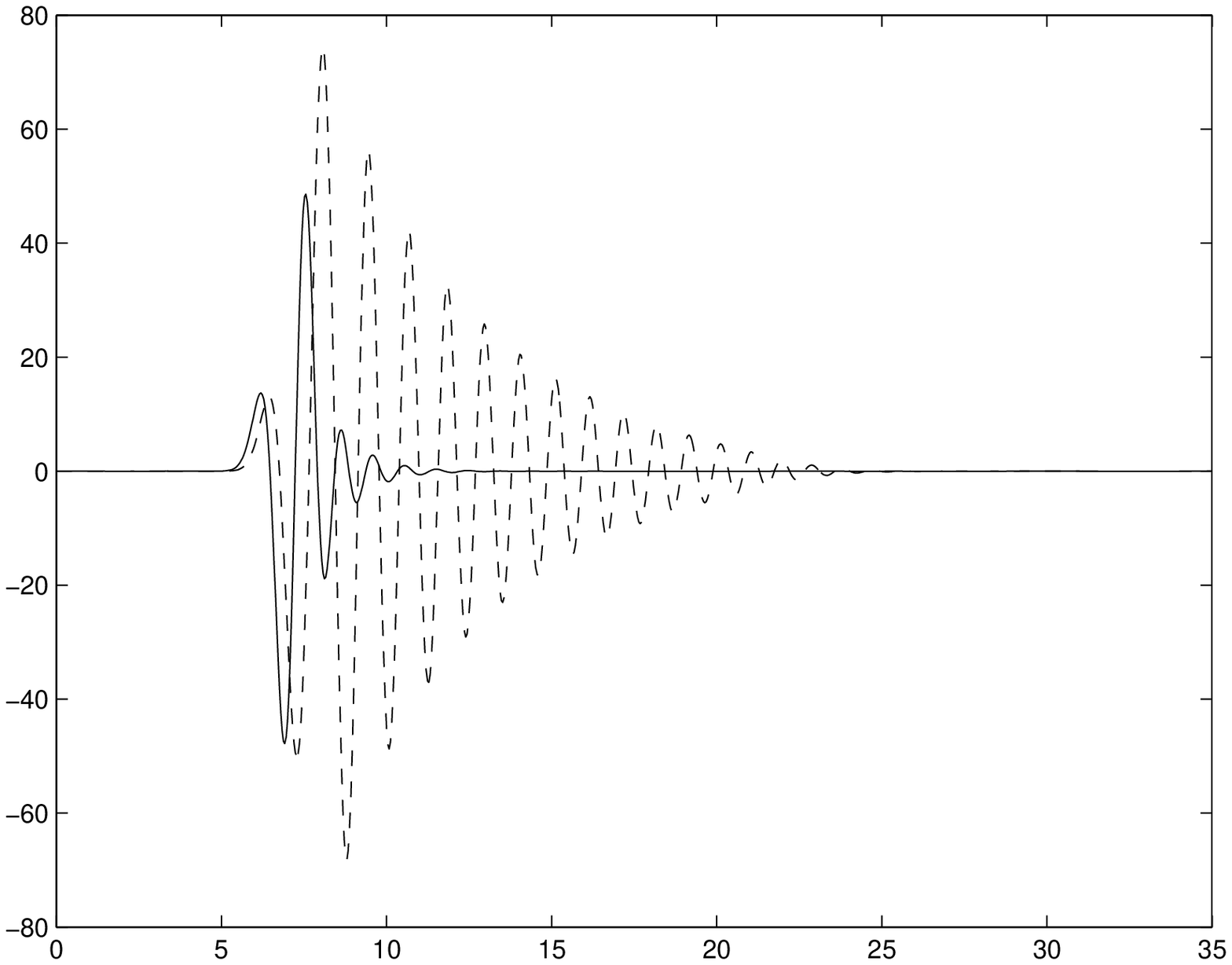}\hfill
\caption{Temporal record of ground response in softer-than-Nice
environment for fixed source-to-observation point distance. The
full curves in left (transfer function) and right (time domain
response to $\nu_{0}=0.5$Hz pulse) panels refer to
$\mathbf{x}^{s}=(0\text{m},3000\text{m})$,
$\mathbf{x}^{s}=(0\text{m},100\text{m})$. The dotted curves in
left (transfer function) and right (time domain response to
$\nu_{0}=0.5$Hz pulse) panels refer to
$\mathbf{x}^{s}=(0\text{m},100\text{m})$,
$\mathbf{x}=(3000\text{m},0\text{m})$.} \label{f9}
\end{figure}
wherein $c^{0}=600$m/s, $\rho^{1}=1300$kg/m$^{3}$,
$c^{1}=200$m/s (fig. \ref{f9}). Again, the two source/observation point couples are:
$\mathbf{x}^{s}=(0\text{m},3000\text{m})$, $\mathbf{x}=(3000\text{m},0\text{m})$
(solid line curve in fig. \ref{f9}  and
$\mathbf{x}^{s}=(0\text{m},100\text{m})$, $\mathbf{x}=(3000\text{m},0\text{m})$
(dotted line curve in fig. \ref{f9}). All that was said in the previous example
concerning
the transfer functions holds in the present case. Likewise, the
repercussions on the temporal signals are the same as in the
previous case (short duration signal for a remote source and
relatively-long signal for a near source.

We next consider  a Mexico-city like site (again, without the buildings)
\begin{figure}
[ptb]
\includegraphics[width=6.0cm]{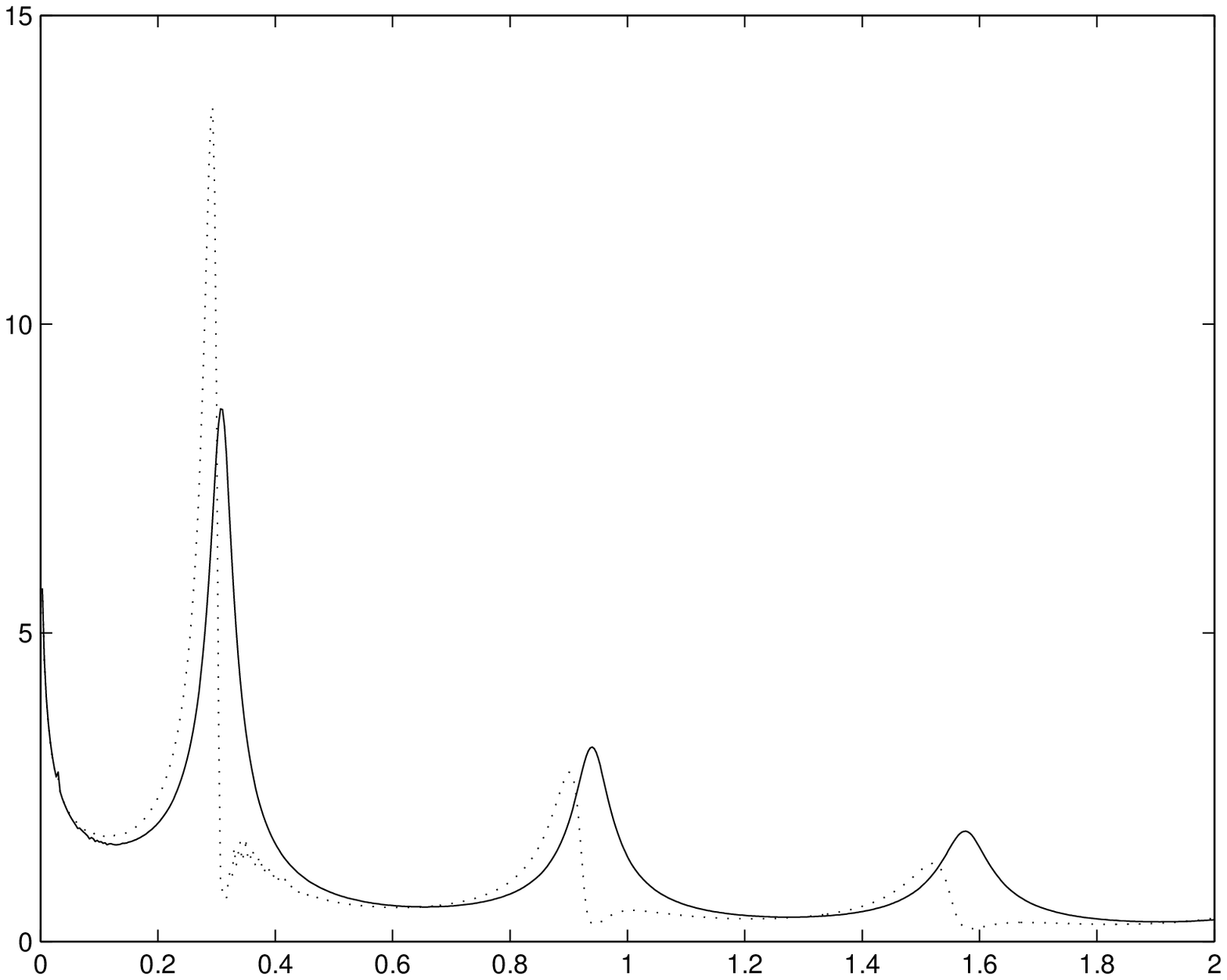}\hfill
\includegraphics[width=6.0cm]{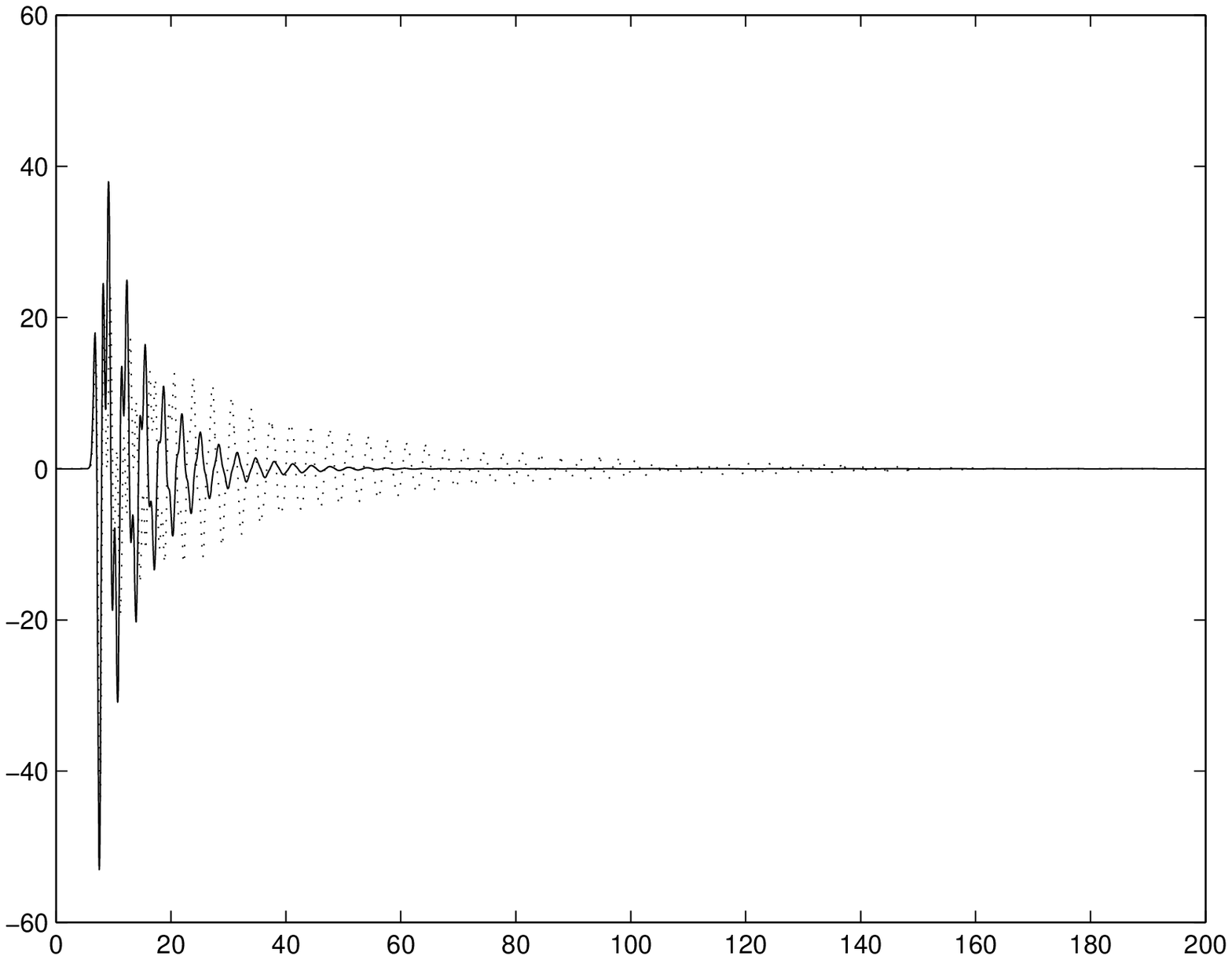}\hfill
\caption{Transfer functions (left panel) and temporal records for $\nu_{0}=0.5$Hz
input pulse (right panel) of ground response in Mexico City-like
environment for various source locations and observation points:
$\mathbf{x}^{s}=(0\text{m},3000\text{m})$,
$\mathbf{x}=(100\text{m},0\text{m})$ (solid line curves),
$\mathbf{x}^{s}=(0\text{m},100\text{m})$,
$\mathbf{x}=(3000\text{m},0\text{m})$ (dotted line curves).}
\label{f10}
\end{figure}
in which $c^{0}=600$m/s,
$\rho^{1}=1300$kg/m$^{3}$, $c^{1}=60$m/s (fig. \ref{f10}).
Again, the two source/observation point couples are:
$\mathbf{x}^{s}=(0\text{m},3000\text{m})$, $\mathbf{x}=(3000\text{m},0\text{m})$
(solid line curve in fig. \ref{f10})  and
$\mathbf{x}^{s}=(0\text{m},100\text{m})$, $\mathbf{x}=(3000\text{m},0\text{m})$
(dotted line curve in fig. \ref{f10}). All that was said in the two previous
examples
concerning
the transfer functions holds in the present case. Likewise, the
repercussions on the temporal signals are the same as in the
previous two cases (short duration signal for a remote source and
relatively-long signal for a near source.
\subsection{Time domain responses for very large and very small
source-to-observation point distances} We again consider  a
Mexico-city like site
\begin{figure}
[ptb]
\includegraphics[width=6.0cm]{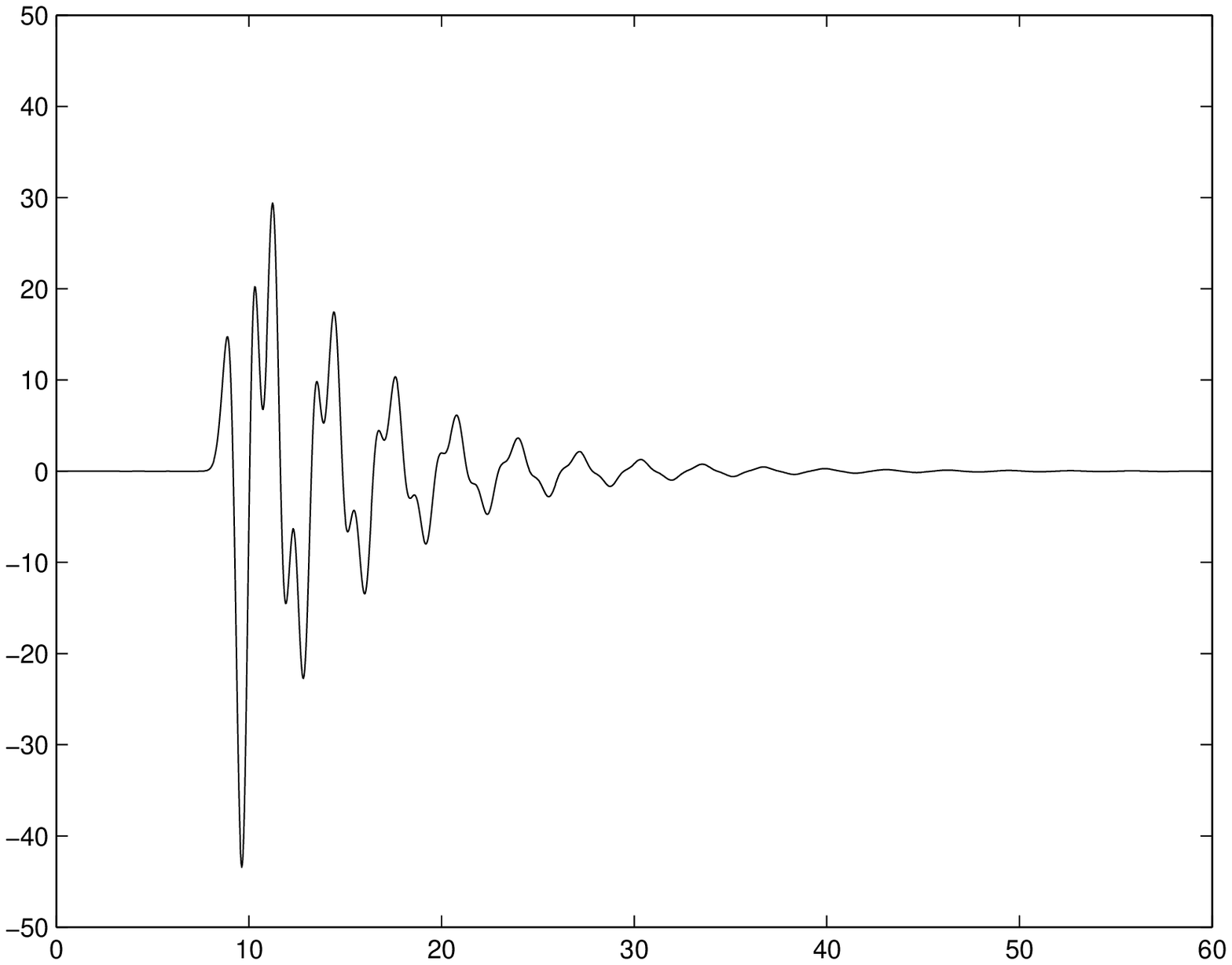}\hfill
\includegraphics[width=6.0cm]{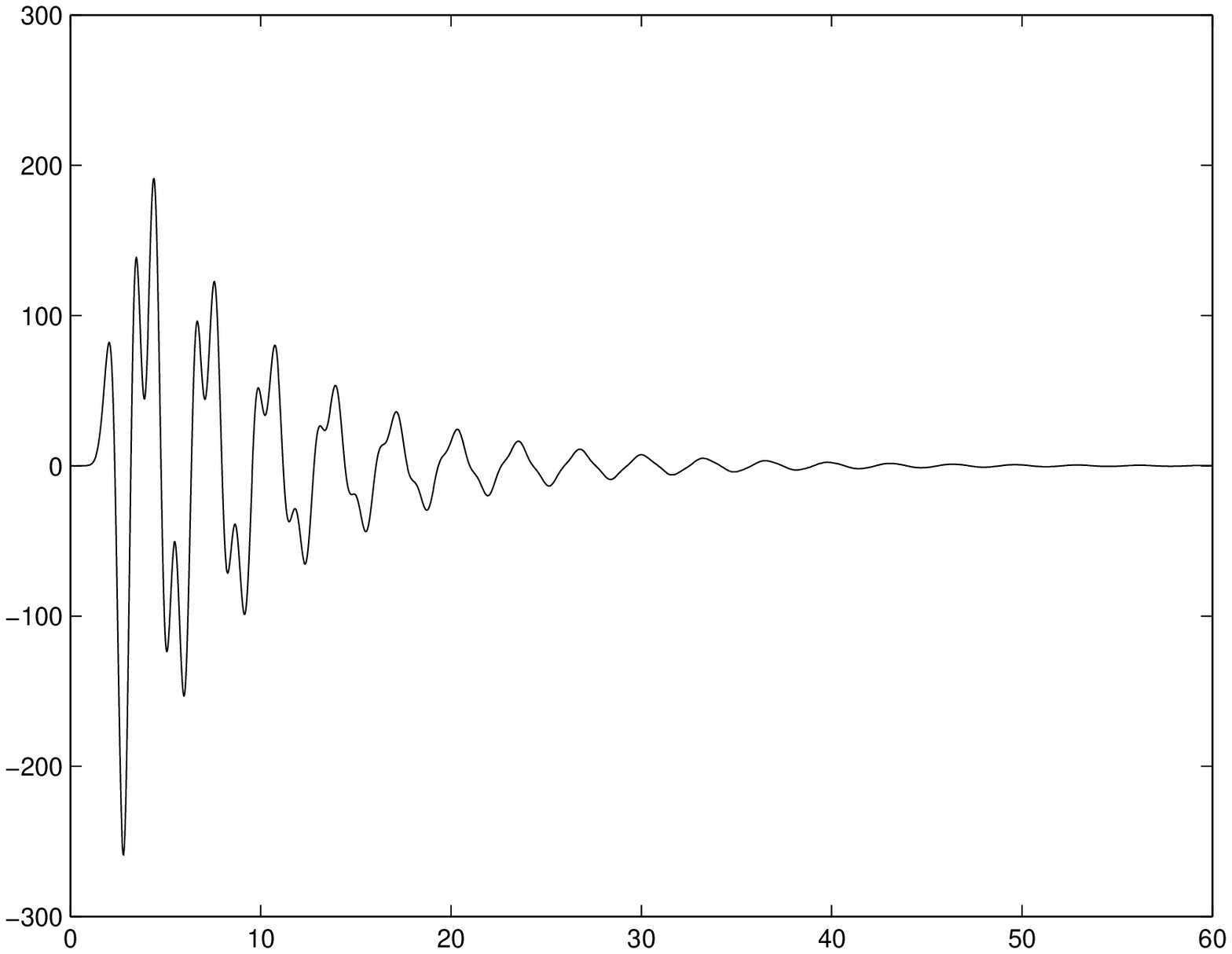}\hfill
\caption{Temporal records of ground response in Mexico City-like
environment for various  very large (left panel:
$\mathbf{x}^{s}=(0\text{m},3000\text{m})$,
$\mathbf{x}=(3000\text{m},0\text{m})$) and very small (right
panel: $\mathbf{x}^{s}=(0\text{m},100\text{m})$,
$\mathbf{x}=(100\text{m},0\text{m})$) source-to-observation point
distances). } \label{f10a}
\end{figure}
at which $c^{0}=600$m/s, $\rho^{1}=1300$kg/m$^{3}$, $c^{1}=60$m/s
(fig. \ref{f10a}). Although  in fig. \ref{f4} upper left and lower
right panels) we observed that the transfer functions for very
large  and very small source-to-observation point distances are
qualitatively very similar, we are somewhat surprized to find that
the two corresponding signals in fig. \ref{f10a} are so
qualitatively  similar, due to the fact that the transfer function
corresponding to the left panel in fig. \ref{f4} is dominated by
the SBW1 contribution, whereas the transfer function corresponding
to the right hand panel in fig. \ref{f4} has strong contributions
from both the SBW1 and SBW2. The clue to this unexpected result
resides in the spectrum of the $\nu_{0}=0.5$Hz input pulse (see
middle panel of fig. \ref{f1a}), since the maximum of the latter
is around $\nu_{0}=0.6$Hz and this frequency is both far-removed
form the peaks of the transfer functions and characterized by a
predominant contribution of the SBW1 (the latter fact providing an
explanantion of the relatively-short duration of the response
signals in fig. \ref{f10a} (note that the intensity of the very
close source-to-observation point signal is much larger than that
of the other signal, as it should be).
\subsection{Time domain responses for different input pulses and
fixed source and observation point coordinates} It is of
considerable interest to ascertain to what extent the spectrum of
the source affects the ground response \cite{bo03}, notably when
the source is near the layer.
\begin{figure}
[ptb]
\includegraphics[width=4.0cm]{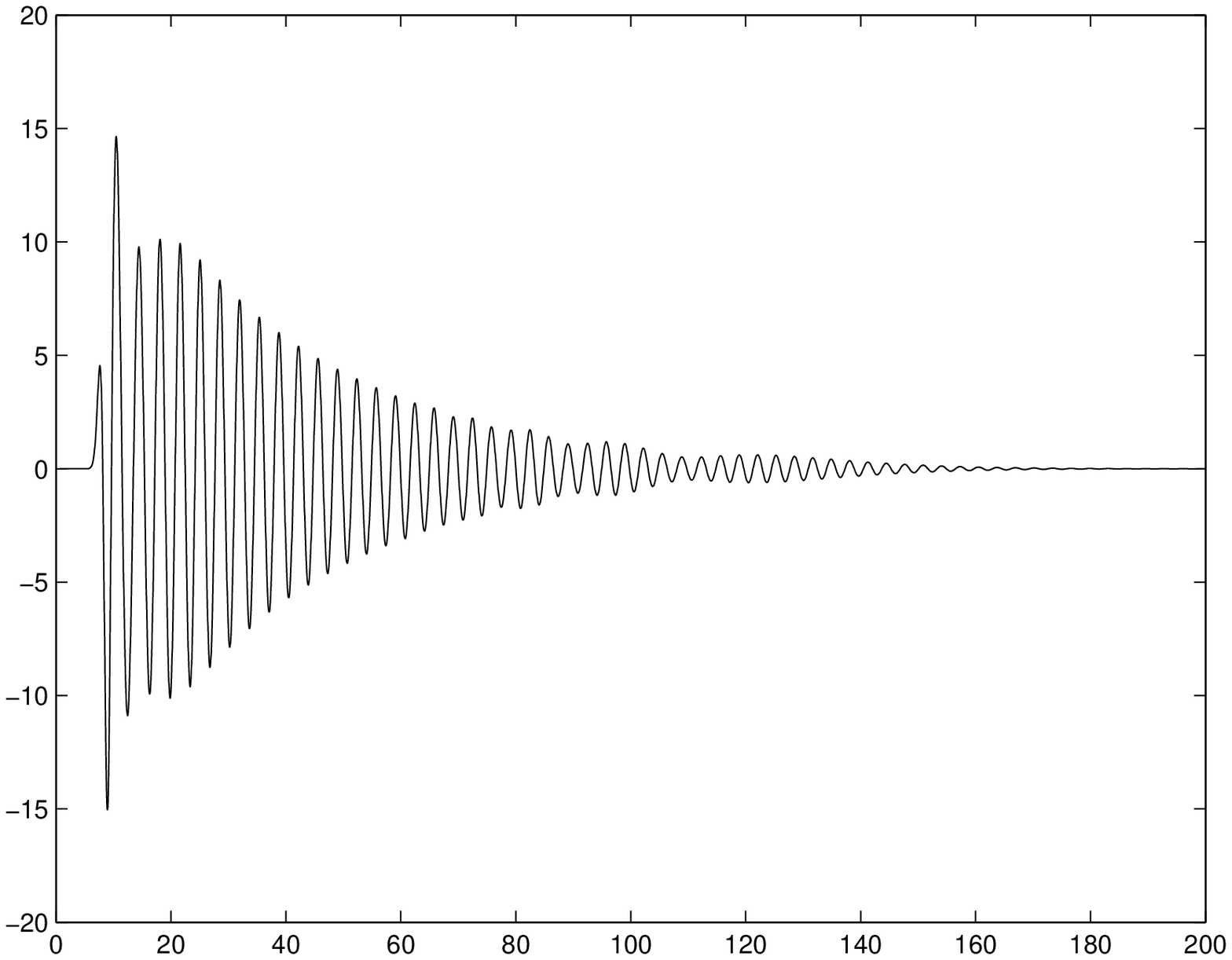}\hfill
\includegraphics[width=4.0cm]{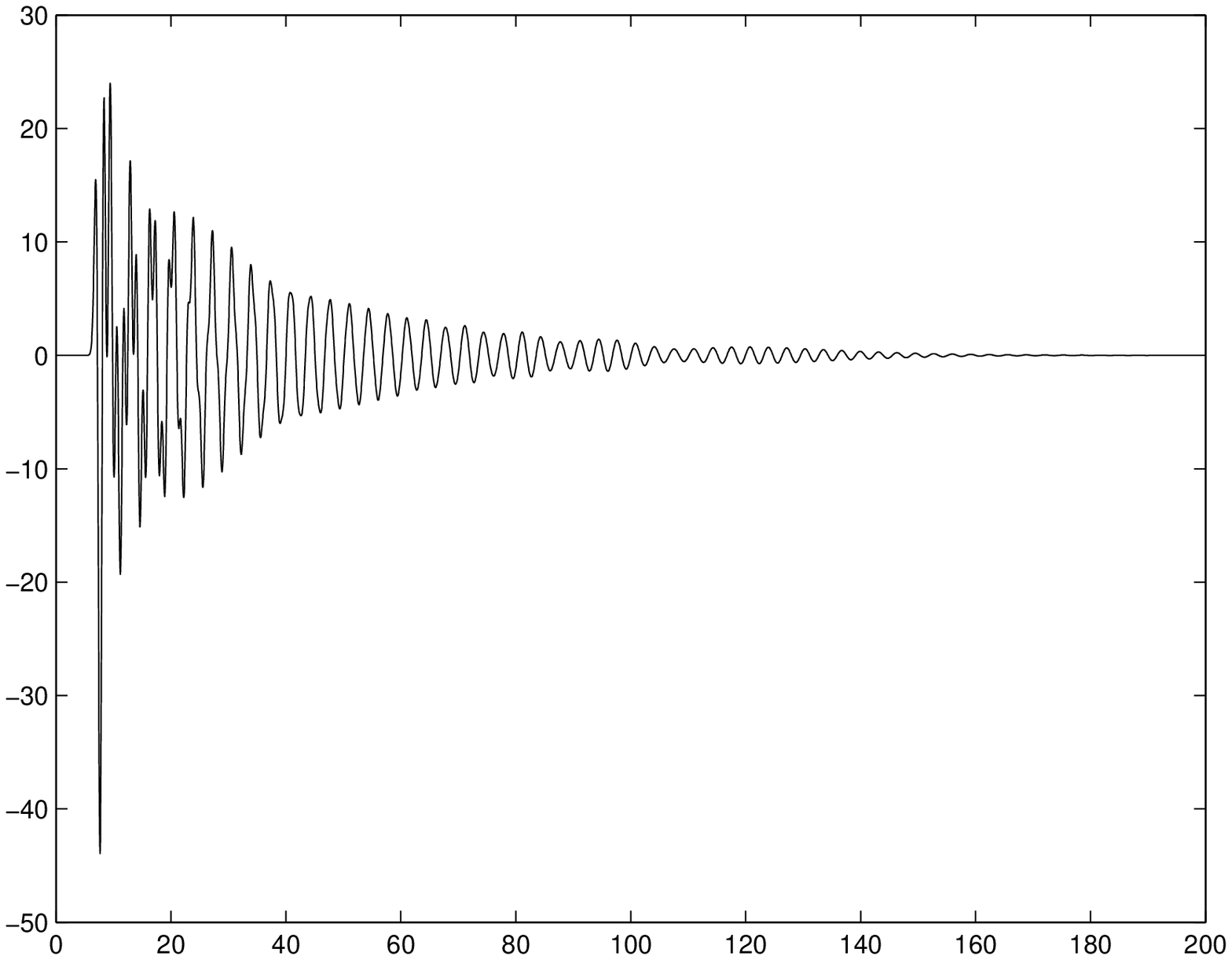}\hfill
\includegraphics[width=4.0cm]{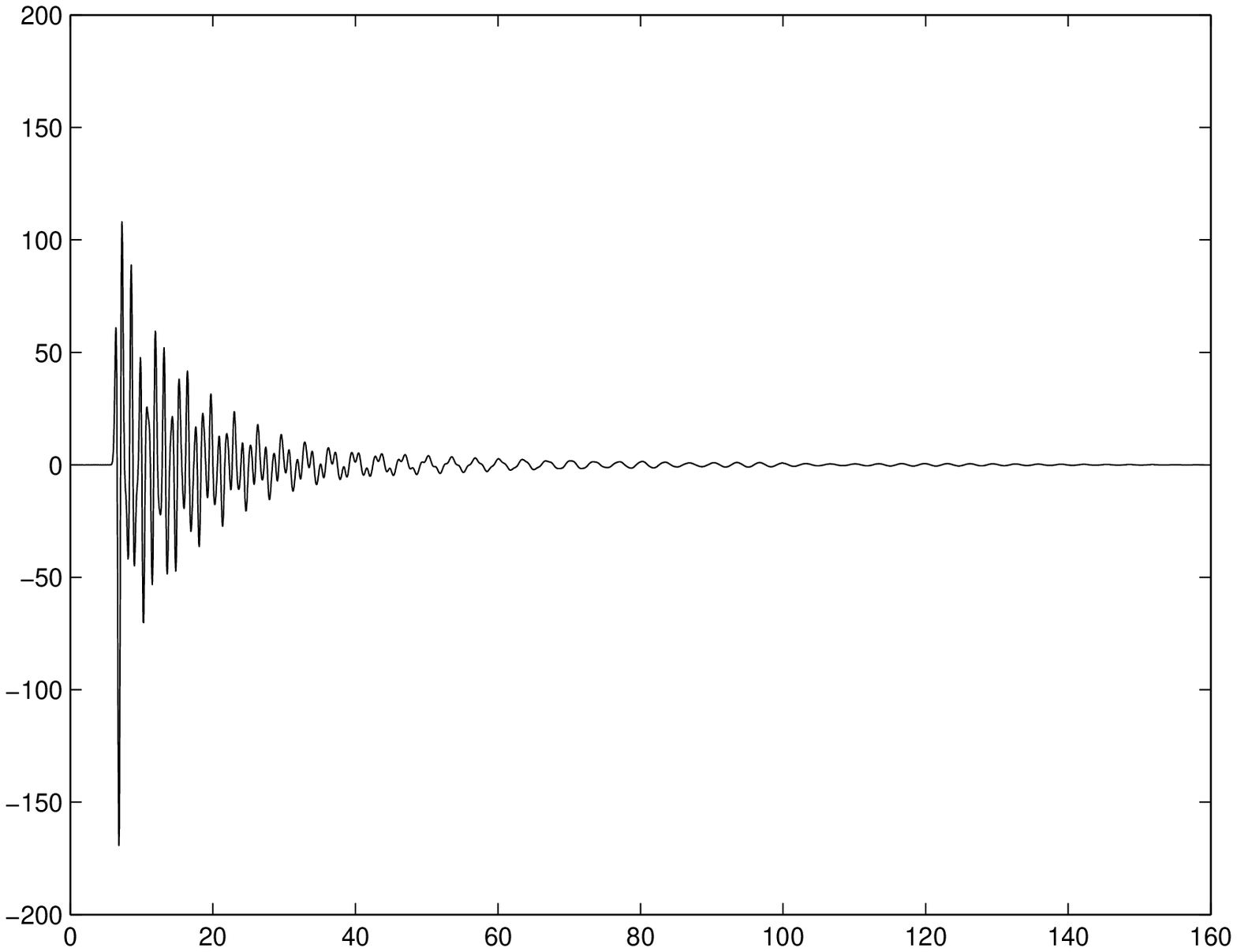}\hfill
\caption{Temporal records of ground response for $\nu_{0}=0.3$Hz
(left panel), $\nu_{0}=0.5$Hz (middle panel), $\nu_{0}=0.9$Hz
(right panel) input pulses
 in Mexico City-like
environment for
$\mathbf{x}^{s}=(0\text{m},100\text{m})$,
$\mathbf{x}=(3000\text{m},0\text{m})$.}
\label{f11}
\end{figure}
In fig. \ref{f11} (which again applies to the Mexico City like
environment), we see, as expected, that when the spectrum of the
input pulse is such as to overlap substantially the frequency band
covered by the fundamental Love mode peak (left and middle panels
in fig. \ref{f11}), the duration of the signal is large. When the
spectrum of the input pulse is such as to overlap substantially
the frequency band covered by the second Love mode peak (right
panel in  fig. \ref{f11}), the duration of the signal is smaller
than in the previous case due to the smaller finesse of the second
Love mode resonance peak. In all three cases, we observe some
beating, presumably due to the proximity of the interference peak
and Love resonance peak in the transfer function.
\subsection{Time domain responses for various layer thicknesses and input pulses}
 A more thorough
study of the influence of the input spectrum must take into
account variations of the layer thickness.
\begin{figure}
[ptb]
\includegraphics[width=4.0cm]{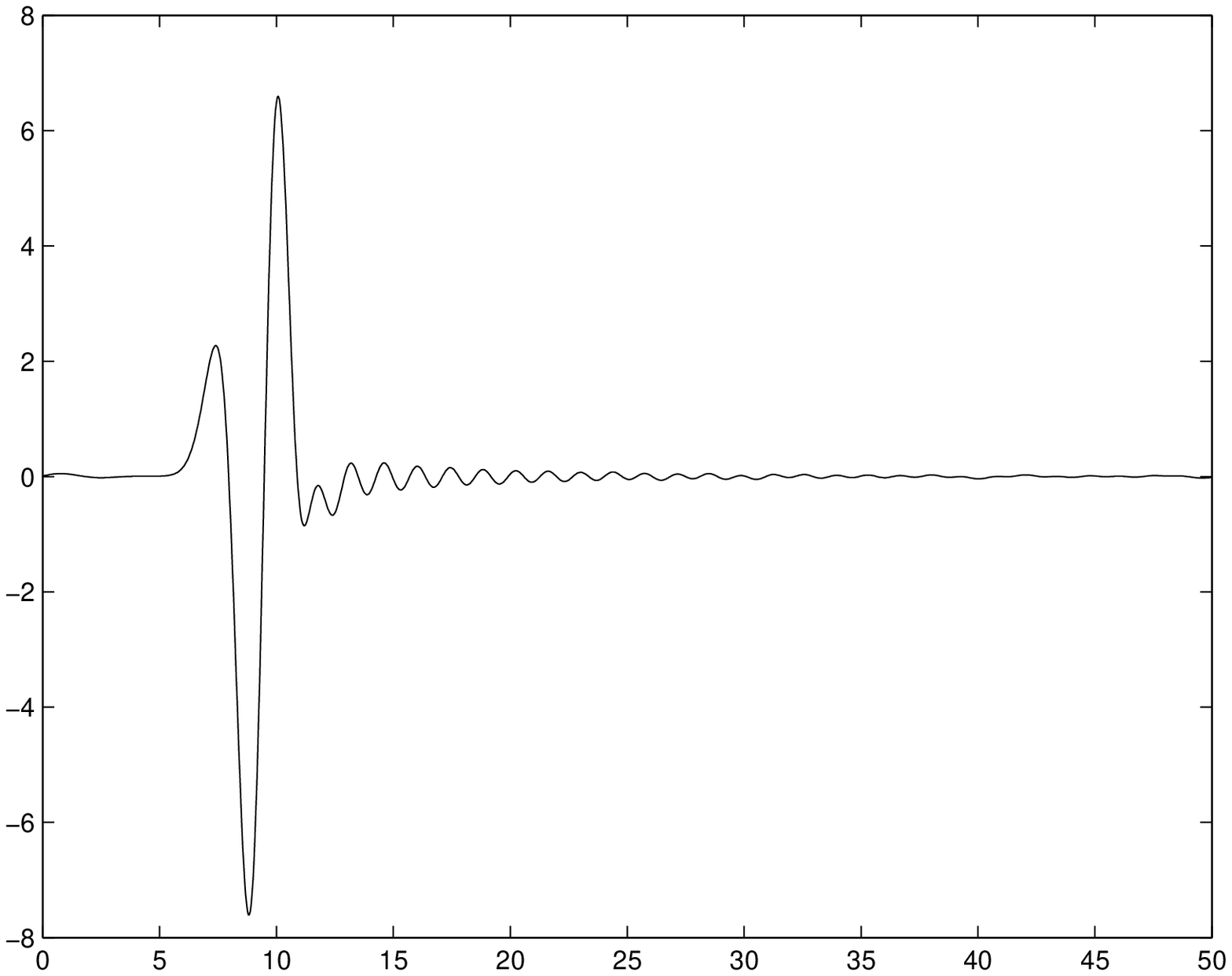}\hfill
\includegraphics[width=4.0cm]{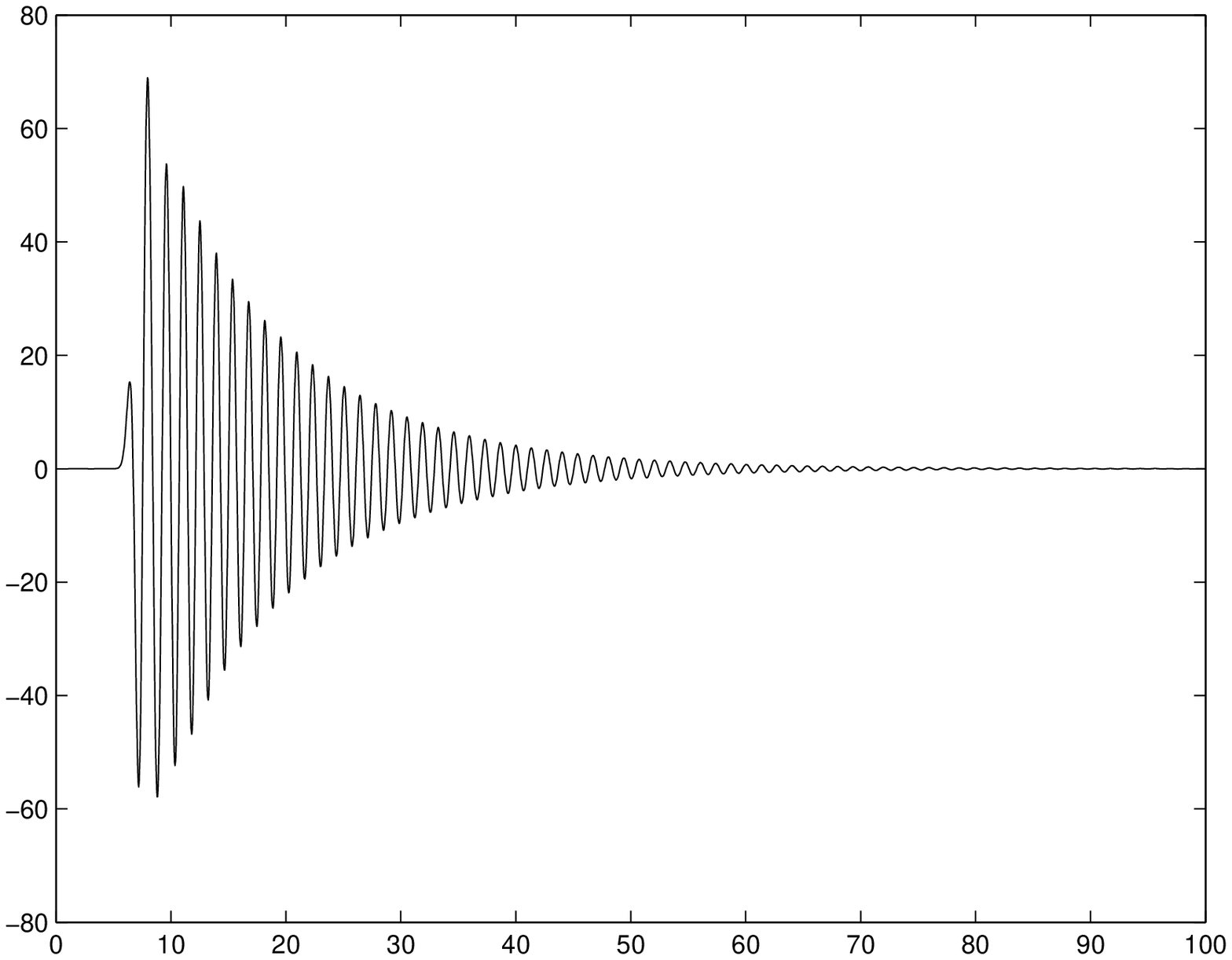}\hfill
\includegraphics[width=4.0cm]{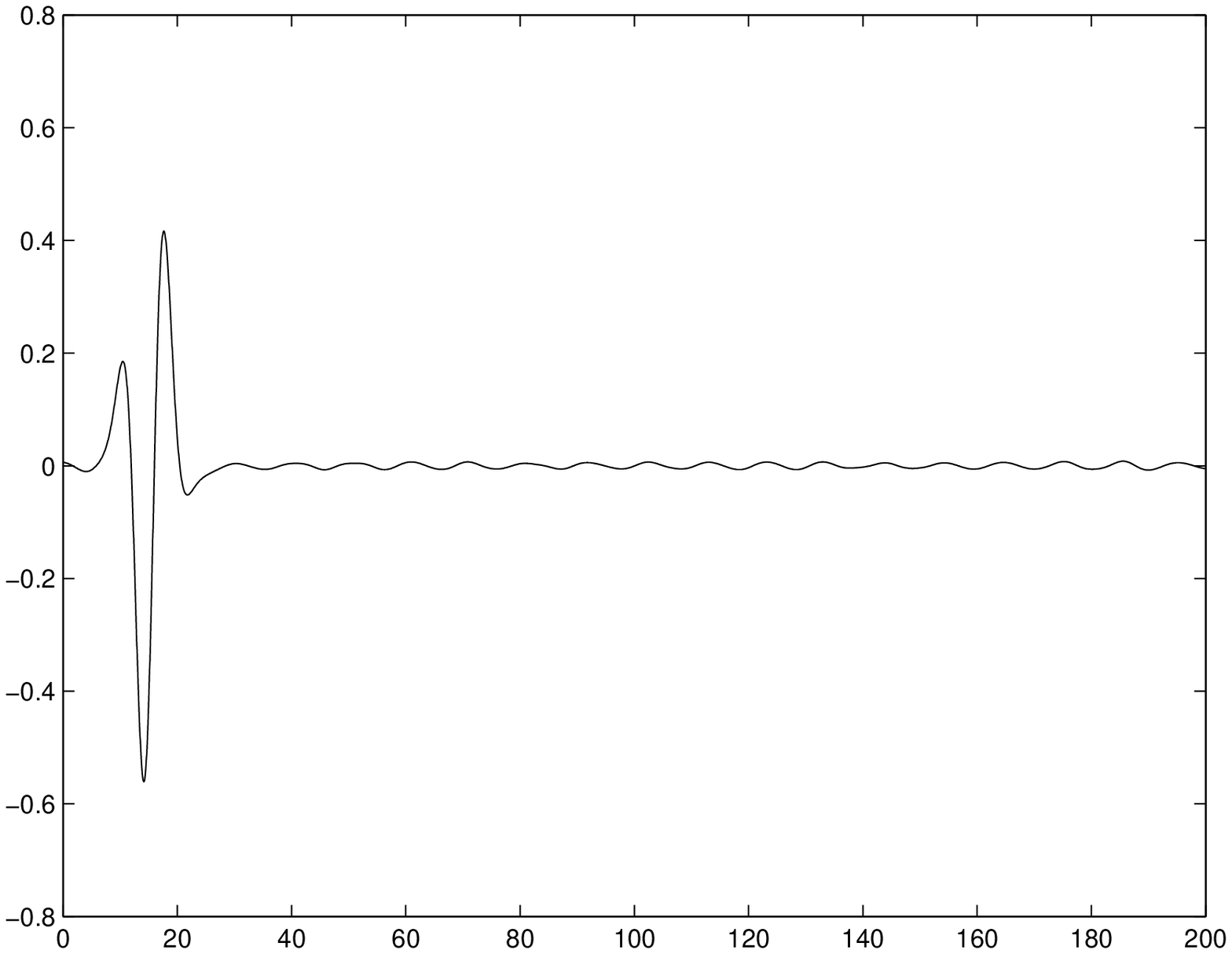}\hfill
\caption{Temporal records of ground response for $\nu_{0}=0.25$Hz
(left panel), $\nu_{0}=0.5$Hz (middle panel), $\nu_{0}=1.0$Hz
(right panel) input pulses
 in Mexico City-like
environment with layer thickness $h=20$m for
$\mathbf{x}^{s}=(0\text{m},100\text{m})$,
$\mathbf{x}=(3000\text{m},0\text{m})$.}
\label{f12}
\end{figure}
\begin{figure}
[ptb]
\includegraphics[width=4.0cm]{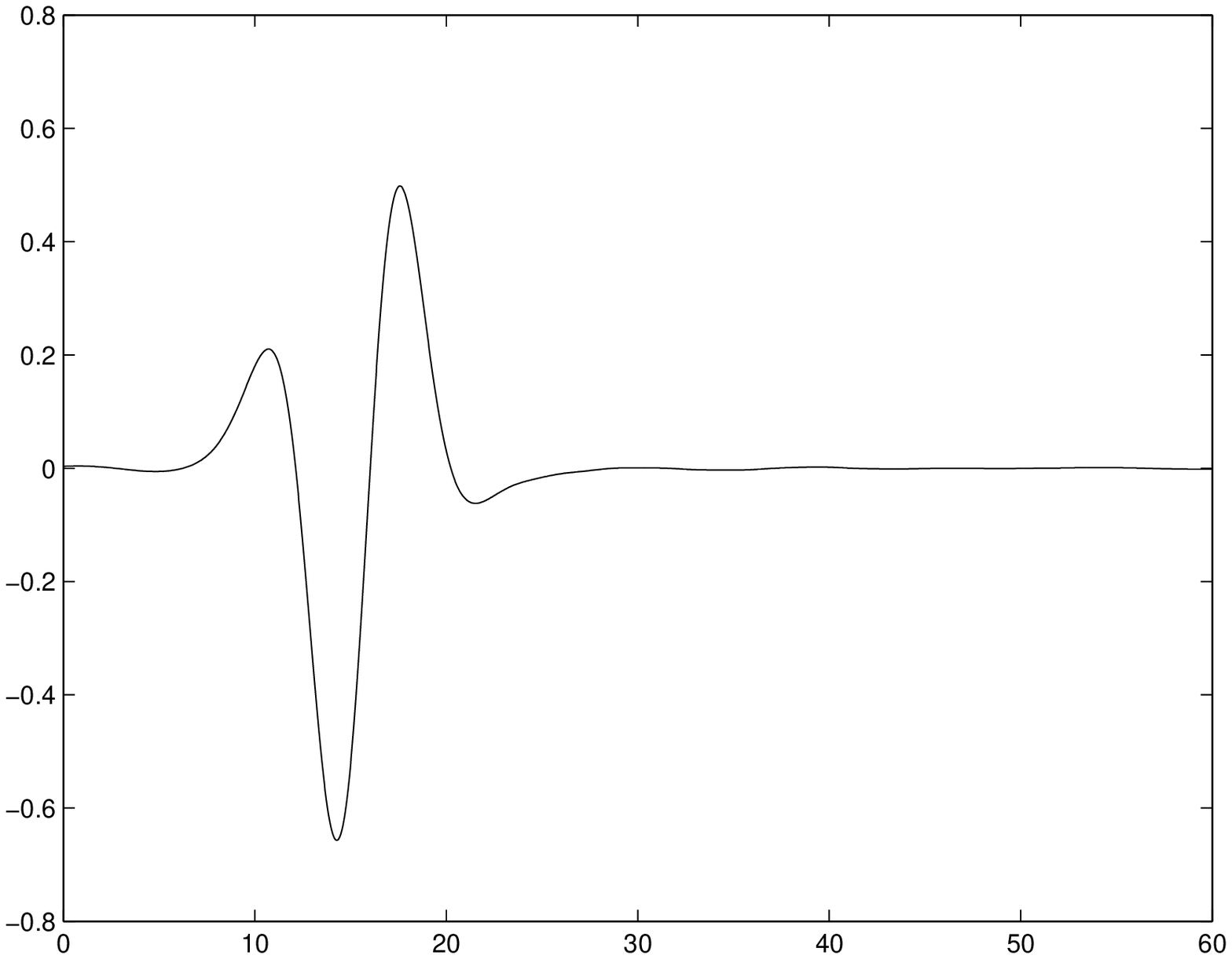}\hfill
\includegraphics[width=4.0cm]{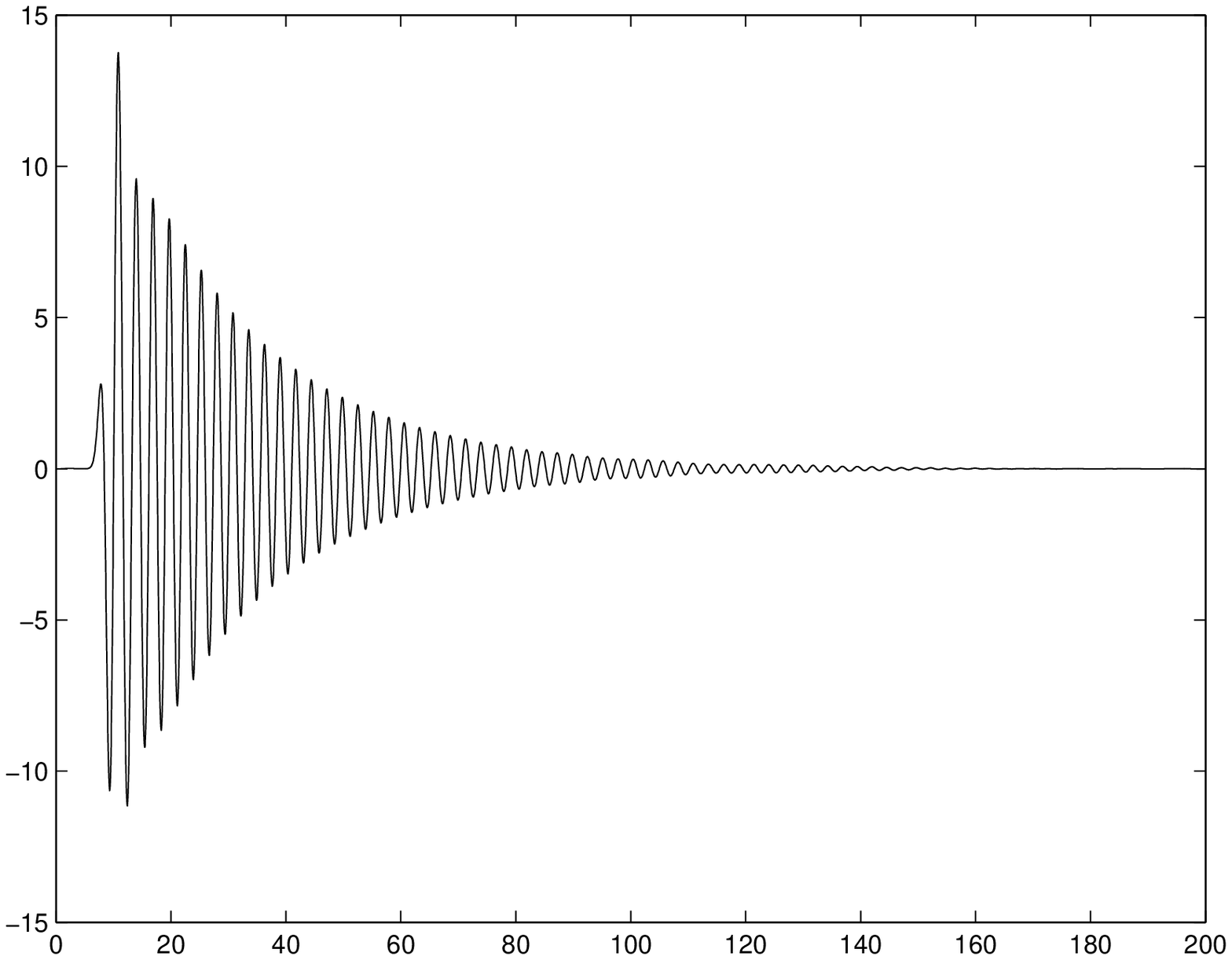}\hfill
\includegraphics[width=4.0cm]{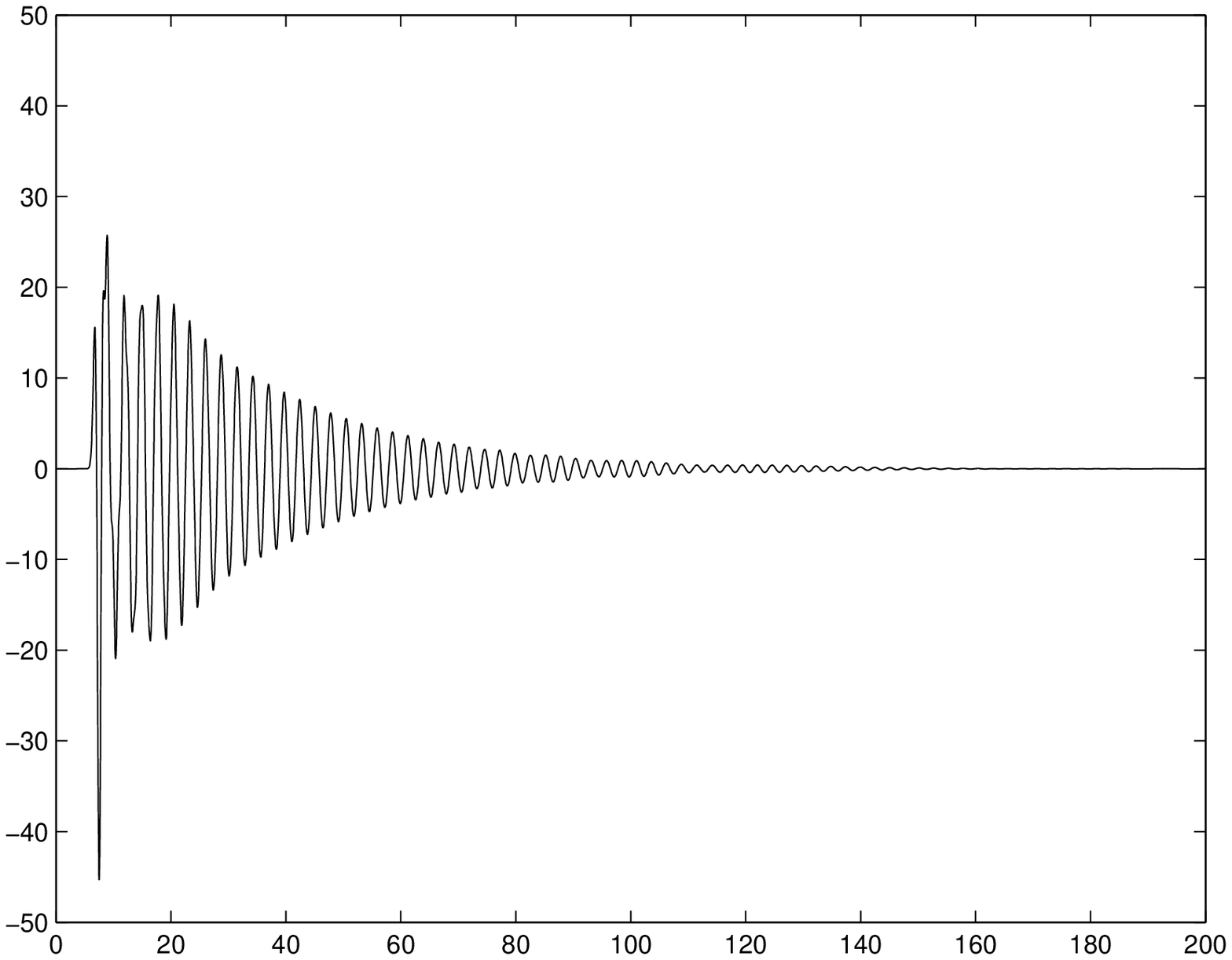}\hfill
\caption{Temporal records of ground response for $\nu_{0}=0.1$Hz
(left panel), $\nu_{0}=0.25$Hz (middle panel), $\nu_{0}=0.5$Hz
(right panel) input pulses
 in Mexico City-like
environment with layer thickness $h=40$m for
$\mathbf{x}^{s}=(0\text{m},100\text{m})$,
$\mathbf{x}=(3000\text{m},0\text{m})$.}
\label{f13}
\end{figure}
\begin{figure}
[ptb]
\includegraphics[width=4.0cm]{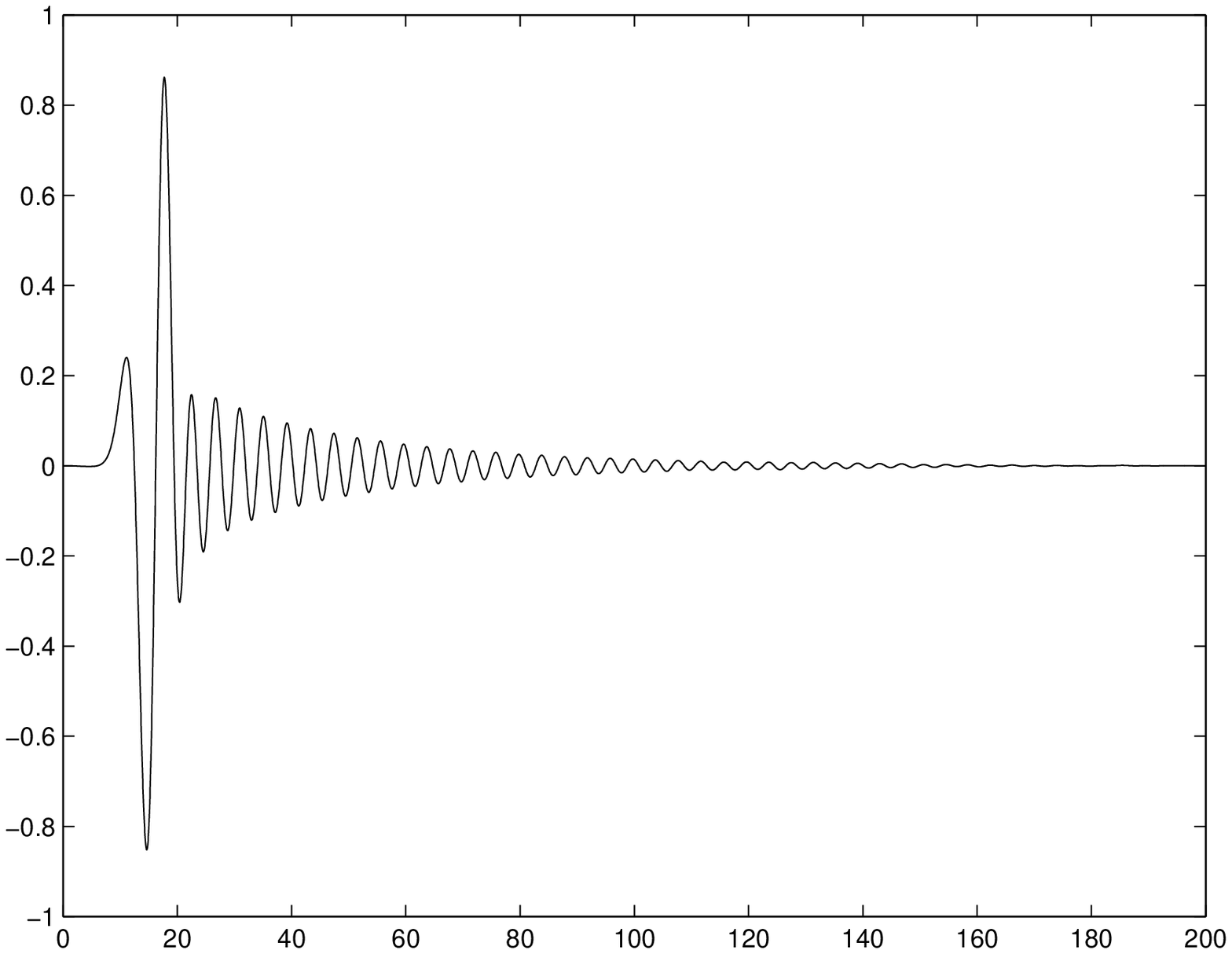}\hfill
\includegraphics[width=4.0cm]{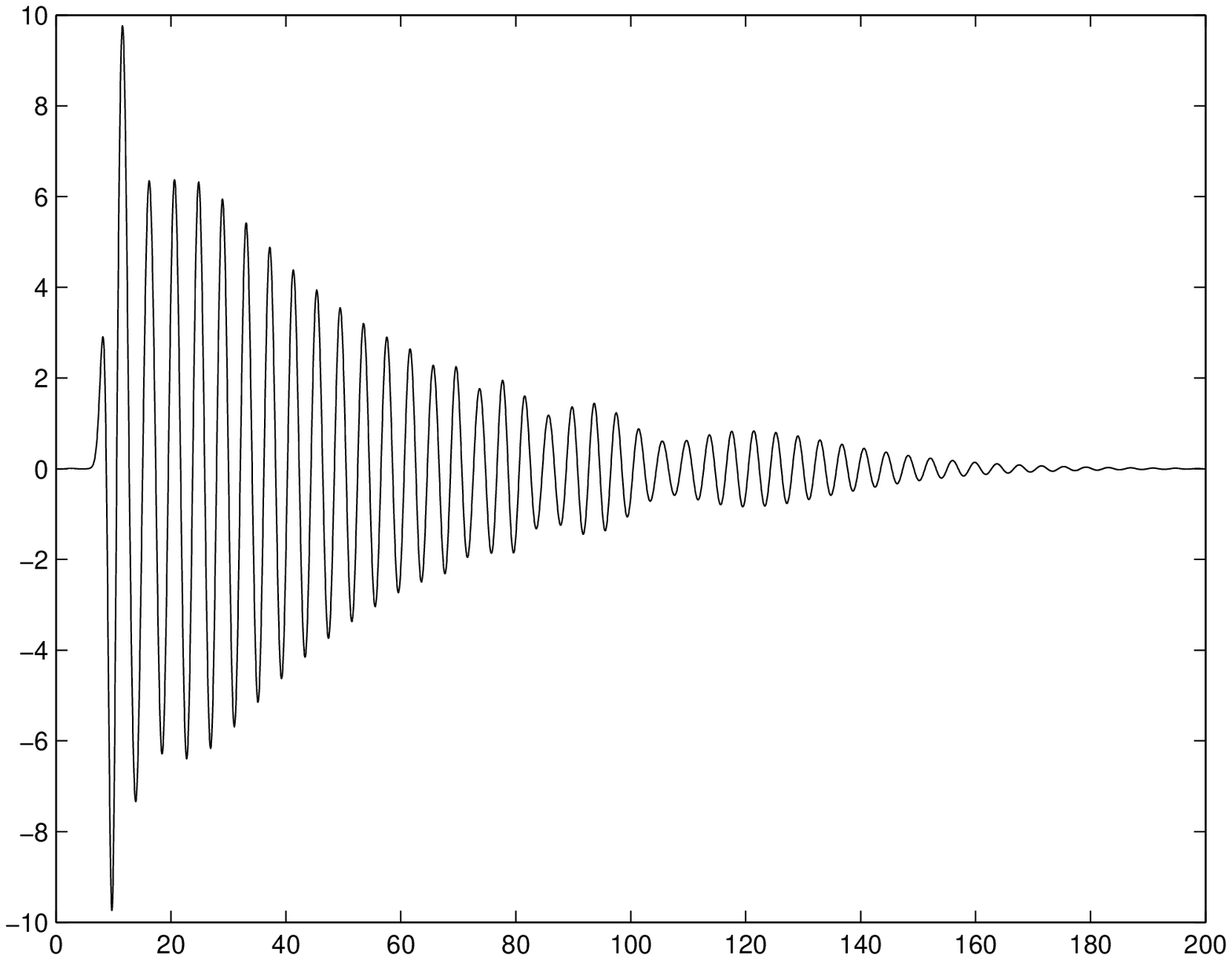}\hfill
\includegraphics[width=4.0cm]{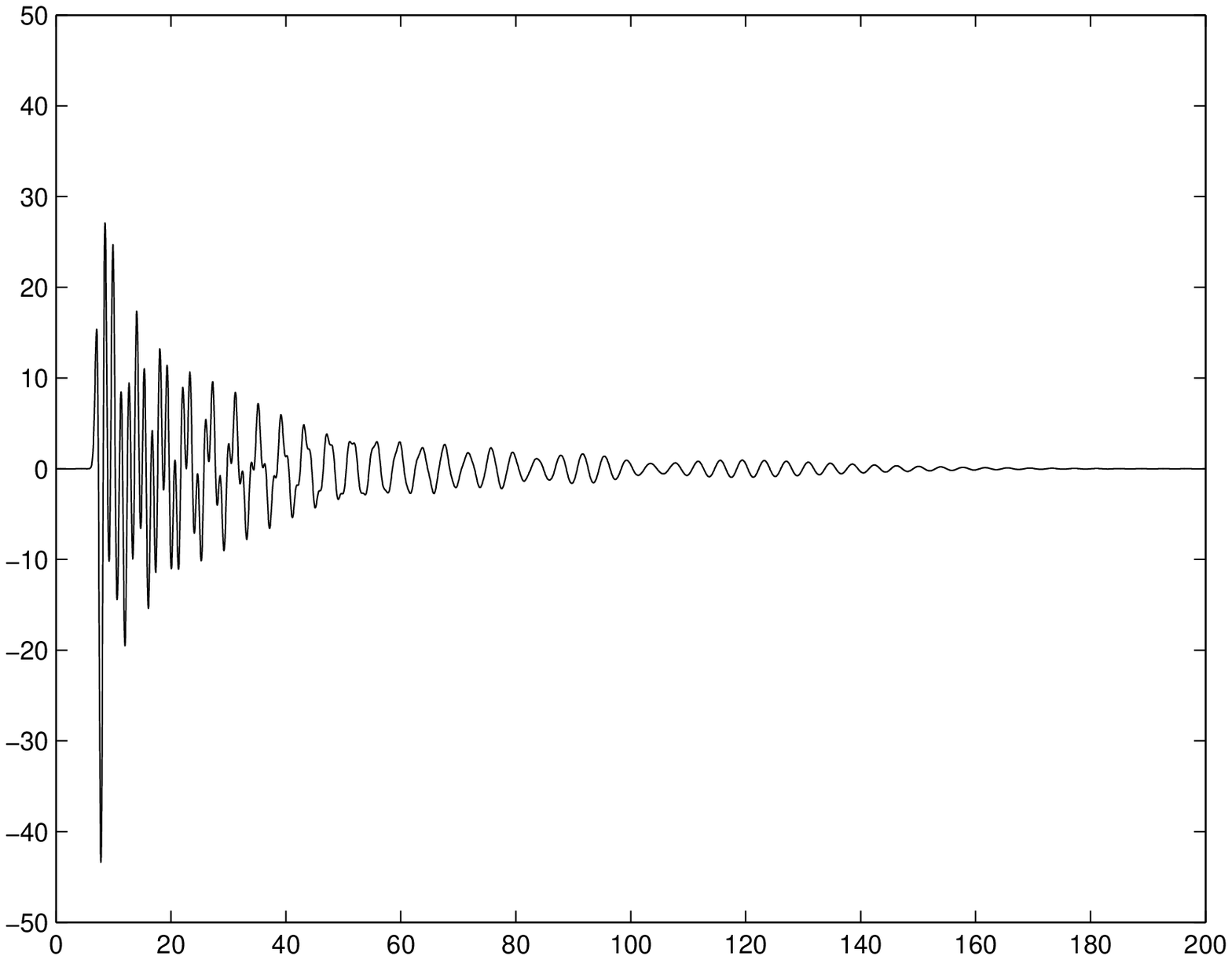}\hfill
\caption{Temporal records of ground response for $\nu_{0}=0.1$Hz
(left panel), $\nu_{0}=0.25$Hz (middle panel), $\nu_{0}=0.5$Hz
(right panel) input pulses
 in Mexico City-like
environment with layer thickness $h=60$m for
$\mathbf{x}^{s}=(0\text{m},100\text{m})$,
$\mathbf{x}=(3000\text{m},0\text{m})$.}
\label{f14}
\end{figure}
\begin{figure}
[ptb]
\includegraphics[width=4.0cm]{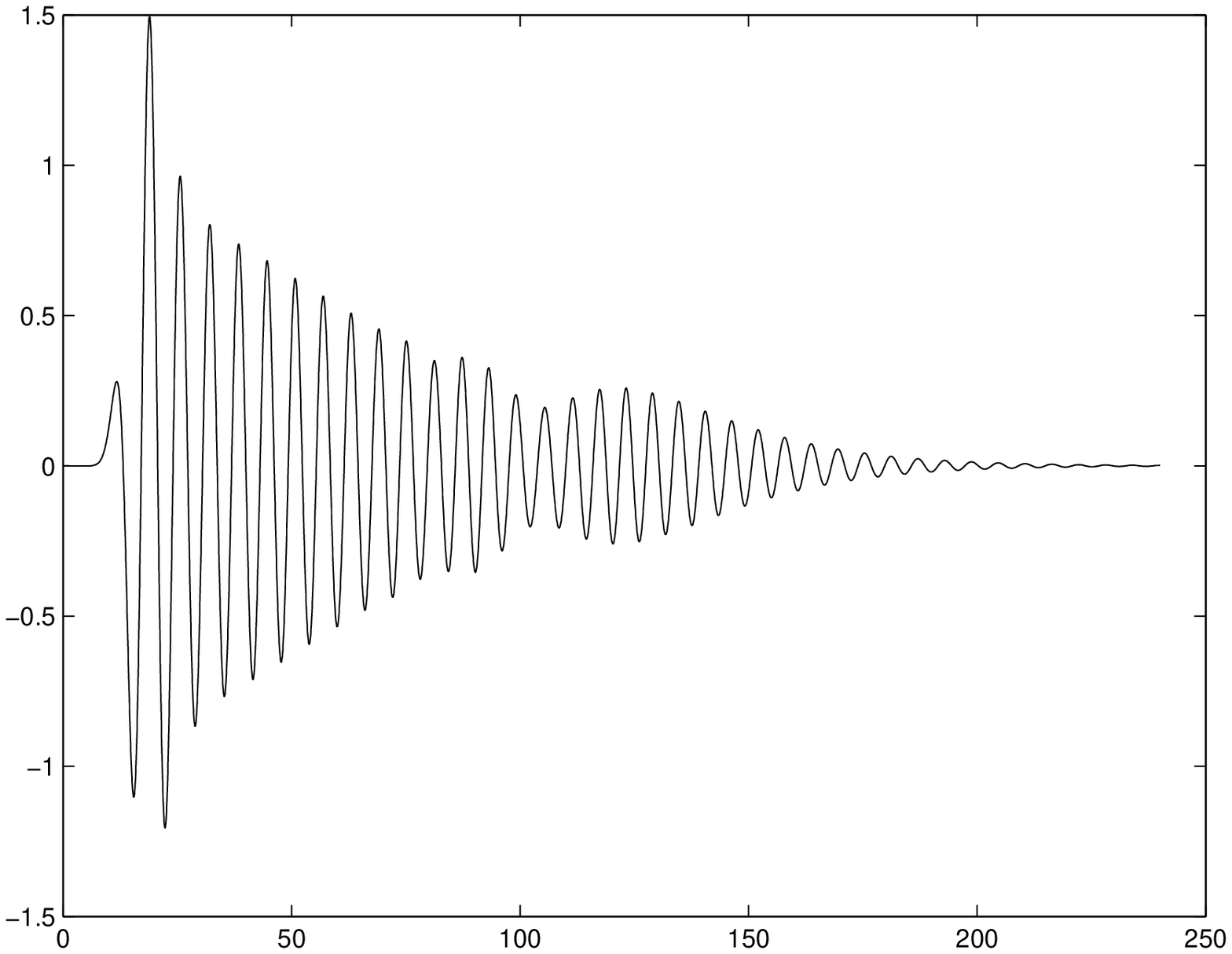}\hfill
\includegraphics[width=4.0cm]{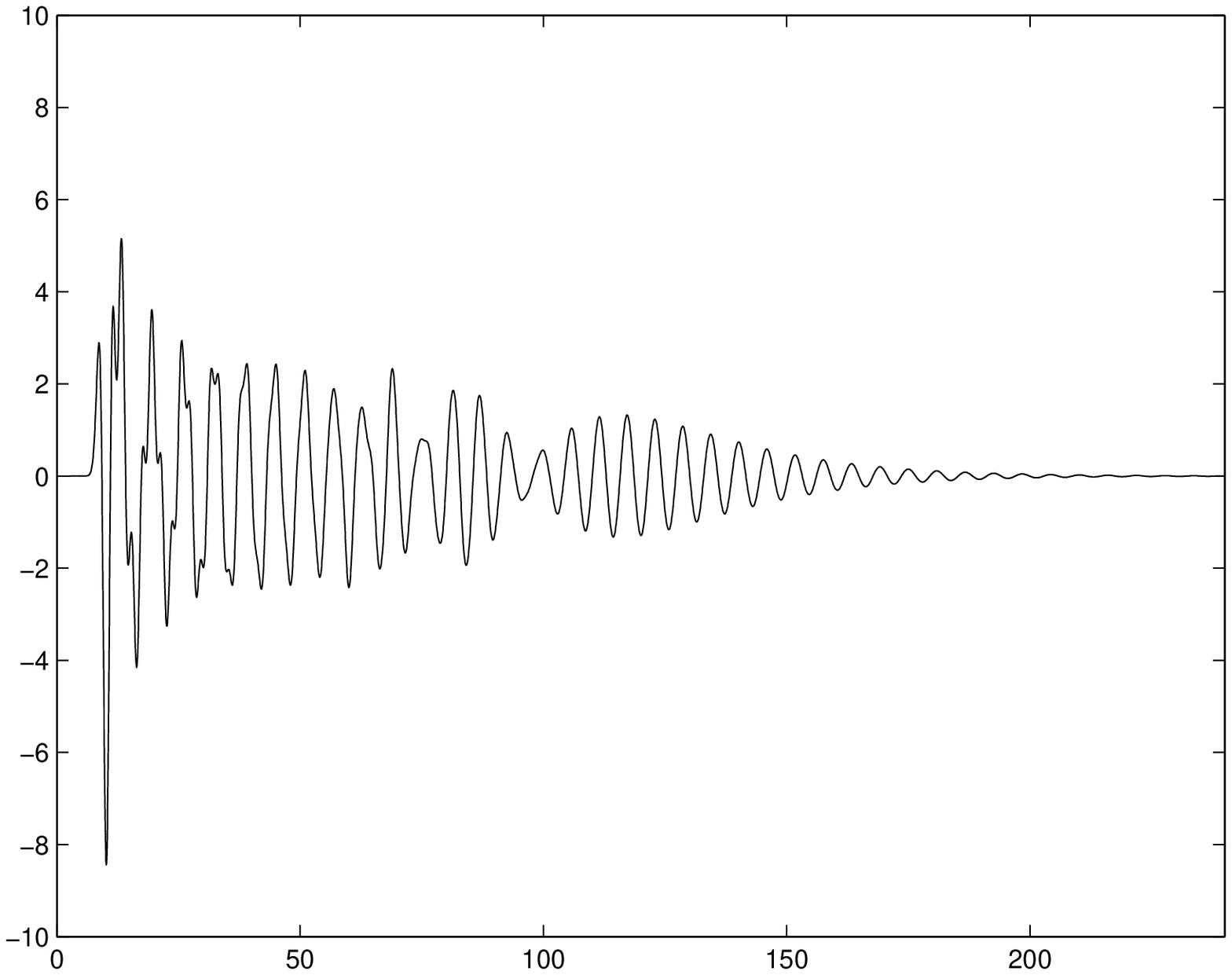}\hfill
\includegraphics[width=4.0cm]{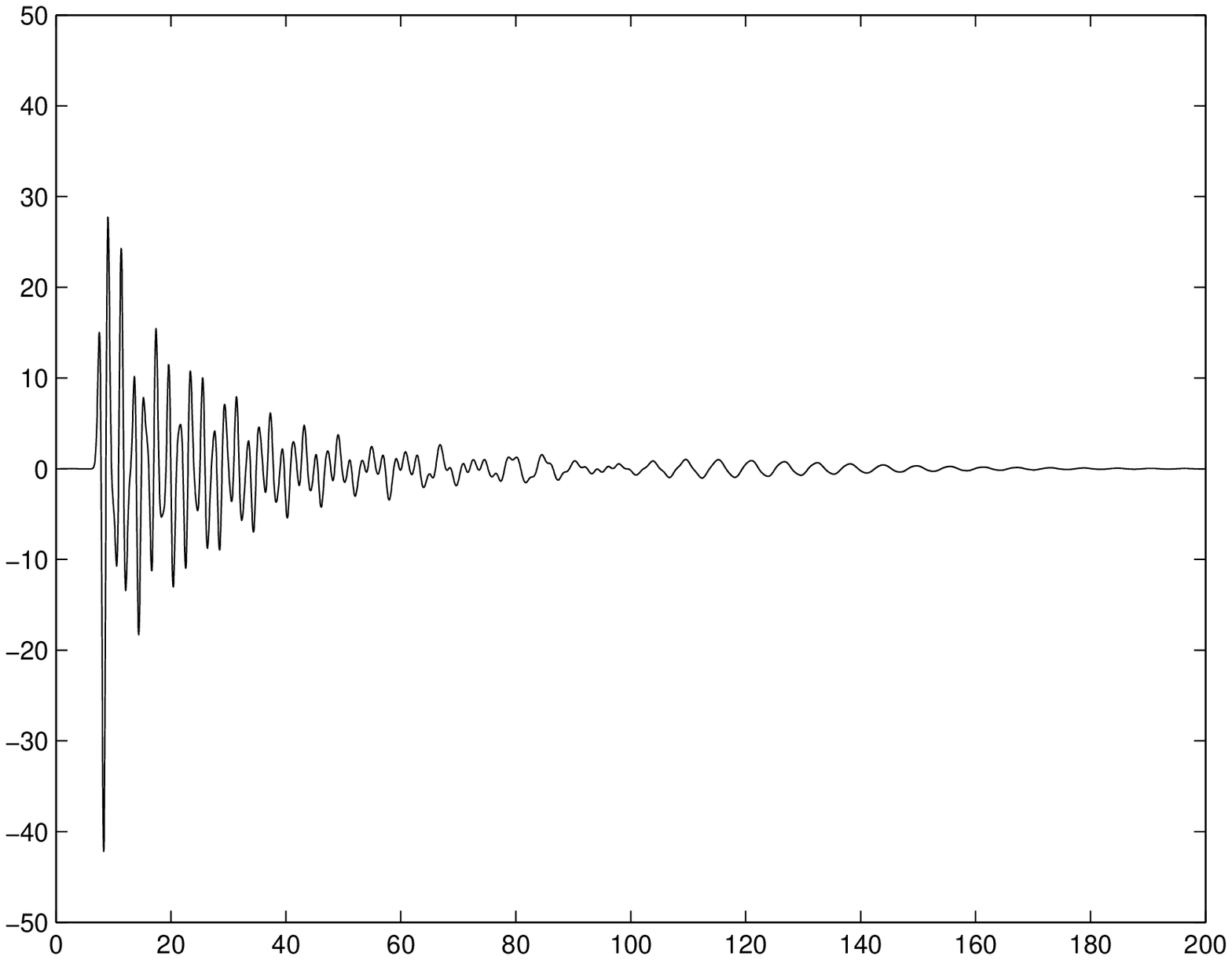}\hfill
\caption{Temporal records of ground response for $\nu_{0}=0.1$Hz
(left panel), $\nu_{0}=0.25$Hz (middle panel), $\nu_{0}=0.5$Hz
(right panel) input pulses
 in Mexico City-like
environment with layer thickness $h=90$m for
$\mathbf{x}^{s}=(0\text{m},100\text{m})$,
$\mathbf{x}=(3000\text{m},0\text{m})$.}
\label{f15}
\end{figure}
This is done (again in the Mexico City-like environment and for a
source that is 100m below the ground) in fig. \ref{f12} for a 20m
thick layer, in  fig. \ref{f13} for a 40m thick layer, in  fig.
\ref{f14} for a 60m thick layer, and in  fig. \ref{f15} for a 90m
thick layer. We observe quite different responses, varying from a
short pulse quite similar to the input pulse (for the thinnest
layer and the lowest frequency input pulse) to a very long
duration pulse (as much as 200s as compared to the the 4s duration
of the input pulse) with pronounced beating (for the thickest
layer and a medium frequency input pulse). Note that the 90m layer
also corresponds to the case in which the source is closest to the
layer (10m from the bottom face of the layer), which may also be a
factor contributing to the pronounced anomalous character of the
response in this configuration.

\begin{figure}
[ptb]
\includegraphics[width=6.0cm]{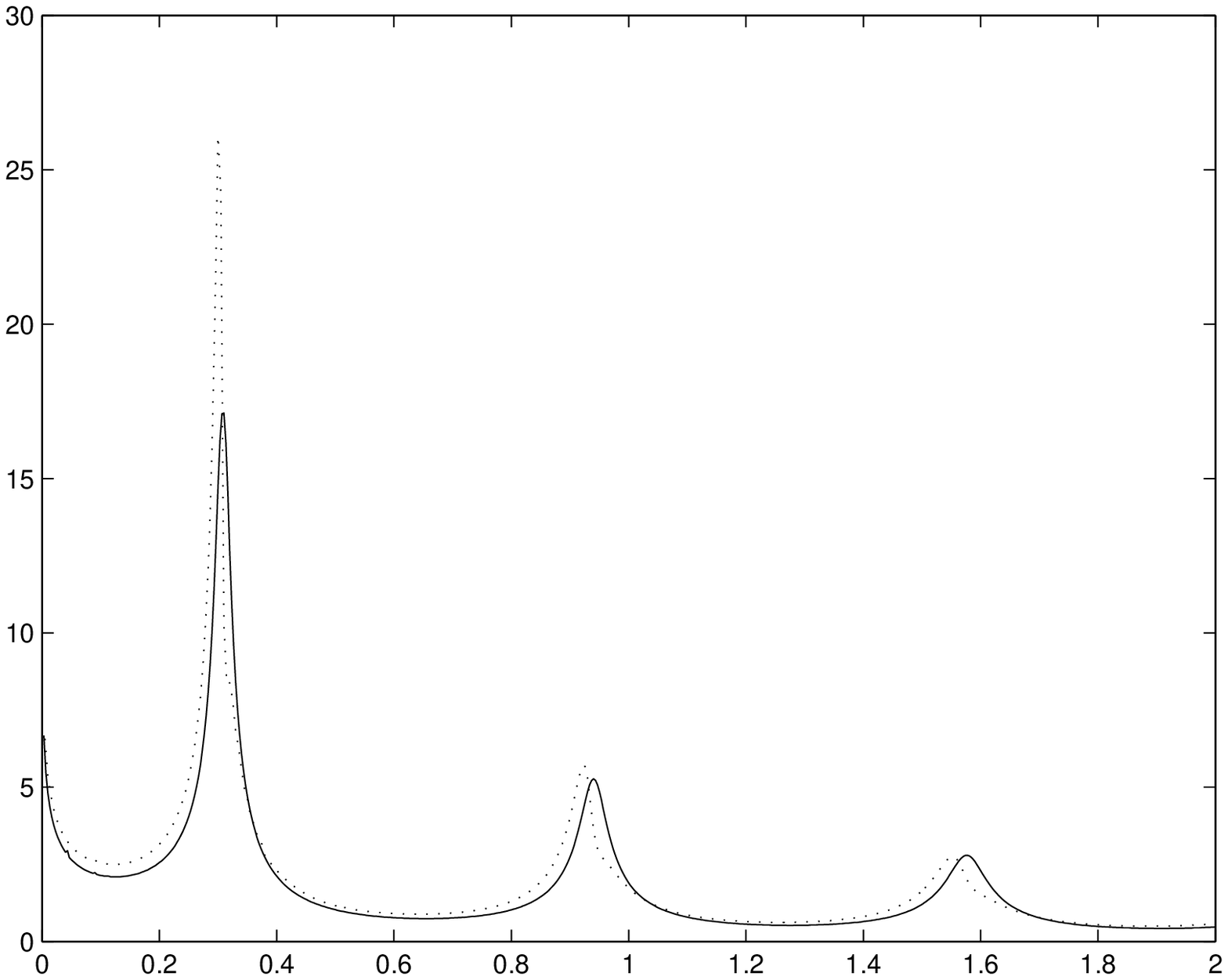}\hfill
\includegraphics[width=6.0cm]{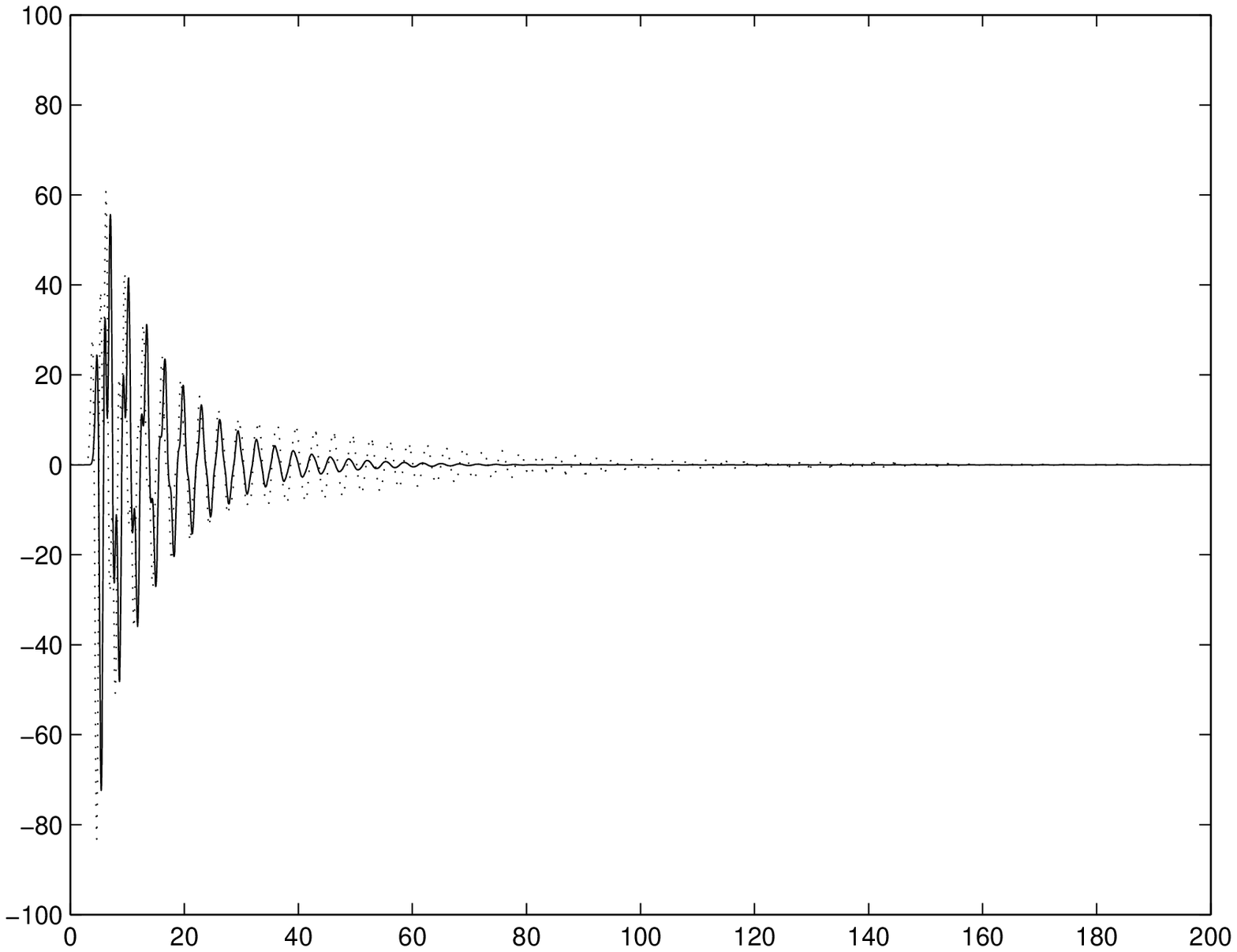}\hfill
\caption{Transfer functions (left panel) and temporal records (right panel)
of ground response to a $\nu_{0}=0.5$Hz input
pulse in Mexico City-like
with harder substratum environment for various source locations
and observation points
$\mathbf{x}^{s}=(0\text{m},3000\text{m})$,
$\mathbf{x}=(100\text{m},0\text{m})$(solid line curves),
$\mathbf{x}^{s}=(0\text{m},100\text{m})$, $\mathbf{x}=(3000\text{m},0\text{m})$.}
\label{f16}
\end{figure}

Finally, we consider a Mexico City-like environment with a
somewhat harder substratum for which $c^{0}=1500$m/s,
$\rho^{1}=1300$kg/m$^{3}$, $c^{1}=60$m/s (fig. \ref{f16}). Two
constant source-to-observation point distance situations are again
considered: a) $\mathbf{x}^{s}=(0\text{m},3000\text{m})$,
$\mathbf{x}=(100\text{m},0\text{m})$ (solid line curves in fig.
\ref{f16}), and b) $\mathbf{x}^{s}=(0\text{m},100\text{m})$,
$\mathbf{x}=(3000\text{m},0\text{m})$ (dotted line curves in fig.
\ref{f16}). For a deep source and large epicentral distance, the
frequency response (in the left hand panel of the figure) is
dominated, as usual, by the fundamental interference peak, and
this results in a relatively-long duration time domain signal
(solid line curve in the right hand panel of the figure) in
response to a $\nu_{0}=0.5$Hz input pulse. When the source is
shallow and the epicentral distance is large the frequency
response is dominated, unsurprizingly, by the fundamental Love
mode resonance peak, and this gives rise to a somewhat
longer-duration signal (dotted line curve in the right hand panel
of the figure) in response to the $\nu_{0}=0.5$Hz input pulse.
This example indicates that it may be difficult to distinguish
between the contributions of the SBW1 and SBW2 when the contrast
of the material properties between the layer and the substratum is
very large.
%
\subsection{Regional path effects}
We now examine the manner in which a wave radiated from a source
located underneath, but close to, the lower crustal boundary
propagates over long distances.

An  example  (real) of such motion, relative to the 11-3-02
(shallow) seismic Denali (Alaska) event recorded at a free-field
ground location (i.e., 1.5km from the buildings of the city of
Anchorage) at an epicentral distance of more than 275km is given
in fig. 10 of \cite{ce04}. A remarkable feature of this motion is
its long duration of over 125s. We will show that this is possible
with our simple model.

We constructed  a hopefully-plausible crustal model starting with
the parameters of a thin layer, softer-than-Nice like
configuration (for which $\rho_{0}=2000$kg/m$^{3}$,
$\rho_{1}=1300$kg/m$^{3}$, $c^{0}=600$m/s, $c^{1}=200$m/s,
$Q^{0}=\infty$, $Q^{1}=30$, $h=80$m), and by assuming conservation
of such quantities as $k^{1}h$, $\rho^{0}/\rho^{1}$, etc. in going
to a much thicker layer. Let us suppose that we have two
configurations, one of which is thin-layered and known
(configuration  with subscript 1), and the other is thick-layered
and unknown (configuration with subscript 2). The layer in the
known configuration is relatively soft and lossy, whereas it is
relatively hard (although always softer than the substratum) in
the unknown configuration. Since harder media are usually less
lossy, we assume rather arbitrarily that the $Q$ of the layer in
the unknown configuration is 20 times larger than the $Q$ in the
known configuration, while the $Q$'s of the substratum remain
infinite in both configurations. Thus, we have:
$Q^{0}_{2}=Q^{0}_{1}=\infty$ and $Q^{1}_{2}=20Q^{1}_{1}=600$.
Conservation of $k^{1}h$ means $k^{1}_{1}h_{1}=k^{1}_{2}h_{2}$,
whence $c^{1}_{2}=c^{1}_{1}\nu_{2}h_{2}/\nu_{1}h_{1}$, or, if we
choose $\nu_{1}=1$Hz, $\nu_{2}=0.08$Hz, and $h_{2}=10km$, then
$c_{2}=2000$m/s. Conservation of $\rho^{0}/\rho^{1}$ means
$\rho^{0}_{1}/\rho^{1}_{1}=\rho^{0}_{2}/\rho^{1}_{2}$, so that if
we choose $\rho^{0}_{2}=2600$kg/m$^{3}$ (close to the density of
granite), then $\rho^{1}_{2}=1690$kg/m$^{3}$. We would also like
to conserve wavespeed proportions, i.e.,
$c^{0}_{1}/c^{1}_{1}=c^{0}_{2}/c^{1}_{2}$, but this turns out to
give  wavespeeds in $M^{0}$ that are much larger than those for
granite (3200m/s) for the choice $c^{1}_{2}=2000$m/s, so that we
arbitrarily chose $c^{0}_{2}=3000$m/s (i.e., close to the
wavespeed in granite). Finally we chose to conserve the relative
distance of the source to the lower boundary of the layer, i.e.,
$(x_{2}^{s}-h)/x_{2}^{s}$, which gives $x_{2}^{s}=12$km.
\begin{figure}
[ptb]
\includegraphics[width=6.0cm]{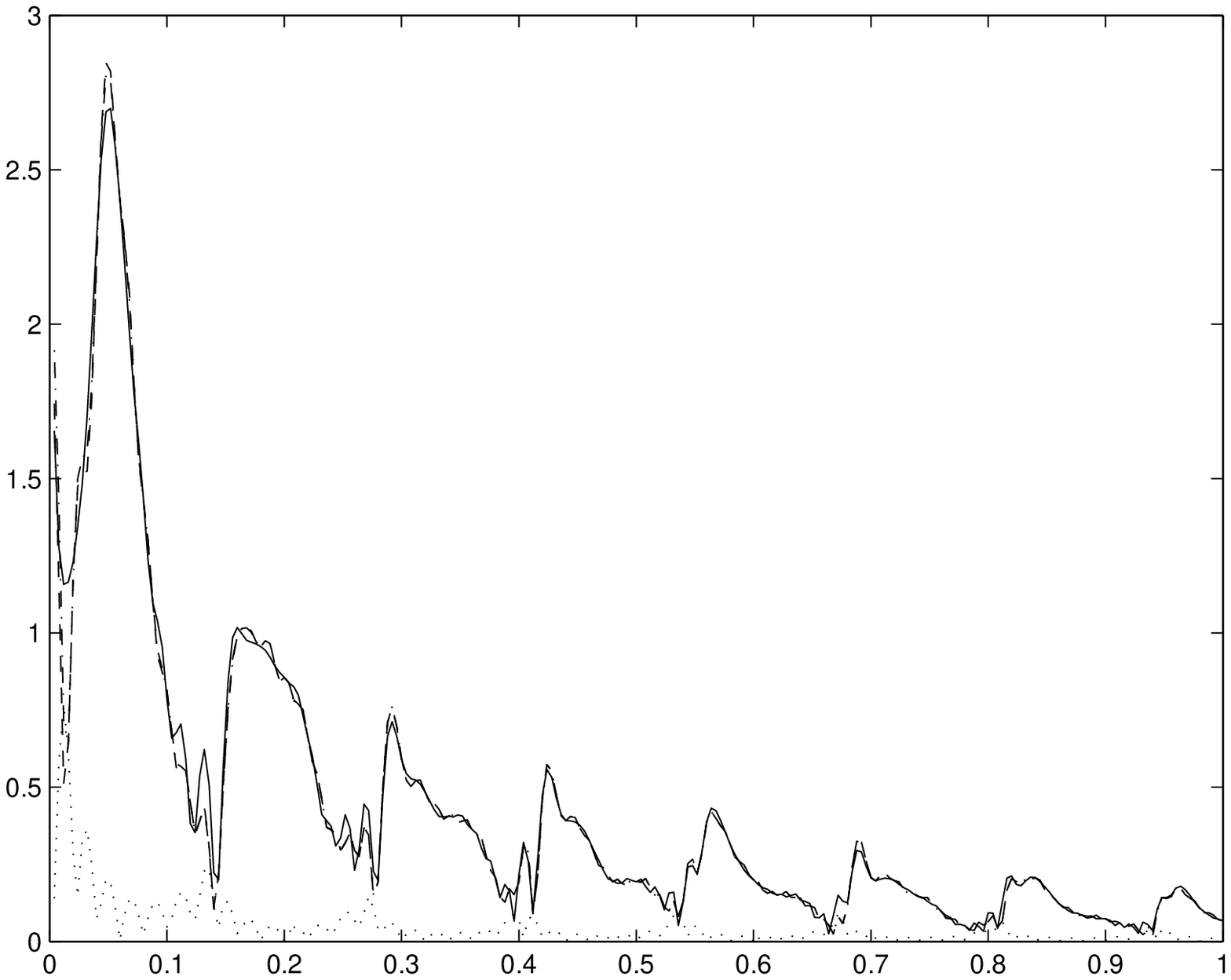}\hfill
\includegraphics[width=6.0cm]{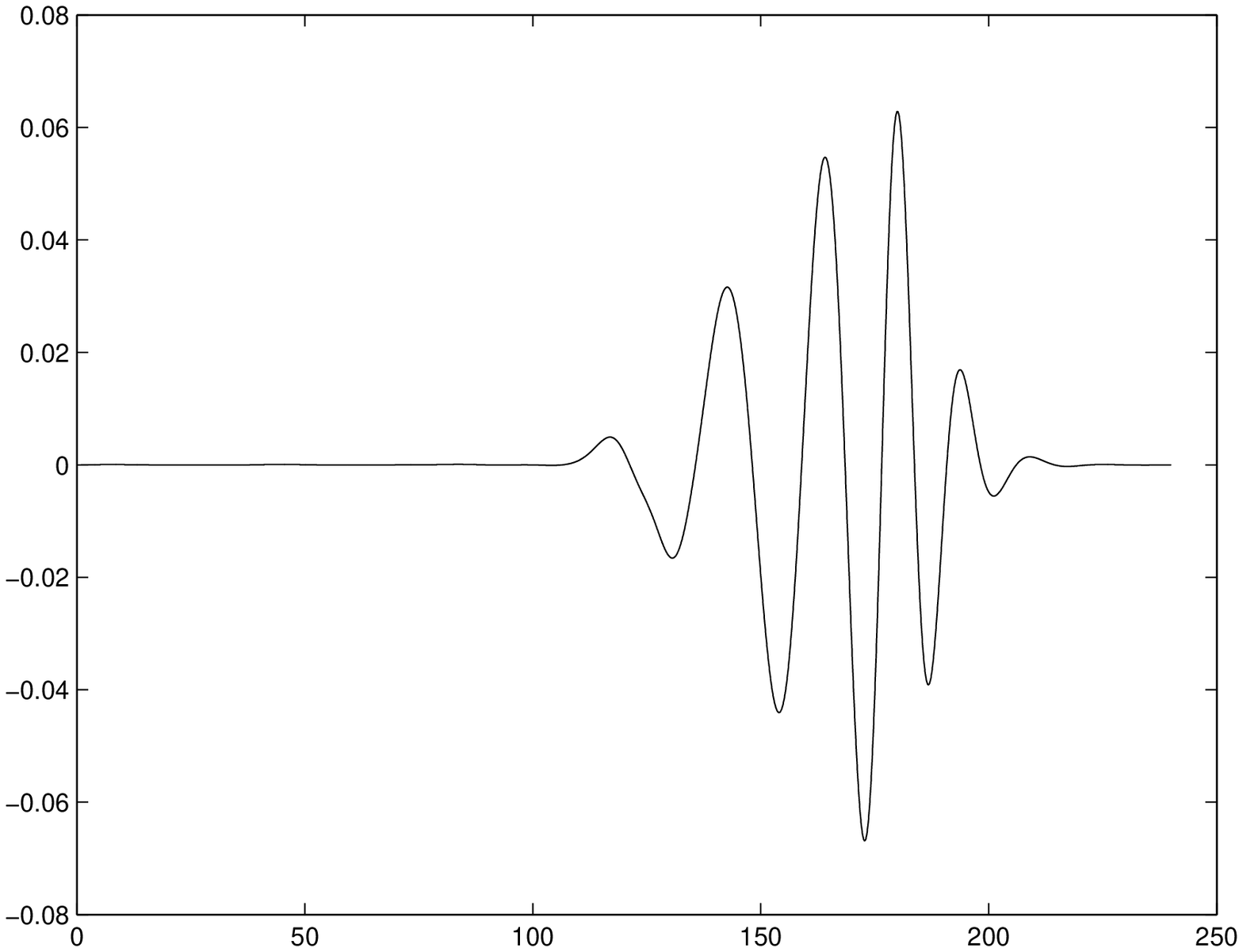}\hfill
\includegraphics[width=6.0cm]{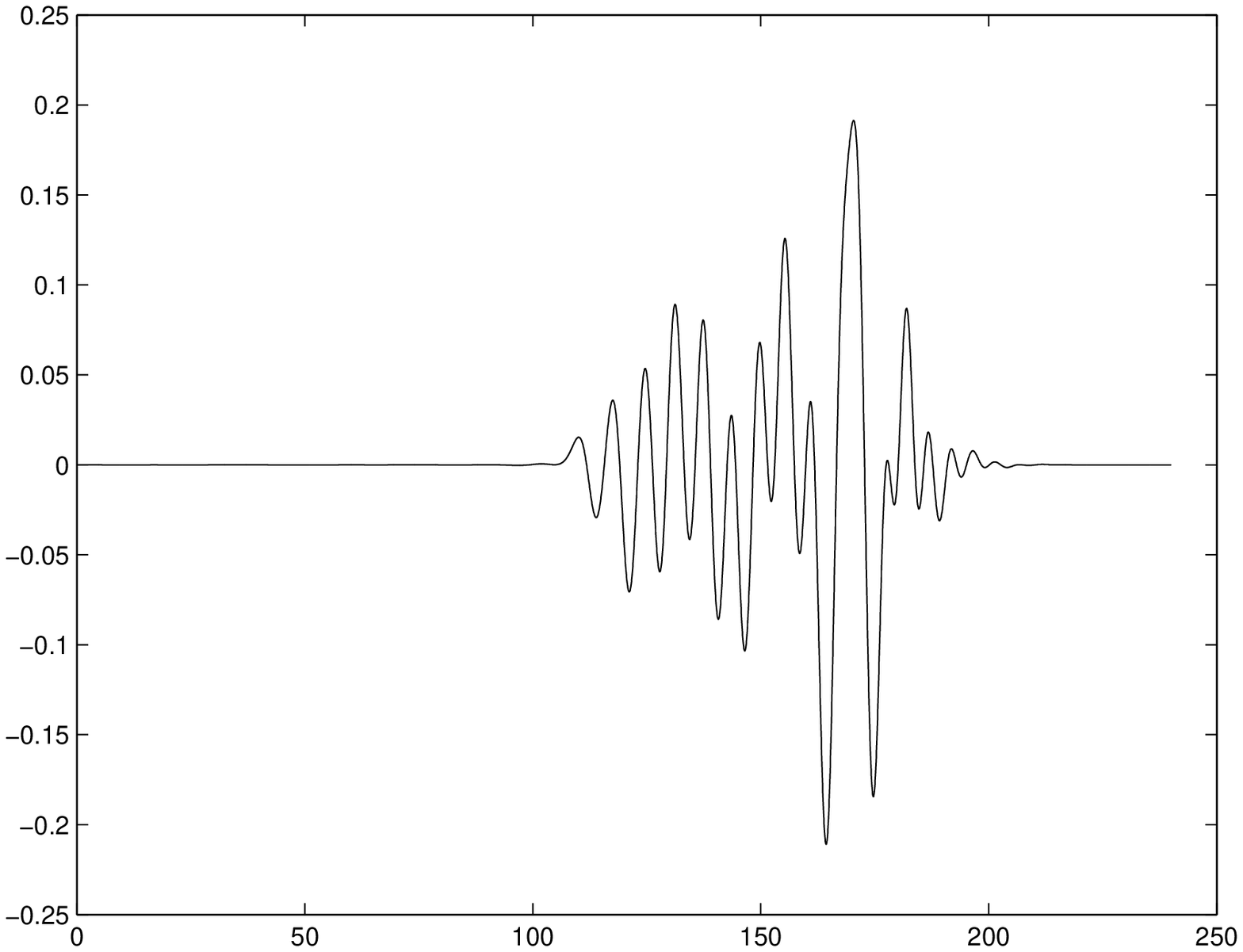}\hfill
\includegraphics[width=6.0cm]{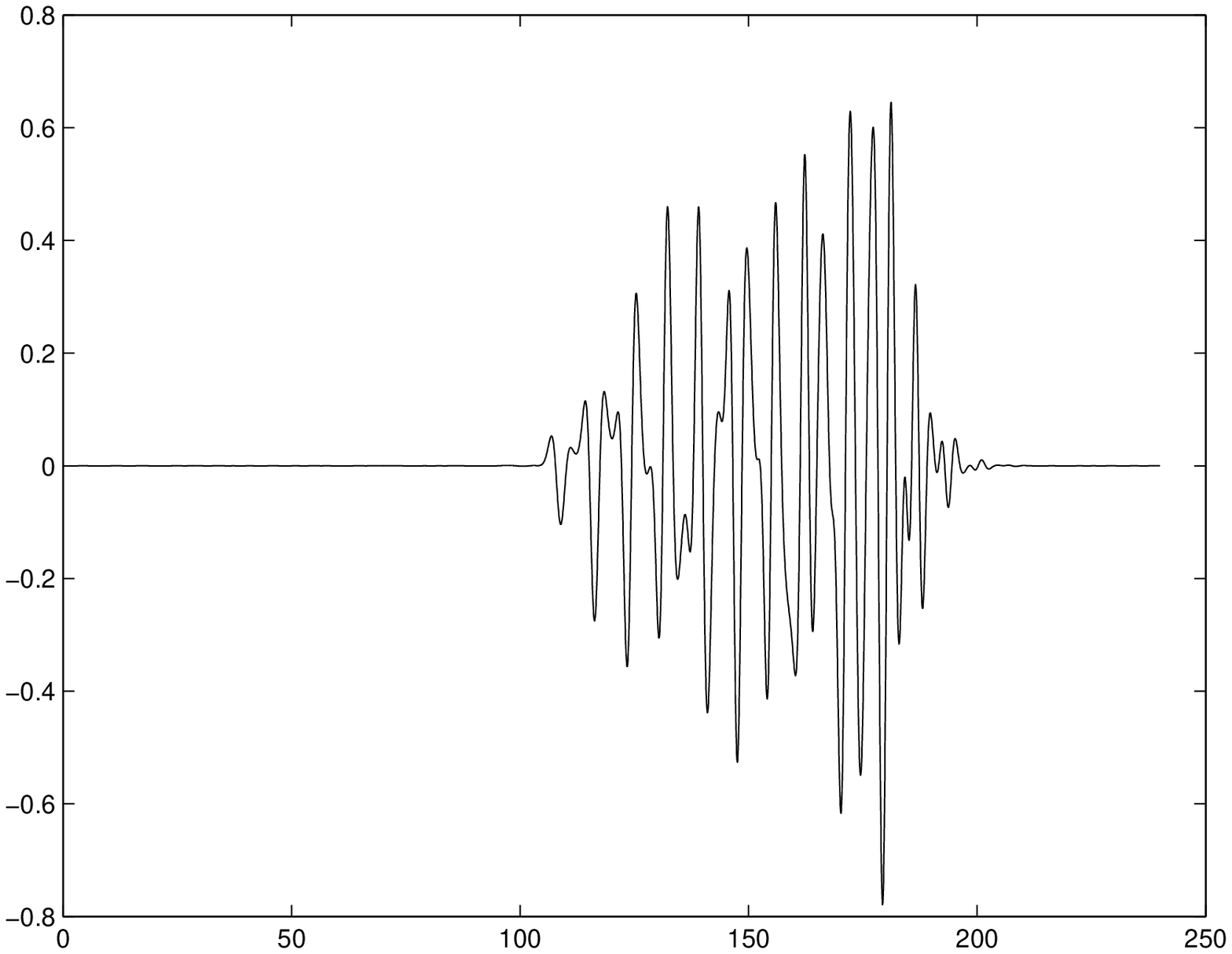}\hfill
\caption{Ground motion (displacement) at large epicentral distance
($x_{1}-x_{1}^{s}=300$km) in response to a pseudo-Ricker pulse
line source underneath, and close to (focal depth
$x^{s}_{2}=12$km), the lower boundary of a thick ($h=10$km),
fairly hard, crust overlying a granite-like substratum
($\rho_{0}=2600$kg/m$^{3}$, $\rho_{1}=1690$kg/m$^{3}$,
$c^{0}=3000$m/s, $c^{1}=2000$m/s, $Q^{0}=\infty$, $Q^{1}=600$,
$h=80$m). Transfer functions, with same notations as in fig.
\ref{f2} (upper left panel) and time histories for: a
$\nu_{0}=0.05$Hz input pulse (upper right panel), a
$\nu_{0}=0.1$Hz input pulse (lower left panel), a $\nu_{0}=0.2$Hz
input pulse (lower right panel).} \label{f17}
\end{figure}
The results for this new (crustal) configuration excited by the
usual pseudo-Ricker pulse line sources are given in fig. \ref{f17}
wherein it can be seen that the ground response far from the
epicenter (300km): 1) is dominated by the excitation of Love modes
(notably the fundamental), 2) takes the form of a pulse which has
 a shape not very different from that in fig. 10 of \cite{ce04} and is of
approximately 100s duration, governed essentially by the
fundamental Love resonance peak. Thus, our theoretical model shows
that it is quite possible for a source underneath, and relatively
close to, a fairly thick, fairly hard crust overlying a very hard
substratum to give rise to a rather long-duration pulse even at
large epicentral distances. What becomes of this pulse when a city
is located at this large lateral distance from the source
constitutes an important, and as yet not fully-elucidated,
question (this meaning, that although studies such as
\cite{fasu94} are designed to take into account all that occurs
between the distant source and the observation point in the basin,
the results that are offered are entirely of numerical nature and
therefore do not provide an explanation of the underlying physical
processes).
\section{Phenomenological  model  of  the  time-domain\\
ground response} In sect. 6 we mentioned the difficulties of
obtaining closed-form expressions of the integrals $I_{1}^{1}$,
$I_{2}^{1}$ and $I_{3}^{1}$ and therefore of the Fourier integral
(\ref{w35.10}) accounting for the time-domain ground response.
Nevertheless, the many numerical examples in sect. 7 of the
frequency-domain ground response all seem to have common features
which we shall attempt to describe in this section in
phenomenological manner. Moreover, this approach will be shown to
provide a simple means of understanding the origin of the main
features of the time domain response.

The principal features of the ground transfer function
$|u(\mathbf{x}_{g},\omega)/u^{i}(\mathbf{x}_{g},\omega)|$ appeared
to be due to interference and resonance causes and manifested
themselves by a series of well-defined, regularly-spaced bumps.
That the nature of these bumps be due either to interference or to
resonance causes is not of primal importance at the present
(phenomenological) level of analysis; the only aspects that
interest us now are the relative widths and heights of the bumps
(recall that, in general, the bumps associated with interferences
(SBW1) are broader and less intense than the corresponding bumps
associated with Love mode resonances (SBW2)).

We represent each of these bumps by a gaussian function of
frequency, which we multiply by the spectrum $S(\omega)$ of the
incident pulse and by an amplitude function
$\mathcal{A}(\mathbf{x}_{g},\mathbf{x}^{s},\omega)$ of
$\mathbf{x}_{g}$, $\mathbf{x}^{s}$ and $\omega$ to take into
account the fact (observed in the numerical results) that the
different bumps of frequency-domain response indeed depend on
these quantitites. Let $G_{l}(\omega)$ be the $l$-th gaussian
function of the form
\begin{equation}\label{w81.1}
  G_{l}(\omega)=\frac{1}{\sqrt{\pi\varepsilon_{l}}}
  e^{-\frac{(\omega-\omega_{l})^{2}}{\varepsilon_{l}}}~.
\end{equation}
The bump connected with this function attains its maximum at
$\omega=\omega_{l}$ and its finesse
is all the larger, the smaller is $\varepsilon_{l}$. In fact
\cite{cakr83} (p. 319), $G_{l}(\omega)$ tends towards the Dirac
delta distribution $\delta(\omega-\omega_{l})$ as
$\varepsilon_{l}\rightarrow 0$.

Thus, we represent the frequency-domain ground response by
\begin{equation}\label{w81.2}
 u^{1}(\mathbf{x}_{g},\omega)\approx \sum_{l=1}^{L}S(\omega)
 \mathcal{A}_{l}(\mathbf{x}_{g},\mathbf{x}^{s},\omega)
 G_{l}(\omega)~,
\end{equation}
wherein $\omega_{l+1}>\omega_{l}$ and $L$ may be a large positive
integer. However, the latter will be taken equal to 4 due to the
fact that we assume that the spectrum $S(\omega)$ of the input
pulse is not too broad and centered at low frequencies (the case
of interest in the applications considered herein). Introducing
(\ref{w81.2}) into (\ref{w35.10}) gives
\begin{equation}\label{w81.3}
 u^{1}(\mathbf{x}_{g},t)\approx 2\Re \sum_{l=1}^{4}\int_{0}^{\infty}S(\omega)
 \mathcal{A}_{l}(\mathbf{x}_{g},\mathbf{x}^{s},\omega)G_{l}(\omega)e^{-i\omega
 t}d\omega~.
\end{equation}
Although we don't know, nor assume, much about $S$ and
$\mathcal{A}_{l}$, it seems reasonable to suppose that these
functions are slowly-varying in comparison to $G_{l}(\omega)$ with
respect to $\omega$  in the neighborhoods $\omega_{l}$ in which
the gaussians are maximal. Consequently, we can make the
approximation
\begin{equation}\label{w81.4}
 u^{1}(\mathbf{x}_{g},t)\approx 2\Re \sum_{l=1}^{4}S(\omega_{l})
 \mathcal{A}_{l}(\mathbf{x}_{g},\mathbf{x}^{s},\omega_{l})\int_{0}^{\infty}G_{l}
 (\omega)e^{-i\omega t}d\omega~.
\end{equation}
By proceeding as in~~ \cite{cakr83}~~ (p.313) ~~and ~~making use
of the identity \cite{mofe53}
$\int_{0}^{\infty}\exp(-\xi^{2})d\xi=\sqrt{\pi}/2$, we find
\begin{equation}\label{w81.5}
 \int_{0}^{\infty}G_{l}
 (\omega)e^{-i\omega t}d\omega=e^{-\frac{\varepsilon_{l}}{4}t^{2}-i\omega_{l}
 t}~,
\end{equation}
so that
\begin{equation}\label{w81.6}
 u^{1}(\mathbf{x}_{g},t)\approx 2\Re \sum_{l=1}^{4}S(\omega_{l})
 \mathcal{A}_{l}(\mathbf{x}_{g},\mathbf{x}^{s},\omega_{l})
 e^{-\frac{\varepsilon_{l}}{4}t^{2}-i\omega_{l} t}~.
\end{equation}
By representing $S$ and $\mathcal{A}_{l}$ in polar form
\begin{equation}\label{w81.6}
 S(\omega_{l})=|S(\omega_{l})|e^{i\sigma_{l}(\omega_{l})}~~~,~~~
 \mathcal{A}_{l}(\mathbf{x}_{g},\mathbf{x}^{s},\omega_{l})=|
 \mathcal{A}_{l}(\mathbf{x}_{g},\mathbf{x}^{s},\omega_{l})|e^{i\alpha_{l}
 (\mathbf{x}_{g},\mathbf{x}^{s},\omega_{l})}~,
\end{equation}
we get
\begin{multline}\label{w81.6}
 u^{1}(\mathbf{x}_{g},t)\approx \Re \sum_{l=1}^{4}\mathcal{B}_{l}
 (\mathbf{x}_{g},\mathbf{x}^{s},\omega_{l})
 e^{-\frac{\varepsilon_{l}}{4}t^{2}-i\left [ \omega_{l}t-
 \beta_{l}(\mathbf{x}_{g},\mathbf{x}^{s},\omega_{l})\right ] }=
 \\
 \sum_{l=1}^{4}\mathcal{B}_{l}
 (\mathbf{x}_{g},\mathbf{x}^{s},\omega_{l})
 e^{-\frac{\varepsilon_{l}}{4}t^{2}}
\cos \left ( \omega_{l}t-\beta_{l}(\mathbf{x}_{g},\mathbf{x}{s},
\omega_{l})\right ) ~,
\end{multline}
wherein
\begin{multline}\label{w81.7}
 \mathcal{B}(\mathbf{x}_{g},,\mathbf{x}^{s},\omega_{l}):= 2|S(\omega_{l})|
 |\mathcal{A}_{l}(\mathbf{x}_{g},\mathbf{x}^{s},\omega_{l})|,
 \\
 \beta_{l}(\mathbf{x}_{g},\mathbf{x}^{s},\omega_{l}):=
 \alpha_{l}(\mathbf{x}_{g},\mathbf{x}^{s},\omega_{l})+\sigma(\omega_{l})~.
\end{multline}
With this in hand, we now try to account for the main features of
the numerical results pertaining to the time records of ground
response.

Assume that {\it one} of the bumps in the ground transfer function
dominates all others.  In the present paradigm, this signifies that one
of the terms, say the $m$-th in the sum
in (\ref{w81.6}) dominates all the others. As observed in the
numerical results, this term should account either for the
fundamental Love mode resonance ($m$=1) or for the fundamental
interference peak ($m$=2), the necessary condition for the Love peak to be
dominant being that the source is close to the bottom boundary of the layer,
and the necessary condition for the interference peak to be dominant being
that the source is far from the bottom boundary of the layer.
In either case, we have:
\begin{equation}\label{w81.8}
 u^{1}(\mathbf{x}_{g},t)\approx \mathcal{B}_{m}
 (\mathbf{x}_{g},\mathbf{x}^{s},\omega_{m})
 e^{-\frac{\varepsilon_{m}}{4}t^{2}}
\cos \left ( \omega_{m}t-\beta_{m}(\mathbf{x}_{g},\mathbf{x}^{s},
\omega_{m})\right ) ~,
\end{equation}
which is indicative of the existence of a monochromatic wave
(angular frequency $\omega_{m}$) whose amplitude $\mathcal{B}_{m}$
and phase $\beta_{m}$ vary with the positions of the source and
the observation point on the ground, and which is {\it
exponentially-attenuated with time}. This attenuation is more or
less pronounced, so that the duration of the signal is large if
$\varepsilon_{m}$ is small (i.e., the finesse of the corresponding
transfer function bump is large) as would occur for a Love mode
resonance, and is relatively small if $\varepsilon_{m}$ is
relatively large (i.e., the finesse of the corresponding transfer
function bump is relatively small) as would generally occur for an
interference peak. The same phenomenon is produced, although with
less intensity due to the lowering of $|\mathcal{B}_{m}|$, if the
spectrum of the incident pulse is such as to favorize either the
second Love mode resonance or the second interference peak rather
than the fundamental Love mode resonance or the fundamental
interference peak.

Consider a different type of situation in which {\it two} bumps in
the ground transfer function dominate all others. In the present
paradigm this means that two terms in (\ref{w81.6}), say the
$m$-th and $n$-th,  dominate all the others, i.e.,
\begin{multline}\label{w81.9}
 u^{1}(\mathbf{x}_{g},t)\approx \mathcal{B}_{m}
 (\mathbf{x}_{g},\mathbf{x}^{s},\omega_{m})
 e^{-\frac{\varepsilon_{m}}{4}t^{2}}
\cos \left ( \omega_{m}t-\beta_{m}(\mathbf{x}_{g},\mathbf{x}^{s},
\omega_{m})\right )+
\\
\mathcal{B}_{n}
 (\mathbf{x}_{g},\mathbf{x}^{s},\omega_{n})
 e^{-\frac{\varepsilon_{n}}{4}t^{2}}
\cos \left ( \omega_{n}t-\beta_{n}(\mathbf{x}_{g},\mathbf{x}^{s},
\omega_{n})\right ) ~.
\end{multline}
The numerical results in the previous section show that this
occurs only when the $m$-th term corresponds to a Love mode
resonance and the $n$-th term to the interference peak nearest the
Love mode resonance peak, so that $n=m+1$ in the present numbering
system. Moreover, the most pronounced phenomena were shown
numerically to be produced when the fundamental Love mode
resonance and fundamental interference peak are involved, so that
$m=1$, $n=2$ is the most interesting case.

We can write (e.g., for $m=1$, $n=2$)
\begin{multline}\label{w81.10}
 u^{1}(\mathbf{x}_{g},t)\approx
\mathcal{B}_{1}e^{-\frac{\varepsilon_{1}}{4}t^{2}}\mathcal{C}_{1}+
\mathcal{B}_{2}
e^{-\frac{\varepsilon_{2}}{4}t^{2}}\mathcal{C}_{2}=
\\
\left ( \mathcal{B}_{1}e^{-\frac{\varepsilon_{1}}{4}t^{2}}+
\mathcal{B}_{2}e^{-\frac{\varepsilon_{2}}{4}t^{2}}\right )
\frac{\mathcal{C}_{1}+ \mathcal{C}_{2}}{2}+ \left (
\mathcal{B}_{1}e^{-\frac{\varepsilon_{1}}{4}t^{2}}-\mathcal{B}_{2}
e^{-\frac{\varepsilon_{2}}{4}t^{2}}\right )
\frac{\mathcal{C}_{1}-\mathcal{C}_{2}}{2}~,
\end{multline}
so that
\begin{multline}\label{w81.11}
 u^{1}(\mathbf{x}_{g},t)\approx
\left ( \mathcal{B}_{1}e^{-\frac{\varepsilon_{1}}{4}t^{2}}+
\mathcal{B}_{2}e^{-\frac{\varepsilon_{2}}{4}t^{2}}\right ) \times
\\
\cos\left [
\frac{(\omega_{1}+\omega_{2})t}{2}-\frac{\beta_{1}+\beta_{2}}{2}\right
] \cos\left [
\frac{(\omega_{1}-\omega_{2})t}{2}-\frac{\beta_{1}-\beta_{2}}{2}\right
] -
\\
\left ( \mathcal{B}_{1}e^{-\frac{\varepsilon_{1}}{4}t^{2}}-
\mathcal{B}_{2}e^{-\frac{\varepsilon_{2}}{4}t^{2}}\right ) \times
\\
\sin\left [
\frac{(\omega_{1}+\omega_{2})t}{2}-\frac{\beta_{1}+\beta_{2}}{2}\right
] \sin\left [
\frac{(\omega_{1}-\omega_{2})t}{2}-\frac{\beta_{1}-\beta_{2}}{2}\right
] ~.
\end{multline}
An interesting case is: $ \mathcal{B}_{1}\approx \mathcal{B}_{2}$,
$\varepsilon_{1}\approx\varepsilon_{2}$, whence
\begin{multline}\label{w81.12}
 u^{1}(\mathbf{x}_{g},t)\approx
 \\
2\mathcal{B}_{1}e^{-\frac{\varepsilon_{1}}{4}t^{2}} \cos\left [
\frac{(\omega_{1}+\omega_{2})t}{2}-\frac{\beta_{1}+\beta_{2}}{2}\right
] \cos\left [
\frac{(\omega_{1}-\omega_{2})t}{2}-\frac{\beta_{1}-\beta_{2}}{2}\right
] ~,
\end{multline}
which is indicative of monochromatic sinusoidal motion of angular
frequency $\frac{\omega_{1}+\omega_{2}}{2}$, amplitude-modulated
by an attenuated sinusoid of frequency
$|\frac{\omega_{1}-\omega_{2}}{2}|$. In this case, the signal
associated with the ground motion exhibits the  beating and
attenuation observed in some of the computed results, with the
duration depending, as in the previous case, on the finesse of the
resonance and interference bumps.

Actually, it is not necessary for $ \mathcal{B}_{1}\approx
\mathcal{B}_{2}$, $\varepsilon_{1}\approx\varepsilon_{2}$ in order
to have beating in the signal, since although the motion
associated with  (\ref{w81.11}) is more complicated than that of
(\ref{w81.12}), in both cases a form of attenuated signal with
more or less regular beating is present. The signal with irregular
beating predicted by (\ref{w81.12}) is the case most commonly
observed in the computed results of the previous section.

Thus, the phenomenological model accounts for all of the features
of the computed time records: quasi-monochromatic attenuated
signals (whose duration is governed by the finesse of the
frequency domain bump) without beating when either the Love mode
resonance or interference peak is involved, an attenuated signal
with regular beating when the frequency domain bumps are of
comparable magnitude and finesse (the latter governing the
duration of the signal), an attenuated signal with irregular
beating when the magnitude and finesse of the fundamental Love and
interference bumps are rather different.
\section{Discussion}
We shall now attempt to provide answers to the questions raised in
sect. 1.

The first question was: is it possible to obtain anomalous
response without any lateral heterogeneity in the underground
medium? The configuration studied herein was laterally-
homogeneous. We have shown that 1D response only accounts for
interference effects (as embodied by the SBW1), but not for
coupling to Love modes (as embodied by the SBW2) in the layer,
which is particularly strong when the source is in the
neighborhood of the lower boundary of the layer. Insofar as
anomalous effects are essentially characterized by long duration
and beating phenomena in the signals (e.g., curves in the middle
and right panels of fig. \ref{f14}), the answer to this question
is negative as concerns 1D response. However, when the contrast of
material properties between the layer and substratum is very
large, it is possible to obtain fairly-long duration signals
(albeit without beating), which are essentially associated with 1D
response, even when the source is far from the lower boundary of
the layer (solid curve in right panel of fig. \ref{f16}).   More
generally, i.e., when coupling to Love modes is achieved, the
answer to the question is positive.

The second question was: what is the relation of 1D to 2D response
and how adequate is it to model the general response of the
configuration by its response to a (nearly) vertically-incident
plane wave? We have shown that not only does the 1D model not give
rise to resonance phenomena, but that truly-resonant phenomena
associated with the excitation of Love modes can only be described
by a fully 2D (or 3D) model. For a source far from the lower
boundary of the layer, the response is essentially due to the
contributions of the SBW1 (more or less equivalent to the 1D
response), but when the source is near this boundary, the waves
(SBW2) not included in the 1D model play a major role in the
overall response in that they either overwhelm the 1D response
(long duration response without beating) or combine with the 1D
response to produce signals with long duration and beating. This
finding should be taken into account in relation to studies (e.g.,
\cite{sabo04}) that attempt to predict seismic response of urban
sites from 1D type of analysis.

The third question was: how does the focal distance of the source
affect the response? The answer to this question was already
provided in the previous two paragraphs. However, it is opportune
to reconsider this question in the light of topic (b) concerning
the effects of underlying soil heterogeneities, lateral variations
of the underlying soil layer, and built environment on seismic
response in urban sites. One can show (\cite{wi02}) that a wave
incident on a heterogeneous medium gives rise to a diffracted wave
which can be considered to be radiated by {\it induced sources}
(as opposed to the {\it active source} associated with the primary
seismic disturbance) located within the medium. These induced
sources can also appear on the boundary of the medium (especially
at endpoints, corners and irregularities of the boundary), so that
the edges of a soft basin or the stress-free ground which includes
the buildings overlying a homogeneous soft layer in a city-like
site, can also constitute the locations of intense induced sources
in response to an incident seismic wave. The fields radiated by
all these induced sources can be represented in a manner identical
(provided the basic geometry of the configuration is similar) to
that of the present work, so that much of what was written and
found above, notably concerning the response to active sources
located outside, and in the vicinity of the soft layer (and, by
extension, to induced sources located {\it within or on the
boundaries of} the soft layer, should apply to city-like sites
built on soft layers or basins. The most important point
(mentioned in references such as \cite{tswi03}, implicit in
\cite{igja02}, \cite{jaig02}, and proven herein as concerns active
sources) is the following: the presence of these active or induced
sources, located {\it near or within} the soft layer overlying a
relatively-hard substratum, enables coupling to Love-type modes
which are responsible for  a part of the anomalous ground response
observed in cities such as Mexico, notably motion characterized by
long durations and beatings.

The fourth question was: how does the epicentral distance affect
the response? We have shown that the epicentral distance
($|x_{1}-x_{1}^{s}|$, for $\mathbf{x}_{g}=(x_{1},0$) is not as
sensitive as other parameters (especially the focal distance
$|0-y_{1}^{s}|$) as concerns its influence on duration and on the
presence or absence of beatings in the ground motion (it should be
mentioned that in \cite{shta04} the duration appears to be a
linearly-increasing function of epicentral distance, but the slope
of this function decreases when the crustal layer is softer).
However, the epicentral distance (more generally: the distance of
the source to the observation point) is a critical factor in
determining the {\it intensity} of the signal on the ground (see,
e.g., figs. \ref{f10}-\ref{f10a}).

The fifth question was: how does the contrast of mechanical
properties between the layer and the half space affect the ground
response? There does not appear to exist a clear-cut answer to
this question (see, however, \cite{shta04} in which it appears
rather systematically that softer layers lead to longer durations
and lower peak response for a given epicentral distance), since
the dependence on the mechanical parameters is very much
intermingled with that on the values of the geometrical and source
parameters. On the whole, most of the answers to the previous four
questions hold in a qualitative sense whatever the contrast of
mechanical properties (see, e.g., the results herein for Nice,
softer Nice, Mexico, Mexico with harder substratum), although
there are some quantitative differences (e.g., the contrast
influences considerably the intensity of the SBW1 contribution).
Naturally, the most remarkable features of the ground response of
our simple configuration, which are due to the excitation of the
fundamental Love mode, can only be observed if the layer is softer
than the substratum in the sense of (\ref{w48.2}) and
(\ref{w48.2a}). These conditions are so broad and widespread (for
city-like sites) as to render the 'anomalous' response described
herein a quite universal phenomenon.

The sixth question was: how does the thickness of the layer affect
the response? In fig. \ref{f6} it was found that increasing the
layer thickness increases the number of peaks, as well as the
finesse of each of the latter, in a given range of frequencies of
the transfer function. This has the effect of lowering the
frequency of occurrence of the first peak so that the time-domain
response will be largely conditioned by the spectrum of the input
pulse, assuming the latter to be centered at a relatively high
frequency.  Thus, a low frequency pulse can produce substantially
the same type of response for a thick layer as a relatively high
frequency pulse in a thin layer. This point is important in
connection with the topic of regional path effects mentioned in
sect. 1.

The seventh question was: how do the spectral characteristics of
the incident pulse affect the response? The answer to this
question can be found by comparing the three subfigures in any one
of figs. \ref{f12}-\ref{f15}. Obviously, the spectrum of the
incident pulse is a key factor (see sect. 8), which: a) if it
overlaps either a constructive interference peak or Love mode
peak, gives rise to attenuated, quasi monochromatic response,
often of long duration (see figs. \ref{f12}-\ref{f15} in which an
example is given of a pulse having a duration of 4s that gives
rise to substantial ground response of 200s duration), b) if it
overlaps both a constructive interference peak and Love mode peak,
gives rise to attenuated, quasi monochromatic  response with more
or less regular beatings, c) if it doesn't overlap significantly
either a constructive interference of Love mode peak, gives rise
to a time domain response that can be qualitatively quite similar
to the input signal (see, e.g., left panel of fig. \ref{f13}).
When the sources are induced, their spectra will be modified with
respect to that of the spectrum of the primary active source due
to diffraction and dispersion, so that an a priori unfavorable
situation for anomalous response from the point of view of the
primary active source may turn out to be favorable from the point
of view of the induced sources.
\section{Conclusion and perspectives}
This work originated in the observation that no satisfactory
physical explanation has been given until now of anomalous seismic
response in urban environments with soft layers or basins
overlying a hard substratum. The principal reason for this
knowledge gap probably lies in the complexity of the sites
examined in previous (essentially-numerical) studies: 1) a
homogeneous or multilayered basin of complicated form not
including buildings (e.g., \cite{sedu00}, \cite{fasu94},
\cite{ol00}), 2) a homogeneous layer overlain by a periodic or
non-periodic set of blocks or buildings (e.g., \cite{tswi03},
\cite{grts04}). The choice was therefore made herein to simplify
as much as possible the characteristics of the site and
sollicitation, while retaining as many as possible of their
essential features. Thus, it was thought that: i) the problem had
to be treated at least as a 2D one, ii) the sollicitation should
not be a plane wave (for which coupling to Love modes is
impossible in the chosen configuration) but rather the wave
radiated by a source which could be as simple as a line source (
this source eventually being able to mimick induced sources in
more complicated configurations), iii) the soft component of the
site could be a layer (rather than a basin) with flat, horizontal
boundaries (i.e., flat rather than irregular ground, as rendered
by the presence of buildings, flat interface with the substratum,
rather than curved or irregular as for a basin or irregular
layer).

In spite of the simplicity of the model, obtaining an explanation
of the principal features of the seismic response turned out not
to be straightforward. The temporal response took the form of a
Fourier transform (with respect to frequency) of a frequency
domain response function which itself is an integral with respect
to the horizontal wavenumber component of plane body and
heterogeneous waves. It was shown that the wavenumber integral
splits quite naturally into three parts comprising either standing
body waves in the layer (SBW1) and propagating body waves (BW) in
the substratum, standing body waves in the layer (SBW2) and
surface waves (SW2) in the substratum, standing surface waves in
the layer (SSW) and surface waves in the substratum (SW3). It
turns out that the amplitudes of the SW and SBW2 diminish
exponentially as the vertical distance of the source to the lower
boundary of the layer increases so as to make the contribution of
the SBW1 preponderant for sources with large focal depths. This
fact provides an explanation of the relative success of the 1D
model (a remote source radiates a wave that has practically all
the attributes of a plane wave when arriving on the layer) for
remote sources, but also of the reason why the 1D model is
inappropriate for active (and, by extension, induced) sources that
are close to (and, by extension, within) the layer, since the SBW2
waves are not included in this model. It was shown that each
SBW1/BW pair is the principal ingredient of the 1D model and that
the maxima of the amplitudes of the SBW1 do not correspond to
resonances, but are rather the result of the constructive
interference of standing waves in the layer. Each SBW2/SW2 pair
turned out to be a Love mode when the frequency satisfies the Love
mode dispersion relation. The amplitudes of each SSW/SW3 pair were
found to be negligible compared to those of the SBW1/BW and
SBW2/SW2 pairs.

The theoretical analysis reached its limits when the horizontal
wavenumber integration was attempted. Thus, the integrals
appearing in the frequency-domain response were carried out
numerically and a parametric study was made of the cumulative
contributions of the SBW1 and SBW2. It was shown, as expected,
that the SBW1 give the preponderant contribution for remote
sources, while both the SBW1 and SBW2 cumulative contributions can
be significant for nearby sources. The interference nature of the
amplitudes of the individual SBW1 was shown to be maintained in
the frequency-domain cumulative response of these waves. The
resonant nature of the amplitudes of the individual SBW2 was shown
to be maintained in the frequency-domain cumulative response of
these waves. However, it was not possible to obtain mathematical
expressions for the integrals of the frequency-domain responses
appearing in the global time-domain ground response.

We also showed that it  possible for a source, underneath, and
relatively close to, a fairly thick (10km), fairly hard crust
overlying a very hard substratum, to give rise to a rather
long-duration pulse even at large (e.g., 300km) epicentral
distances, and that this finding is in agreement with what has
been observed in connection with real earthquakes (see, e.g.,
\cite{ce04},\cite{fasu94}). We did not carry out an extensive
analysis of this finding, nor address the issue of what becomes of
this pulse when it enters an urban center located at large lateral
distances from the source (as was done numerically in works such
as \cite{fasu94},\cite{fapa94},\cite{paro00a},\cite{paro00b}).

In the last section of this investigation, a phenomenological
model was introduced based on the observation that the
frequency-domain cumulative response components of both the SBW1
and SBW2 appear as a series of regularly-spaced bumps which were
modeled as gaussians. This enabled a closed form expression of the
integral of the frequency-domain responses to be obtained which
revealed and accounted for the type of time-domain response
obtained by purely numerical means, notably, its attenuated, quasi
monochromatic character, with regular or irregular beatings,
governed by the finesse and relative position of the frequency
domain bumps. This type of response has often been observed in
earthquake-prone cities built on soft soil, so that it may be that
some of the causal agents inherent in our simple model are
operative in more complicated sites.

A question that naturally arises is whether the type of analysis
carried out herein can be extended to more realistic
configurations in which induced sources are likely to play a major
role. Our feeling is that this can be done provided some clever
approximations are made in the expressions for the response of
these configurations.

Another question (alluded-to in one of the previous paragraphs) is
that of regional path effects on global response in cities such as
Mexico subject to earthquakes arising from laterally-remote
sources. This very important theoretical issue, already considered
in works such as \cite{robi96}, will have to be treated in more
depth, first in the manner of the present contribution, to examine
how the wave radiated by the source reaches the city site, what
the nature of the waves are when they arrive in the city, and how
they are converted therein into the form they have been observed
to take (quasi Love or Rayleigh waves giving rise to high
intensity, extremely long (even longer than what was found herein)
duration ground motion, accompanied by beatings).

Most of the extensions of the present work will have to be carried
out first in the 2D, shear horizontal wave context in order to
discern the essential issues. The extensions to  the 2D- P/SV (as
in e.g., \cite{fasu94}) and 3D (as in e.g., \cite{ol00}) cases
with more general types of sources \cite{fasu94}, \cite{fapa94}
are, of course, the requisites for a full understanding of what
happens when a seismic wave hits a realistic urban site.
\section*{Acknowledgements}
This research was carried out within the framework of the Action
Concertée Incitative "Prévention des Catastrophes Naturelles"
entitled "Interaction 'site-ville' et aléa sismique en milieu
urbain" of the French Ministry of Research.
\bibliographystyle{unsrt}
\bibliography{gji_180108_18}
\end{document}